\documentclass[usenatbib]{mnras}

\usepackage[T1]{fontenc}

\DeclareRobustCommand{\VAN}[3]{#2}
\let\VANthebibliography\thebibliography
\def\thebibliography{\DeclareRobustCommand{\VAN}[3]{##3}\VANthebibliography}

\usepackage{graphicx}	% Including figure files
\usepackage{amsmath}	% Advanced maths commands
\usepackage{amssymb}	% Extra maths symbols
\usepackage{hyperref} %When citations are broken [draft]
\usepackage{pdflscape} % 
\usepackage{newtxtext,newtxmath}
\title[Chemical abundances in the Palomar Survey]{Chemical abundances in the nuclear region of nearby galaxies from the Palomar Survey}

\author[B. P\'{e}rez-D\'{\i}az et al.]{
B. P\'{e}rez-D\'{\i}az,$^{1}$\thanks{E-mail: bperez@iaa.es}
J. Masegosa,$^{1}$
I. M\'{a}rquez,$^{1}$
and E. P\'{e}rez-Montero$^{1}$
\\
$^{1}$Instituto de Astrof\'{\i}sica de Andaluc\'{\i}a (IAA-CSIC), Glorieta de la Astronom\'{\i}a s/n, 18008 Granada
}

\date{Accepted XXX. Received YYY; in original form ZZZ}

\pubyear{2021}
\begin{document}
\label{firstpage}
\pagerange{\pageref{firstpage}--\pageref{lastpage}}
\maketitle

\begin{abstract}
We estimate chemical abundances and ionization parameters in the nuclear region of a sample of 143 galaxies
from the Palomar Spectroscopic Survey, composed by Star-Forming Galaxies (87), Seyferts 2 (16) and LINERs (40) using the \textsc{Hii-Chi-mistry} code. We also study for each spectral type the correlation of the derived quantities with other different  properties of the host galaxies, such as morphology, stellar mass, luminosity and mass of their Supermassive Black Holes. The results obtained for Star-Forming Galaxies are used to check the soundness of our methodology. Then, we replicate a similar study for our sample of AGN, distinguishing between Seyferts 2 and LINERs. We report a saturation of Oxygen abundances for the nuclear regions of SFG. The correlations between chemical abundances and their host galaxy properties for SFG are in good agreement with previous studies. We find that Seyferts 2 present slightly higher chemical abundances but this result must be reexamined in larger samples of Seyfert galaxies. In contrast, we obtain lower chemical abundances for LINERs than for SFG. We confirm these relatively lower abundances for another sample of infrared luminous LINERs  in the same stellar mass range. Our analysis of AGNs (both LINERs and Seyferts) shows that their host galaxy properties are not correlated with our estimated chemical abundances.
\end{abstract}


%%%%%%%%%%%%%%%%%%%%%%%%%%%%%%%%%%%%%%%%%%%%%%%%%%

%%%%%%%%%%%%%%%%% BODY OF PAPER %%%%%%%%%%%%%%%%%%

\section{Introduction}
\label{sec1}
Gas-phase metallicity (Z) in different regions of galaxies is key for the study of their evolution: the enrichment of the interstellar medium (ISM) is caused by the metals formed in the cores of stars, that are lately driven to their surfaces by convective flows and finally ejected into the ISM by stellar winds and supernovae.

As a proxy for the gas-phase Z, most studies use the abundances of elements such as Oxygen or Nitrogen relative to Hydrogen. The most widely used chemical abundance ratio is $12+\log \left( O/H \right) $, since Oxygen is the most abundant metal in mass in gas-phase of the ISM \citep{Asplund_2009, Maiolino_2019} and because it presents very prominent lines in optical spectra emitted by the ionized gas, easily detectable in bright objects \citep{Osterbrock_book, Kewley_2019}. In addition, the chemical abundance ratio $\log \left( N/O \right) $ can be also used \citep{Edmunds_1978, Garnett_1990, Thuan_1995, Zee_1998}  since it provides additional information about the process of metal enrichment: distinguishing between a primary origin, if the metals are produced due to the helium burning in stars; and, a secondary origin, if the metals are already present during the star formation in the ISM.

For decades, the metallicity and the above chemical abundance ratios have been estimated in samples of star-forming galaxies (SFG), by using their emission line ratios to calculate the abundances of the different elements that originate them. Many techniques were developed for such studies, such as the use of the $T_{e}$-method, photoionization models or optical calibrations based on strong emission lines (see the review by \citealt{Maiolino_2019} for a more detailed discussion). By using these techniques, some studies show a correlation between both $12+\log \left( O/H \right) $ and $\log \left( N/O \right) $ \citep{Vila-Costas_1993, Masegosa_1994, Andrews_2013}. On the other hand, it is also important how the different properties of the host galaxies correlate with their chemical abundances. For instance, preliminary studies of the Local Group of galaxies \citep{McClure_1968, Lequeux_1979} revealed the existence of a relation between the luminosity of a SFG and its metallicity, the so-called \textit{luminosity-metallicity relation} (the most luminous galaxies have the higher chemical abundances), which has been later studied in further detail \citep{Garnett_1987, Skillman_1989, Brodie_1991, Vilchez_1995, Mateo_1998}. Later on it was probed that the luminosity-metallicity relation was in fact the result of a more fundamental \textit{mass-metallicity relation} \citep{Garnett_2002, PG_2003, Pilyugin_2004}. This relation has been probed in large sample of SFG both at low-redshift \citep{Contini_2002, Melbourne_2002, Lamareille_2004, Tremonti_2004} and at high-redshift \citep{Erb_2006, Perez-Montero_2009b, Perez-Montero_2013, Izotov_2015, Gao_2018, Torrey_2019}. In addition, it has been reported that the star-formation rate is also related to gas-phase Z in SFG \citep{Mannucci_2010, Lara-Lopez_2010, Yates_2012}. Morphological type may also be related with the chemical abundances, as early-type galaxies, despite some of them can present episodes of on-going star-formation \citep{Zhu_2010a}, tend to have on average older stellar populations than late-type galaxies \citep{Kennicutt_1998, Gebhardt_2003, Trager_2000, Schiavon_2007}, being thus more chemically evolved. 

In the last decades, similar studies have been performed on Active Galactic Nuclei (AGN), analyzing the metallicity of the Narrow Line Region (NLR), which is photoionized by the accretion disk surrounding the Supermassive Black Hole \citep{Shuder_1981, Alloin_2006, Nagao_2006, Dors_2019, Thomas_2019, Perez-Montero_2019}. The study of gas-phase Z in AGN is based on the same methodologies applied for SFG, but considering a different source of photoionization, i.e., accretion disks in AGN and hot massive stars in SFG \citep{Storchi_1998, Dors_2015, Castro_2017, Dors_2019, Thomas_2019, Carvalho_2020, Dors_2020, Flury_2020}. However, these studies have shown that the $T_{e}$-method, which requires the measurement of faint emission lines such as [O\textsc{iii}]$\lambda$4363\r{A}, can lead to an underestimation ($\sim 0.8$ dex) of their chemical abundances \citep{Dors_2015, Dors_2019, Maiolino_2019, Dors_2020}. One important limitation of many of these models is that they assume a relation between the chemical abundance ratios $12+\log \left( O/H \right) $ and $\log \left( N/O \right) $ which is measured in SFG, what can lead to non-negligible deviations in the derivation of the total oxygen abundance in AGN as based on nitrogen lines, such as [N\textsc{ii}] $\lambda$6584\r{A} \citep{Perez-Montero_2009}. Moreover, all of these studies are focused on the analysis of the NLR of high ionization AGN (Seyferts 2), being remarkable the absence of low-ionization AGN: only the study by \citet{Storchi_1998} presents the estimation of the metallicity in 4 LINERs. Recently, \citet{Flury_2020} published a study of chemical abundances in AGN. However, rather than estimating chemical abundances in the sample of LINERs, they extrapolate their calibration obtained for Seyferts 2 to the sample of low-luminosity AGN. The work by \citet{Thomas_2019} indicates that the mass-metallicity relation is also obtained in Seyferts 2, but  more flattened than for SFGs, and, there is still a lack of these studies for LINERs. Moreover, other integrated properties of the host galaxies as morphology, luminosity or mass of the supermassive black hole must be explored in both types of AGN to better check their influence on the chemical abundances derived in the central region of these galaxies.

In this work we perform a study through photoionization models of Z in the nuclear region of galaxies in the Palomar Spectroscopic Survey \citep{Ho_II_1995}, composed by nearby SFG, Seyferts 2 and LINERs. This survey was studied in great detail by \citeauthor{Ho_II_1995} \citeyearpar{Ho_II_1995, Ho_III_1997, Ho_IV_1997, Ho_V_1997}, including the presence of AGN and the environment of their sample galaxies. Although there are larger spectroscopic surveys such as SDSS or MaNGA, they cover a larger region of the galaxy. The median redshift of $z \sim 0.104$ and the aperture of 3'' used in SDSS \citep{Strauss_2002} imply that we are analyzing the spectra from regions of $\sim 2.9$ kpc. Despite MaNGA improving this value to 1.3 kpc \citep{Manga_2017}, these larger samples of galaxies are tracing a bigger nuclear region. Regarding LINERs, whose emission is weaker \citep{Marquez_2017}, these implies that the contribution from the host galaxy plays a mayor role in the spectra retrieved, which could lead to estimations of chemical abundances more related to the host galaxy than the NLR itself. To avoid this problem, we use the Palomar Spectroscopic Survey range of the nuclear radius between  6.8 pc and 1055.0 pc.

By using this sample of nuclear regions, we have also the information of several physical properties of their host galaxies, allowing us to carry out a detailed analysis of $12+\log \left( O/H \right) $ and $\log \left( N/O \right) $ for both SFG and the NLR of AGN. We use the code \textsc{Hii-Chi-mistry} (hereafter \textsc{HCm}), developed by E. P\'{e}rez-Montero \citeyearpar{Perez-Montero_2014, Perez-Montero_2019} and based on photoionization models, to estimate chemical abundances in SFG and AGN. The subsample of star-forming galaxies allows us to check the soundness of our methodology. Then, we carry out our study in Seyferts 2 and, for the first time in this kind of studies, in LINERs. We also perform the same analysis in an additional sample of luminous infrared LINERs from \citet{Povic_2016}.

This paper is organized as follows. In Sec. \ref{sec2} we describe the sample of galaxies selected for this work and their classification, we analyze the possible existence of biases during the selection and we discuss the methodology used. In Sec. \ref{sec3} we present the main results of the chemical abundances and ionization parameters for the three spectral type of galaxies (SFG, Seyferts 2 and LINERs). In Sec. \ref{sec4} we analyze the possible relation between different host galaxy properties (morphology, luminosity, stellar mass and mass of the supermassive black hole) and the estimated chemical abundances. In Sec. \ref{sec5} we discuss the results obtained from this study and we compare them to other derived in previous works. Finally, Sec. \ref{sec6} provides an overview of our main results.

\begin{figure*}
	\centering
	\includegraphics[width=\hsize]{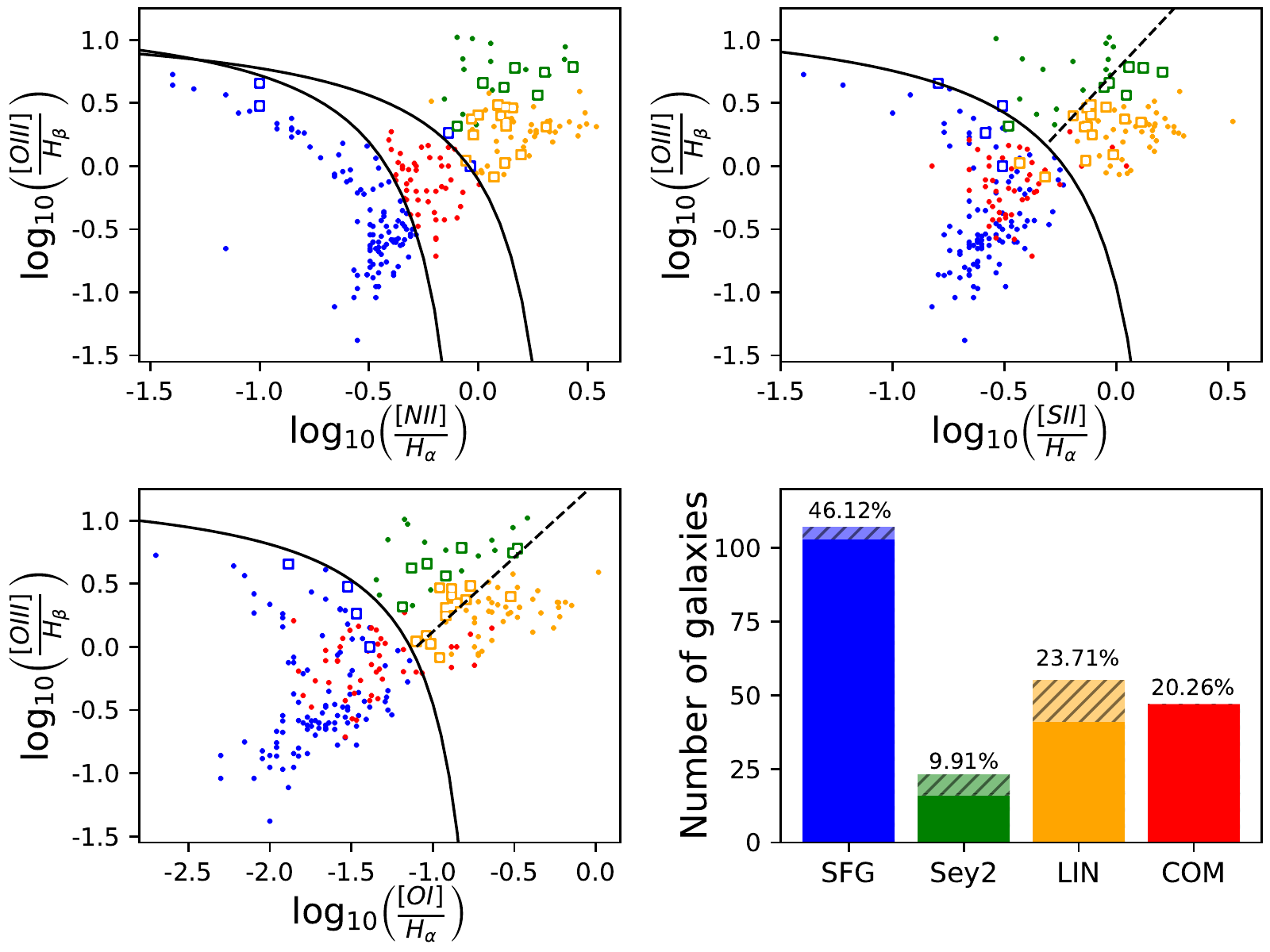}
	\caption{Diagnostic diagrams (two top panels and bottom-left)  used to  classify the galaxies of our sample. The black lines are the limits established by Kewley \citeyearpar{Kewley_2006}. The dot symbols are used for non-ambiguous galaxies and the square-empty symbols for ambiguous ones. The bottom-right panel shows the statistics of each spectral class and the color codification used in this work, being the dashed-filled bar the additional contribution from ambiguous galaxies.}
	\label{BPT}
\end{figure*}

\section{Sample and methodology}
\label{sec2}

\subsection{The sample}
\label{subsec21}
We compile our sample of galaxies from those observed in the Palomar Spectroscopic Survey \citep{Ho_II_1995, Ho_III_1997}. The Palomar Survey constitutes a magnitude-limited spectroscopic optical survey of the nuclear emission for galaxies in the local Universe (the median distance is $17.9$ Mpc and the maximum is $108.8$ Mpc). The nuclear spectra from 486 galaxies, brighter than $B_{T} \leq 12.5$ and declination $\delta > 0$, were retrieved with the Double Spectrograph mounted on the Hale 5m telescope, using an effective aperture of 2''x4''\citep{Ho_II_1995}. The Palomar Spectroscopic Survey is composed by galaxies from the Revised Shapley-Ames Catalog of Bright Galaxies \citep{Sandage_1981}. Hence we calculate the fraction of completeness $f$ by the approximation \citep{Sandage_1979}:
\begin{equation}
\label{ec: aux1} f \left( m \right) = \frac{1}{\exp \left[ \left( m - 12.72 \right) / 0.19 \right] + 1}
\end{equation}
being $m$ the magnitude in the $B_{T}$ system. Thus, for the limit magnitude of $B_{T} = 12.5$, the Palomar Spectroscopic Survey has a completeness fraction of $f \approx 0.76$. We recall then that the Palomar sample is a magnitude-limited sample, but it is not complete strictly speaking.

Considering the distance range and the aperture, we are analyzing regions whose nuclear radii cover a range between  6.8 pc and 1055.0 pc, below the radius of normalization $r \sim 3$ kpc proposed by \citet{Zaritsky_1994}, a smaller central region than that observed in larger samples of galaxies (such as SDSS with a radius of $\sim 2.9$ kpc or MaNGA with $\sim 1.3$ kpc) which allow us to avoid contribution from the host galaxy. Moreover, we also analyze the statistics of the nuclear region observed for each spectral type, as shown in Tab. \ref{Apertures}. Although there is a high dispersion of values, with standard deviations above $> 100$ pc and large ranges of values, the median values of each spectral type are identical. Therefore, no biased related with the observed region is introduced among the three types of galaxies.
		
\begin{table}
	\caption{Statistics of the nuclear radii (in pc) corresponding to the observed apertures, for SFG, Seyferts and LINERs in our sample. See Sec. \ref{subsec23} for details in the classification.}
	\label{Apertures}     
	\centering          
	\begin{tabular}{lccc}
		
		\multicolumn{4}{c}{\textbf{Nuclear radius (pc)}} \\ \hline \textbf{Type} & \textbf{Median} & \textbf{Std. Dev.} & \textbf{Range} \\ \hline
		\textbf{SFG} & 164.8 & 138.5 & [12.6, 1055.0] \\
		\textbf{Seyferts 2} & 164.8 & 141.9 & [65.9, 633.2] \\
		\textbf{LINERs} & 164.8 & 140.787 & [6.8, 888.2] \\ \hline
	\end{tabular}      
\end{table}

In order to subtract the stellar contribution in the nuclear spectra from the galaxies in their sample, \citet{Ho_III_1997} used galaxies whose spectra do not present emission lines as templates (a total of 79 galaxies). For a given galaxy whose spectra is corrected from reddening, a single template or a combination of templates (whose spectra is already corrected from reddening) are selected considering the best match for the velocity dispersions, metallicity and range in stellar populations for both, templates and input galaxy. This process was implemented in a $\chi^{2}$-minimization algorithm \citep{Ho_III_1997}. 

We take from Tab. 2 in \citet{Ho_III_1997} all the nuclear spectroscopic data, and their corresponding emission line fluxes once the stellar continuum is subtracted. Moreover, a considerable amount of ancillary data \citep{Ho_III_1997, Ho_VI_2003, Ho_VII_2009} can be used to study chemical abundances of the nuclear region in relation to the properties of their host galaxies (listed in Tab. \ref{TabA1}).

Both star-formation and AGN activity lead to the presence of emission lines which are key to estimate chemical abundances in the gas-phase of nuclear regions of galaxies. The galaxies that do not show any intrinsic emission once the stellar contribution is subtracted, are omitted hence from the sample. The original sample of 486 galaxies is then reduced to 418 \citep{Ho_III_1997}.

For the AGN sample, we will only consider Type-II AGN (Seyferts 1.8, 1.9 and 2 and LINERs 1.9 and 2 in the notation from \citet{Osterbrock_1977} to avoid the important contamination from the Broad Line Region. For that purpose, we consider the original classification from \citet{Ho_III_1997}. Thus, the 10 Seyferts, with types lower than 1.5, have been excluded \citep{Ho_IV_1997}. 

Finally, to get reliable chemical abundances and a good spectral classification only emission lines with accurate measures can be used. We hence only consider galaxies with relative errors up to 20$\% $ in the flux of the emission lines H$_{\alpha }$, H$_{\beta }$, [O\textsc{iii}]$\lambda $5007\r{A}, [N\textsc{ii}]$\lambda$6584\r{A} and [S\textsc{ii}]$\lambda\lambda$6717,6731\r{A}. The choice of these emission lines
is based on: 1) they are needed to spectroscopically classify the galaxies (see Sec. \ref{subsec23}); and, 2) they are used for the code \textsc{HCm} to estimate chemical abundances (see Sec. \ref{subsec25}). By applying this last criterion, a total of 176 galaxies are excluded from our study, thus reducing our sample to 232 objects. %In addition we apply other filters to exclude 47 composite galaxies (see Sec. \ref{subsec23}) and other 42 galaxies with unreliable results from \textsc{HCm} (see Sec. \ref{subsec26}). Once applied these two additional filters, the final sample amounts to 143 galaxies.

\begin{table*}
	\caption{Statistics of the spectral classification of the sample shown in Fig. \ref{BPT}. The percentages are referred to the sample made of 232 galaxies (without subtracting composite galaxies).}
	\label{Table stats}     
	\centering          
	\begin{tabular}{lcccc}
		
		\textbf{Spectral type} & \textbf{Non Ambiguous} & \textbf{Ambiguous} & \textbf{Total} & \textbf{Percentage (\%)} \\ \hline \textbf{Star-forming} & $103$ & $4$ & $107$ & $46.2$ \\ 
		\textbf{Seyfert 2} & $16$ & $7$ & $23$ & $9.9$ \\
		\textbf{LINER} & $41$ & $14$ & $55$ & $23.7$ \\ 
		\textbf{Composite} & $47$ & $-$ & $47$ & $20.2$ \\ \hline
		\textbf{Sum} & $206$ & $26$ & $232$ & $100.0$ \\  \hline
	\end{tabular}        
\end{table*}

\subsection{Classification of the sample}
\label{subsec23}

Prior to the chemical abundance estimation, we must distinguish the source of ionization, and this may be achieved by using diagnostic diagrams \citep{Veilleux_1987, Kauffmann_2003, Kewley_2006}. In spite of the fact that Ho et al. presented a spectral classification of their sample \citeyearpar{Ho_III_1997, Ho_IV_1997}, based on the Veilleux and Osterbrock's criteria \citeyearpar{Veilleux_1987}, we use the Kewley's criteria \citeyearpar{Kewley_2006}, tested in order of magnitudes larger samples, to classify our sample of 232 galaxies.

The Kewley's criteria allow us to distinguish among SFG, Seyferts 2, LINERs and composite galaxies. However, it also contemplates the possibility that some galaxies may fall in different regions from one diagram to another. These galaxies are called \textit{ambiguous galaxies} \citep{Kewley_2006}. In order to perform our study for these galaxies, we classify them with a given spectral type when this is assigned by, at least, two out of the three diagrams. Fig. \ref{BPT} shows the classification of the galaxies, and its corresponding statistics is presented in Tab. \ref{Table stats}.

The true nature of composite galaxies, supposed to be a mixture between star-forming galaxies and low-luminosity AGN by some authors \citep{Veilleux_1987, Ho_III_1997}, remains unclear \citep{Ho_vol1, Panessa_2005, Davies_2014b, Davies_2014}. Therefore, this spectral type of galaxies is omitted in our study and will be presented in a later work. The omission of 47 composite galaxies leads to a sample of 185 galaxies: 107 SFG, 23 Seyferts 2 and 55 LINERs.

\subsection{Reddening correction}
\label{subsec24}

The emission line ratios required for the code \textsc{HCm} need to be corrected for reddening. \citet{Ho_III_1997} present some of the emission line ratios already corrected for reddening following the Cardelli's extinction curve \citeyearpar{Cardelli_1989}: Nevertheless, we decided to apply our own correction motivated upon two reasons. Firstly, \citet{Ho_III_1997} did not present the reddening correction for all the emission lines, as it is the case of [O\textsc{iii}]$\lambda$4363\r{A}. Secondly, the extinction correction depends on the spectral classification of each galaxy, which is slightly different in our study because we use the Kewley's criteria \citeyearpar{Kewley_2006}. We therefore guarantee the consistency of the extinction correction for all galaxies and all emission line ratios in our sample of galaxies.

We assume the extinction parametrization \citep{Lequeux_1979, Howarth_1983, Cardelli_1989, Osterbrock_book}:
\begin{equation}
\label{1} \frac{I_{\lambda }}{I \left( H_{\beta } \right) } = \frac{I_{\lambda ,0}}{I \left( H_{\beta } \right) _{0}} 10^{- C \left[ f \left( \lambda \right)  - f \left( H_{\beta } \right) \right] } = \frac{I_{\lambda ,0}}{I \left( H_{\beta } \right) _{0}} 10^{- c \left( H_{\beta } \right) \left[ \frac{f \left( \lambda \right)  }{ f \left( H_{\beta } \right) } - 1\right] }
\end{equation}
where $c \left( H_{\beta } \right) $ characterizes the reddening in each galaxy and $f$ is the Howarth's extinction curve \citeyearpar{Howarth_1983}, with $R_{V} = 3.1$. To calculate $c \left( H_{\beta } \right) $ we assume the electronic temperature and density typical for star-forming regions and NLR of AGN, i.e. $T_{e} \sim 10^{4} \ \mathrm{K}$ and $n_{e} \sim 10^{3} \ \mathrm{cm}^{-3}$, and the case B recombination: the Balmer ratio $H_{\alpha }/H_{\beta }$ should be 2.86 in SFG and 3.1 in AGN. If this ratio is below the theoretical values, then $c \left( H_{\beta } \right) $ is set to 0. Otherwise, $c \left( H_{\beta } \right) $ is calculated from Eq. \ref{1} and then used to correct all emission lines from reddening. Tab. \ref{TabA2} shows the emission line ratios corrected for reddening for our sample of galaxies.

\subsection{Chemical abundances determination}
\label{subsec25}

We do not apply the direct method ($T_{e}$-method) for the determination of chemical abundances as a consequence of the lack of measurements for the auroral line [O\textsc{iii}]$\lambda$4363\r{A} (only measured in 5 galaxies, i.e., $3.5\% $ of the sample). In addition, given the limitations of the calibrations of the optical strong emission lines (only available for SFG, \citealt{Kobulnicky_1999, Dors_2015, Strom_2017}, and Seyferts 2, \citealt{Storchi_1998, Dors_2015, Dors_2019}), we use photoionization models to estimate chemical abundances. Particularly, we use the \textsc{HCm} program, developed by P\'{e}rez-Montero \citeyearpar{Perez-Montero_2014, Perez-Montero_2019}, which uses the code \textsc{cloudy v17.01} \citep{Ferland_2017} to generate a grid of models covering a large range of chemical abundances and ionization parameters. 

We compare the O/H and N/O determinations from \textsc{HCm} with those obtained by using empirical calibrations based on optical strong emission lines \citep{Perez-Montero_2009, Pilyugin_2016, Curti_2017} for SFG. Considering the results shown in Appendix \ref{s: A0}, we find a good agreement between our results for SFG and the calibrations of \citet{Curti_2017}  for O/H (a median offset of 0.07 dex and RMSE of 0.06 dex) and \citet{Perez-Montero_2009} for N/O (a median offset of 0.04 dex and RSME of 0.07 dex). We also calculate oxygen abundances $12+\log\left( O/H \right) $ with three other methods for AGN based on the use of strong emission lines \citep{Storchi_1998, Carvalho_2020, Flury_2020}, and conclude that our results are in good agreement with those obtained using the calibration from \citet{Storchi_1998}: we found a median offset of -0.04 dex and RSME of 0.15 dex for Seyferts; and an offset of 0.04 dex and RSME 0.21 dex for LINERs. It must be noticed that we obtain a higher dispersion than that obtained when comparing results for SFG. While the calibration from \citet{Flury_2020} clearly underestimates our results for both Seyferts and LINERs (with median offsets of -0.48 and -0.51 dex and RMSE of 0.17 and 0.22 dex respectively), the calibration from \citet{Carvalho_2020} shows the highest RMSE of all methods ($\sim 0.32$ dex).The details of this comparison are provided in Appendix \ref{s: A}.

\begin{table*}
	\caption{Summary of the input SED and grids that the \textsc{HCm} code uses to calculate the chemical abundances. Information taken from \citet{Perez-Montero_2014, Perez-Montero_2019}.}
	\label{models_hcm}     
	\centering          
	\begin{tabular}{l|l|lll|l}
		& & \multicolumn{3}{c|}{\textbf{Grids}} & \\ \cline{3-5} \textbf{Spectral type} & \multicolumn{1}{c| }{\textbf{SED}} & \boldmath$12+\log_{10} \left( O/H \right) $ & \boldmath$\log_{10} \left( N/O \right) $ & \boldmath{$\log_{10} \left( U \right) $} & \textbf{ N\boldmath$^{\circ}$  Models} \\ & & Step: 0.1 dex & Step: 0.125 dex & Step: 0.25 dex & \\ \hline  & POPSTAR & & & & \\ \textbf{SFG} & Burst of 10$^{6}$ yr & [6.9, 9.1] & [-2.0, 0.0] & [-4.0, -0.5] & 5865 \\ & Ratio Dust/Gas $7.5 \cdot10^{-3}$ & & & & \\ \hline & BBB max. at 13.6 eV & & & & \\ \textbf{Seyferts} & $\alpha_{X} = -1$ & [6.9, 9.1] & [-2.0, 0.0] & [-2.5, -0.5] & 2816 \\ & $\alpha_{OX} = -0.8$ & & & & \\ \hline & BBB max. at 13.6 eV & & & & \\ \textbf{LINERs} & $\alpha_{X} = -1$ & [6.9, 9.1] & [-2.0, 0.0] & [-4.0, -2.5] & 2346 \\ & $\alpha_{OX} = -0.8$ & & & & \\ \hline
	\end{tabular}      
\end{table*}

\begin{table*}
	\caption{Morphological classes used in this study and their correspondence with the morphological type.}
	\label{Class_tab}     
	\centering          
	\begin{tabular}{lcl}
		\textbf{Morphological class} & \textbf{T} & \textbf{Standard types} \\ \hline Elliptical (E) & $\left[ -6, -4 \right] $ & cE, E, E$^{+}$ \\  Lenticular (S0) & $\left[ -3, -1 \right] $ & S0$^{-}$, S0$^{0}$, S0$^{+}$ \\  Early spiral (ES) & $\left[ 0, 2 \right] $ & S0/a, Sa, Sab  \\  Intermediate spiral (IS) & $\left[ 3, 4 \right] $ & Sb, Sbc \\  Late spiral (LS) & $\left[ 5, 8 \right] $ & Sc, Scd, Sd, Sdm \\  Magellanic irregular (Irr) &  $\left[ 9, 10 \right] $ & Sm, Im \\  Non-Magellanic irregular (Irr Non Mag) & $90$ & Non-Magellanic irregular \\  Peculiar (Pec) & $99$ & Peculiar \\ \hline
	\end{tabular}
\end{table*}

\begin{figure*}
	%\resizebox{\hsize}{!}
	\begin{tabular}{cccc}
		\begin{minipage}{0.02\hsize}\begin{flushright}\textbf{(a)} \end{flushright}\end{minipage}  &  \begin{minipage}{0.42\hsize}\centering{\includegraphics[width=1\textwidth]{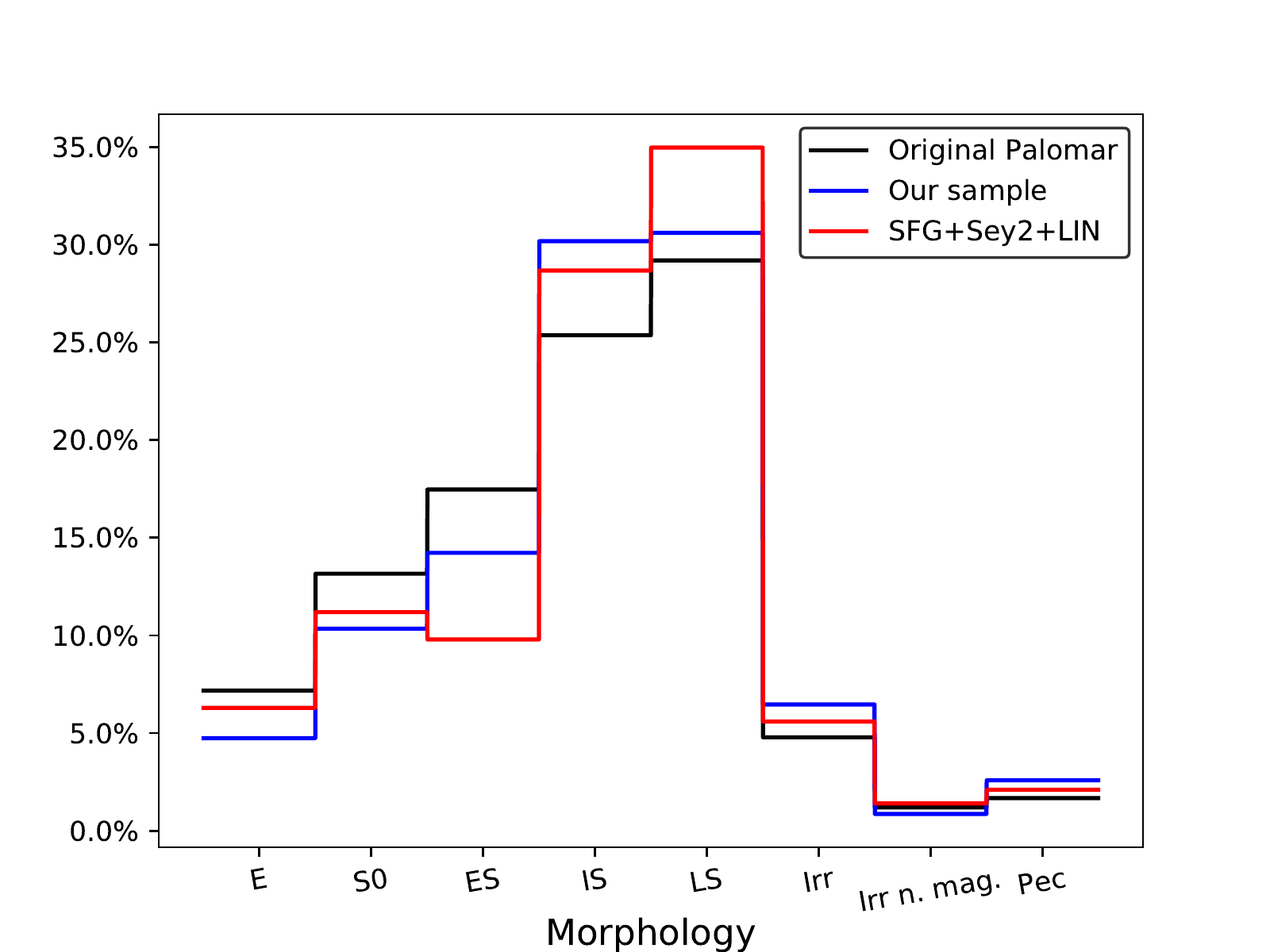}} \vspace{-0.22in} \end{minipage} & \begin{minipage}{0.02\hsize}\begin{flushright}\textbf{(b)} \end{flushright}\end{minipage}  &  \begin{minipage}{0.42\hsize}\centering{\includegraphics[width=1\textwidth]{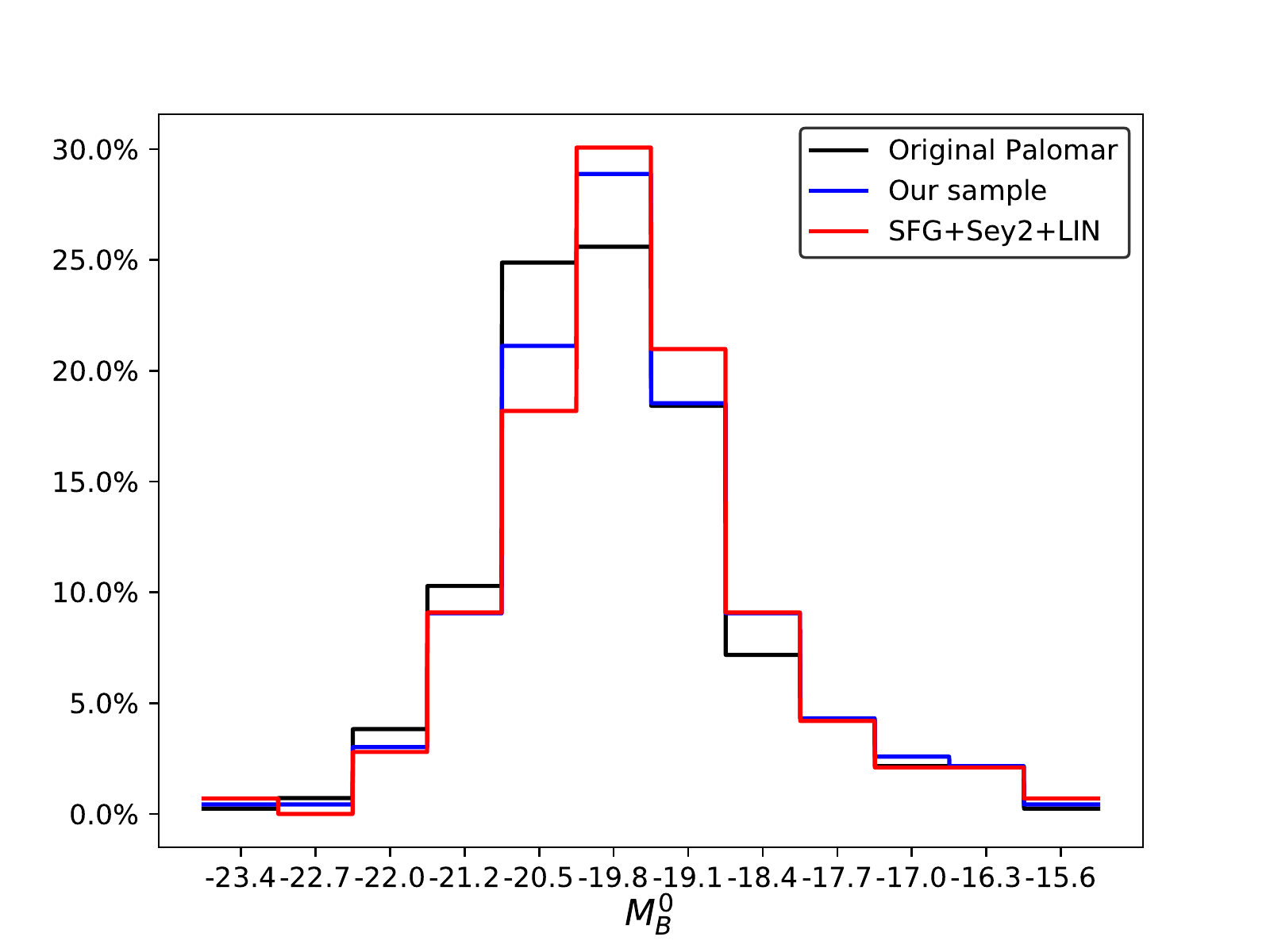}} \vspace{-0.22in} \end{minipage} \\ & \\
		\begin{minipage}{0.02\hsize}\begin{flushright}\textbf{(c)} \end{flushright}\end{minipage}  &  \begin{minipage}{0.42\hsize}\centering{\includegraphics[width=1\textwidth]{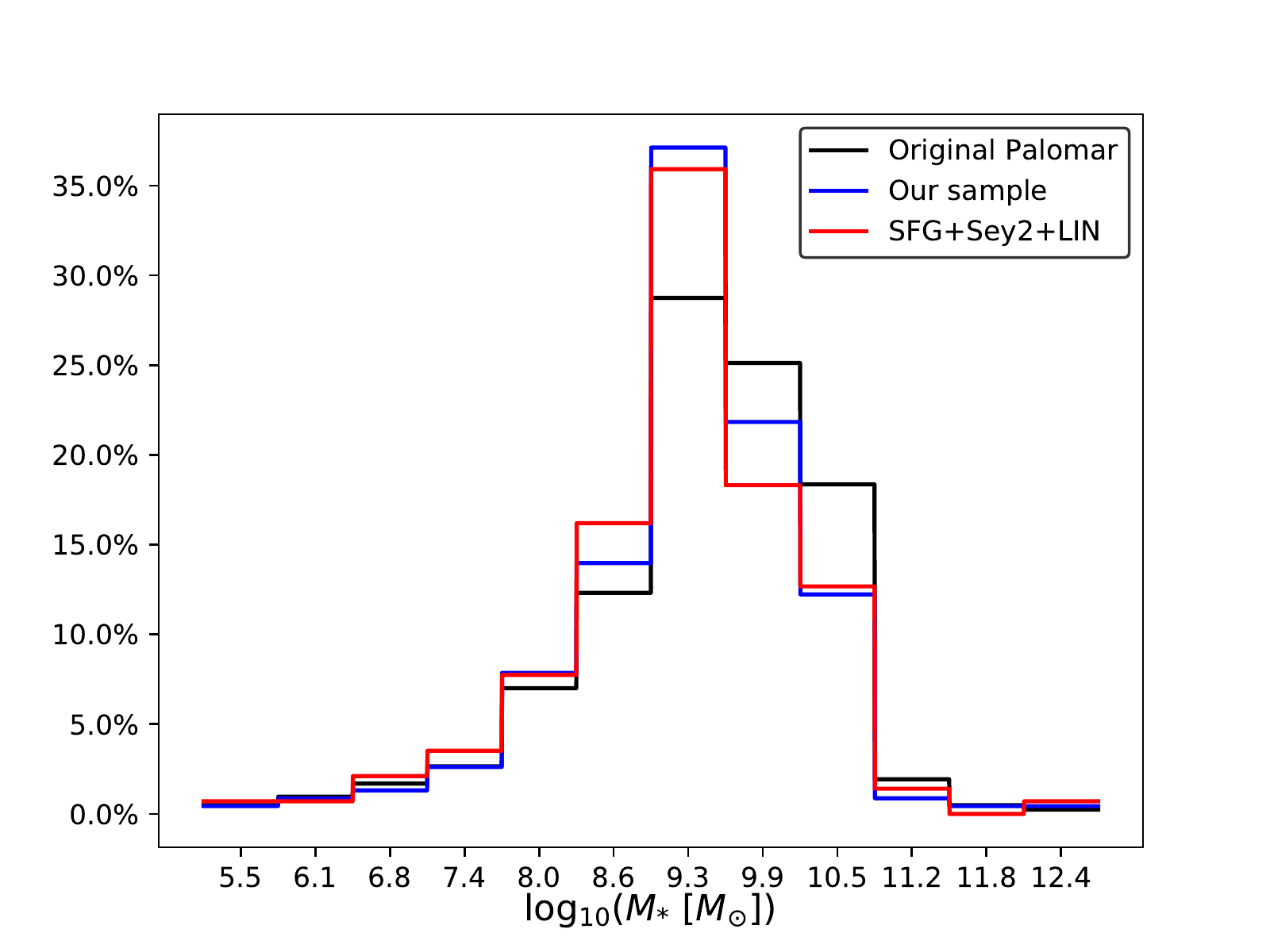}} \vspace{-0.2in} \end{minipage} & \begin{minipage}{0.02\hsize}\begin{flushright}\textbf{(d)} \end{flushright}\end{minipage}  &  \begin{minipage}{0.42\hsize}\centering{\includegraphics[width=1\textwidth]{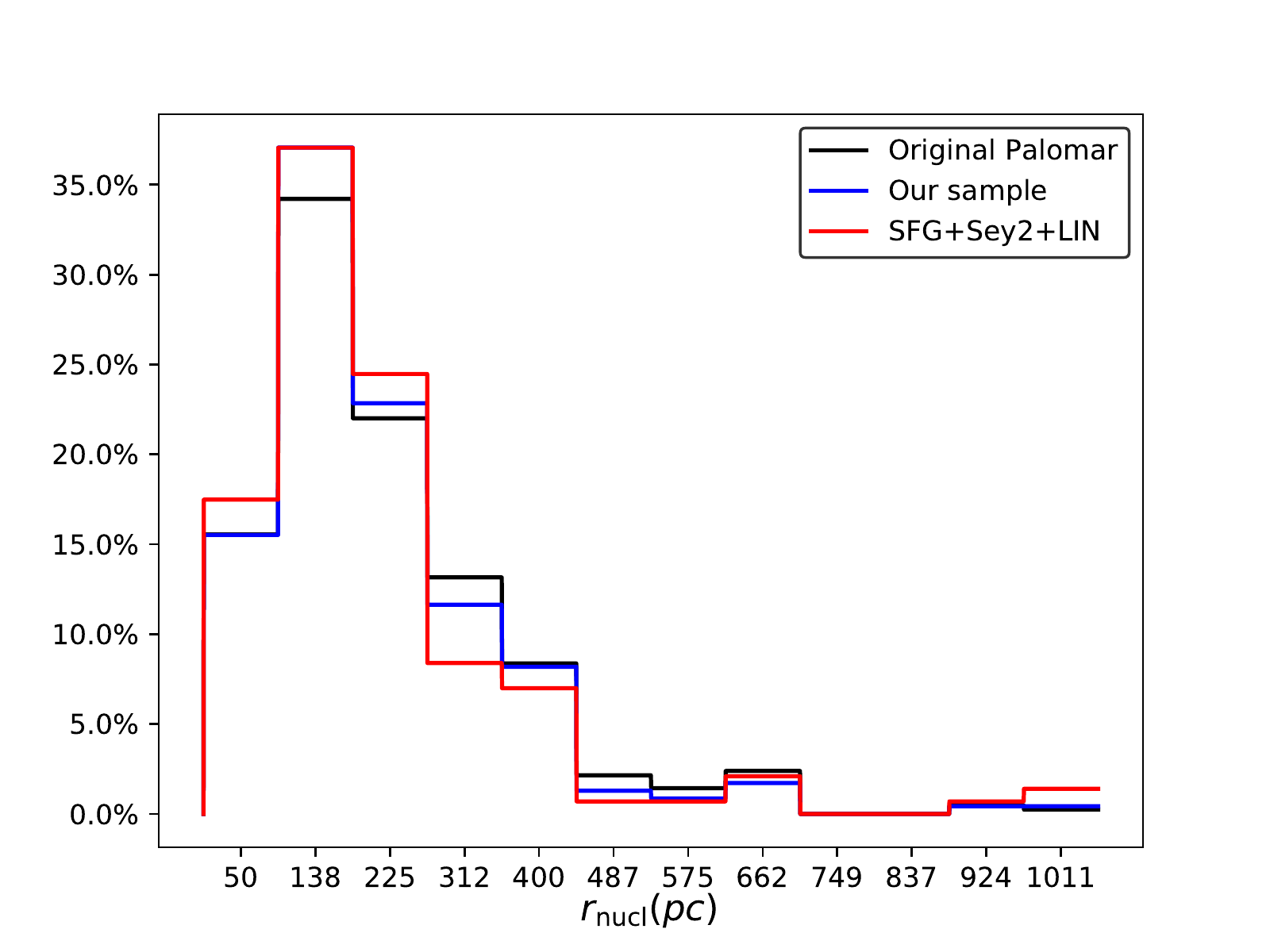}} \vspace{-0.2in} \end{minipage} \\ 
	\end{tabular}
%	\begin{tabular}{cc}
%		\begin{minipage}{0.05\hsize}\begin{flushright}\textbf{(a)} \end{flushright}\end{minipage}  &  \begin{minipage}{0.91\hsize}\centering{\includegraphics[width=1\textwidth]{Images/Morphology_general_relative.pdf}} \vspace{-0.2in} \end{minipage} \\ \begin{minipage}{0.05\hsize}\begin{flushright}\textbf{(b)} \end{flushright}\end{minipage}  &  \begin{minipage}{0.93\hsize}\centering{\includegraphics[width=1\textwidth]{Images/Bmagnitude_general_relative.pdf}} \vspace{-0.2in} \end{minipage} \\ 
%		\begin{minipage}{0.05\hsize}\begin{flushright}\textbf{(c)} \end{flushright}\end{minipage}  &  \begin{minipage}{0.93\hsize}\centering{\includegraphics[width=1\textwidth]{Images/Stellar_mass_general_relative.pdf}} \vspace{-0.2in} \end{minipage}  \\ 
%	\end{tabular}
	\caption{Relative distribution of different host galaxy properties for the original sample of galaxies from the Palomar Survey (418 in total; black line), our sample of galaxies (232 in total; blue line) and the final sample made of SFG, Seyferts 2 and LINERs (143 in total; red line). (a) Distribution of the morphology according to Tab. \ref{Class_tab}. (b) Distribution of the corrected absolute B-magnitude $M_{B}^{0}$. (c) Distribution of the stellar mass $M_{*}$. (d) Distribution of the nuclear radius $r_{nucl}$.}
	\label{Bias_prop}
\end{figure*}

\begin{table*}
	\caption{Distribution of galaxies for each morphological type. It is shown the number of objects for each morphological type and, within parenthesis, the percentage of the given morphological type for each spectral type.}
	\label{morpho_distr}
	\centering
	\begin{tabular}{lccccc}
		\textbf{Morphology} & \textbf{Star-forming} & \textbf{Seyferts 2} & \textbf{LINERs} &  \textbf{Total} & \textbf{Percentage ($\% $)}  \\ \hline
		\textbf{E} & 0 (0.0$\% $) & 0 (0.0$\% $) & 9 (22.5$\% $) & 9 & 6.3 \\  \textbf{S0} & 2 (2.3$\% $) & 3 (18.8$\% $) & 11 (27.5$\% $) & 16 & 11.2 \\  \textbf{ES} & 6 (6.9$\% $) & 3 (18.8$\% $) & 5 (12.5$\% $) & 14 & 9.8 \\  \textbf{IS} & 21 (24.2$\% $) & 8 (50.0$\% $) & 12 (30.0$\% $) & 41 & 28.6 \\  \textbf{LS} & 47 (54.1$\% $) & 2 (12.4$\% $) & 1 (2.5$\% $) & 50 &  35.0 \\  \textbf{Irr} & 7 (8.0$\% $) & 0 (0.0$\% $) & 1 (2.5$\% $) & 8 & 5.6 \\  \textbf{Irr. Non. Mag.} & 1 (1.1$\% $) & 0 (0.0$\% $) & 1 (2.5$\% $) & 2 & 1.4 \\  \textbf{Pec} & 3 (3.4$\% $) & 0 (0.0$\% $) & 0 (0.0$\% $) & 3 & 2.1 \\ \hline  \textbf{Sum} & 87 (100$\% $) & 16 (100$\% $) & 40 (100$\% $) & 143 & 100.0 \\  \hline
	\end{tabular}
\end{table*}

First of all, the code \textsc{HCm} requires the selection of a spectral energy distribution (SED), which accounts for difference between the photoionization in star-forming regions and AGN.

\begin{itemize} 
	\item For SFG, the SED is generated with the code \textsc{popstar} \citep{Molla_2009}, calculated for an instantaneous burst at an age of 1 Myr and a mass ratio of dust/gas $7.5\cdot10^{-3}$ \citep{Perez-Montero_2014}. 
	\item For AGN, the SED is composed by a Big Blue Bump originated by the thermal emission of the accretion disk and a power law with spectral index $\alpha_{X} = -1$ accounting for the non-thermal X-ray emission. The transition between the X-ray and ultraviolet range is modeled also with a power law with spectral index $\alpha_{OX} = -0.8$, for consistency with previous studies \citep{Perez-Montero_2019, Dors_2019}. The dust/gas distribution is that characterizing the NLR of AGN \citep{Perez-Montero_2019}.
\end{itemize}

The code can admit as input six emission line ratios: [O\textsc{ii}]$\lambda$3727\r{A}, [Ne\textsc{iii}]$\lambda$3868\r{A}, [O\textsc{iii}]$\lambda$4363\r{A}, [O\textsc{iii}]$\lambda$5007\r{A}, [N\textsc{ii}]$\lambda$6584\r{A} and [S\textsc{ii}]$\lambda\lambda $6717,6731\r{A}; all of them referred to the Balmer line $H_{\beta }$. However, the code can also be run for a smaller subset of emission lines. Due to the spectral coverage of the Palomar Survey, the two first emission lines are missing for all galaxies. Therefore, we use as input for the program the emission lines [O\textsc{iii}]$\lambda$5007\r{A}, [N\textsc{ii}]$\lambda$6584\r{A}, [S\textsc{ii}]$\lambda\lambda $6717,6731\r{A} and [O\textsc{iii}]$\lambda$4363\r{A} (if measured).

By applying a $\chi ^{2}$-methodology, the code selects the best values of $12+\log \left( O/H \right) $, $\log \left( N/O \right) $ and $\log \left( U \right) $ among a grid of models (see Tab. \ref{models_hcm}), without assuming any underlying relation among them for AGN. In the case of SFG, when [O\textsc{iii}]$\lambda$4363\r{A} auroral line is not measured, \textsc{HCm} assumes a relation between $12+\log \left( O/H \right) $ and $\log \left( U \right) $ \citep{Perez-Montero_2014}, but the determination of $\log \left( N/O \right) $ is independent.
\\

As summarize in Tab. \ref{models_hcm}, the code uses the same grids for the chemical abundances ratios $12 + \log \left( O/H \right) $ and $\log \left( N/O \right) $ for all spectral types: the Oxygen abundance\footnote{From now on, we assume as solar abundance the chemical abundance the values obtained by \citet{Asplund_2009}, $12+\log \left( O/H \right)_{\odot } = 8.69$.} expands from $12+\log \left( O/H \right) = 6.9$ to $9.1$ in steps of 0.1 dex, while the Nitrogen to Oxygen abundance ratio has a range $\left[ -2.0, 0.0 \right] $, with values separated by 0.125 dex. However, the range of the ionization parameter is different for each spectral type. For SFG, the range of $\log \left( U \right) $ is $\left[ -4.0, -0.5 \right]$. Regarding AGN, we split the range of the ionization parameter: for Seyferts 2 only the upper range $\left[ -2.5, -0.5 \right] $ is considered, as discussed in \citet{Perez-Montero_2019}; for LINERs, we consider the range $\left[ -4.0, -2.5 \right] $. This was implemented since the first values of $\log \left( U \right) $ obtained for LINERs were unrealistically high, contrasting with previous studies \citep{Ferland_1983, Halpern_1983, Binette_1985, Kewley_2006}.

In the case of AGN, we check that values of O/H, N/O and U obtained using \textsc{HCm} when the complete grid of models is used, regardless of their position in the diagnostic diagrams, are compatible within the errors with the same quantities derived using only the sets of models that verify the criteria from \citet{Kewley_2006} for both Seyferts 2 and LINERs.

%For the AGN case, we obtain that some values of (O/H, N/O, U) trace regions in the diagnostic diagrams that are characteristic of SFG. Therefore, we check if the use of the whole grids of models for Seyferts 2 and LINERs introduces changes in the determination of chemical abundance ratios. We omit from the grids of models those tuples of values that do not verify Kewley's criteria \citeyearpar{Kewley_2006}. Our results show that chemical abundance ratios obtained using the whole grids of models are compatible within the errors with those obtained after limiting the grids of models by Kewley's criteria \citeyearpar{Kewley_2006}.

\subsection{Filtering reliable results}
\label{subsec26}

Due to the fact that the reddening correction (see Sec. \ref{subsec24}) and the code \textsc{HCm} contribute to the uncertainty of the estimated chemical abundances, we apply another criterion to filter our sample. The grids of models used in the code have a step of $\sigma $, which is 0.1 dex in the case of $12+\log \left( O/H \right) $, 0.125 dex for $\log \left( N/O \right) $ and 0.25 dex for $\log \left( U \right) $. Therefore, any galaxy with a chemical abundance ratio or ionization parameter, whose uncertainty is greater than $3\sigma $, is omitted in this study. A total of 42 galaxies (20 SFG, 7 Seyferts 2 and 15 LINERs) are omitted in the following studies. Therefore, our final sample of galaxies amounts to 143: 87 SFG, 16 Seyferts 2 and 40 LINERs.

The application of these selection criteria, along with those described in Sec. \ref{subsec21} and \ref{subsec23}, largely reduces the number of Seyferts 2 (16 in total) in our sample, so, the results we obtain for this spectral type should be carefully re-examined in larger samples of AGN.

\begin{figure*}
	%\resizebox{\hsize}{!}
	\begin{tabular}{cccc}
		\begin{minipage}{0.05\hsize}\begin{flushright}\textbf{(a)} \end{flushright}\end{minipage}  &  \begin{minipage}{0.4\hsize}\centering{\includegraphics[width=1\textwidth]{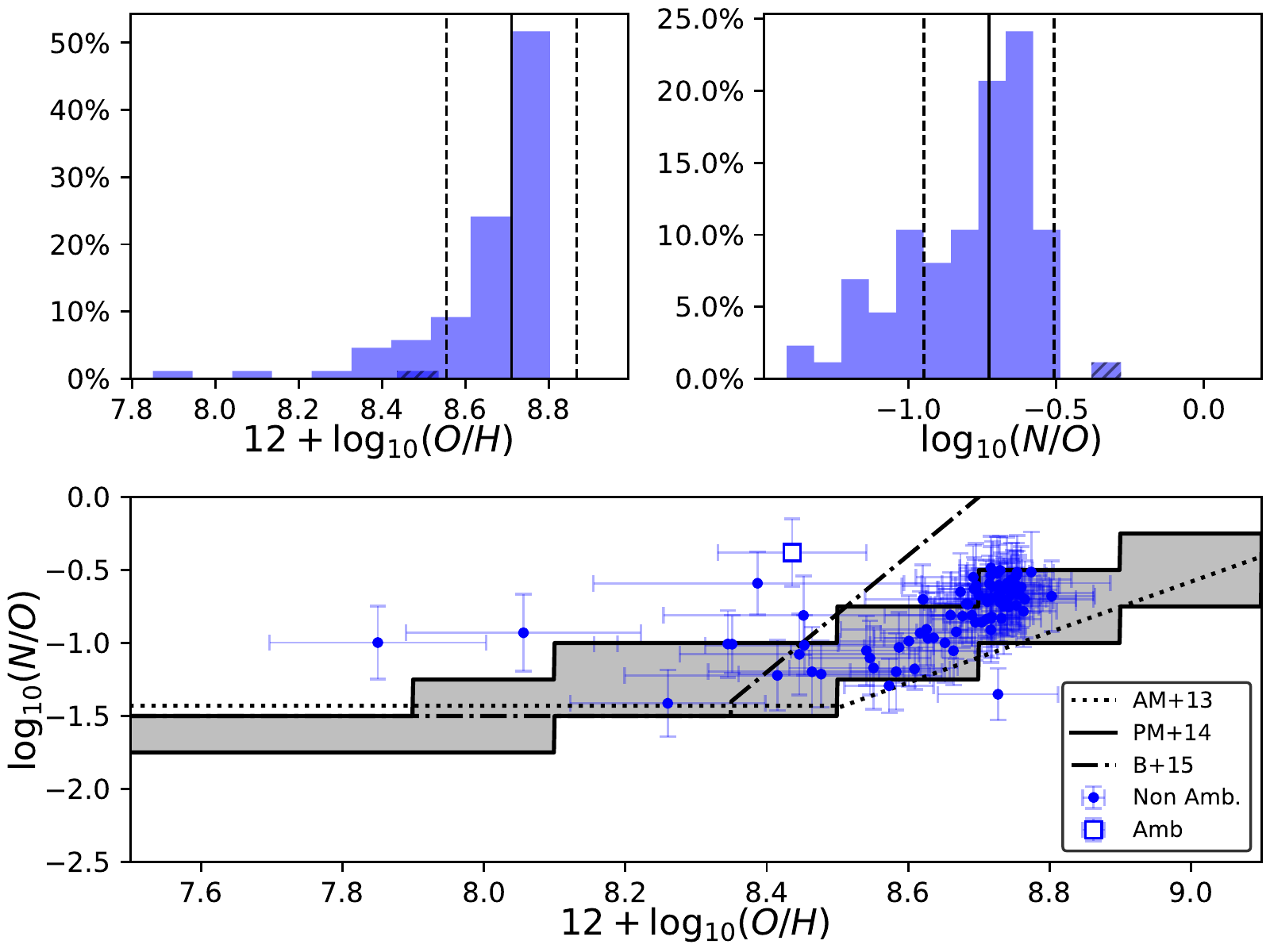}} \vspace{-0.2in} \end{minipage} & \begin{minipage}{0.05\hsize}\begin{flushright}\textbf{(b)} \end{flushright}\end{minipage}  &  \begin{minipage}{0.4\hsize}\centering{\includegraphics[width=1\textwidth]{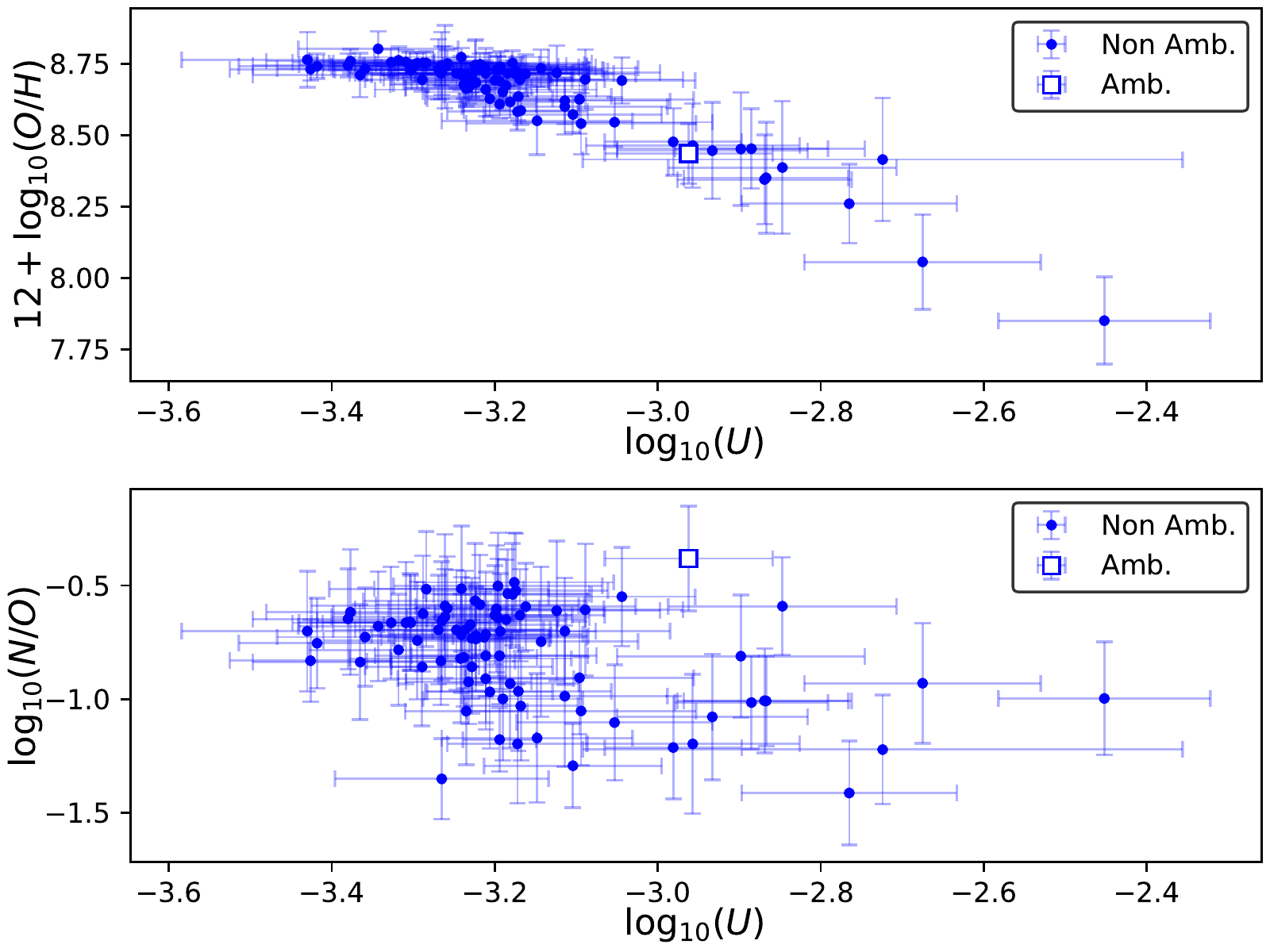}} \vspace{-0.2in} \end{minipage} 
	\end{tabular}
	\caption{Chemical abundances and ionization parameters obtained for star-forming galaxies. (a) The top left plot shows the histogram of the chemical abundance $12 + \log_{10} \left( O/H \right) $. The top right plot shows the histogram of $\log_{10} \left( N/O \right) $. For both histograms, the solid lines represent the median values and the dashed lines the standard deviations. The dashed-filled histograms are associated to ambiguous galaxies. The bottom plot shows the relation between the chemical abundance ratio $\log_{10} \left( N/O \right) $ and $12 + \log_{10} \left( O/H \right) $, using dots for non-ambiguous galaxies and square-empty symbols for ambiguous galaxies. We also present the relations obtained by \citet{Andrews_2013} (AM+13, dotted line), \citet{Perez-Montero_2014} (PM+14, shaded area delimited by a solid line) and \citet{Belfiore_2015} (B+15, dot-dashed line). (b) Variation of the chemical abundances $12+\log_{10} \left( O/H \right) $ and $\log_{10} \left( N/O \right) $ with the ionization parameter $\log_{10} \left( U \right) $ in star-forming galaxies.}
	\label{Chem_SFG}
\end{figure*}

\begin{figure*}
	%\resizebox{\hsize}{!}
	\begin{tabular}{cccc}
		\begin{minipage}{0.05\hsize}\begin{flushright}\textbf{(a)} \end{flushright}\end{minipage}  &  \begin{minipage}{0.4\hsize}\centering{\includegraphics[width=1\textwidth]{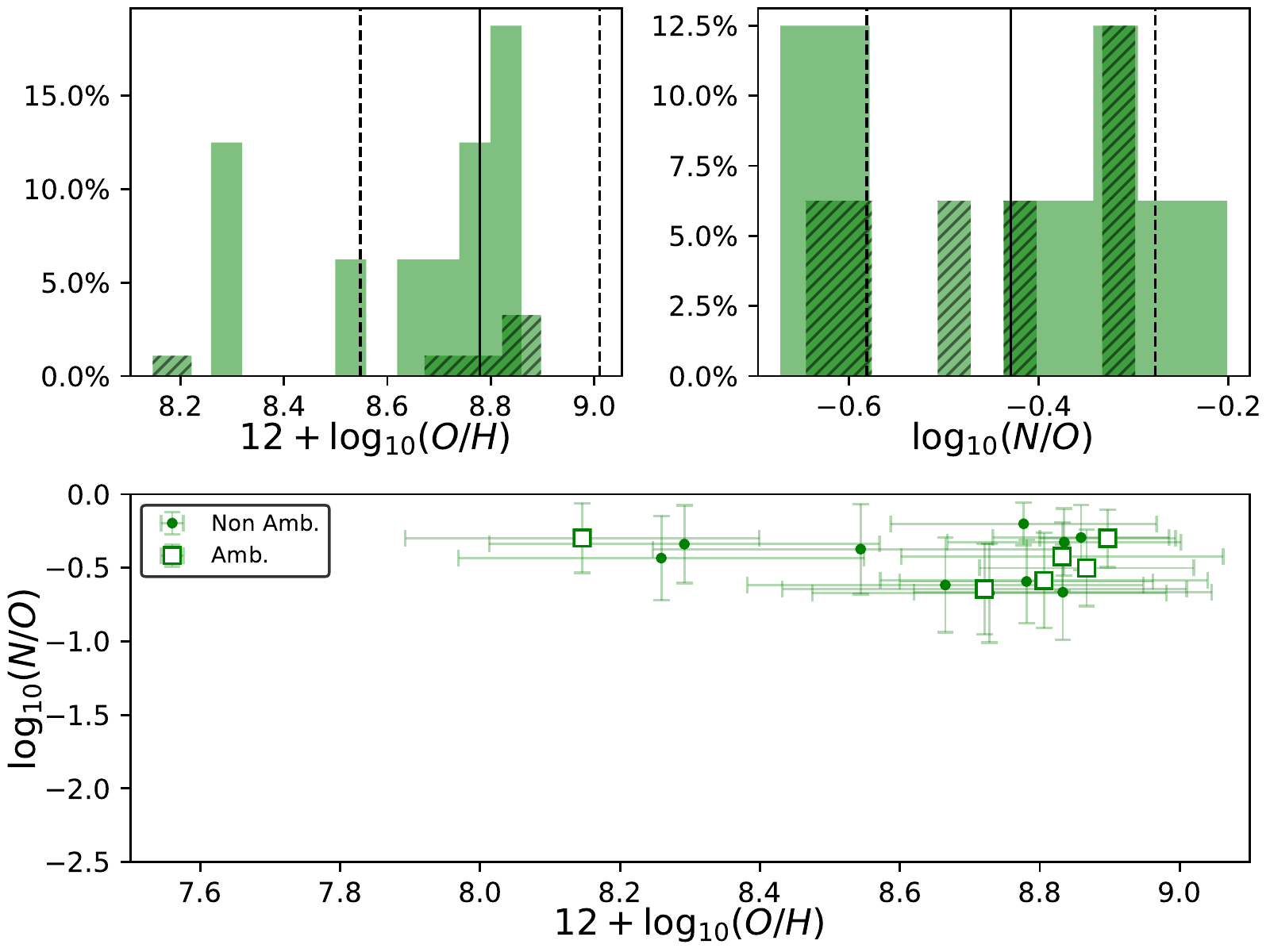}} \vspace{-0.2in} \end{minipage} & \begin{minipage}{0.05\hsize}\begin{flushright}\textbf{(b)} \end{flushright}\end{minipage}  &  \begin{minipage}{0.4\hsize}\centering{\includegraphics[width=1\textwidth]{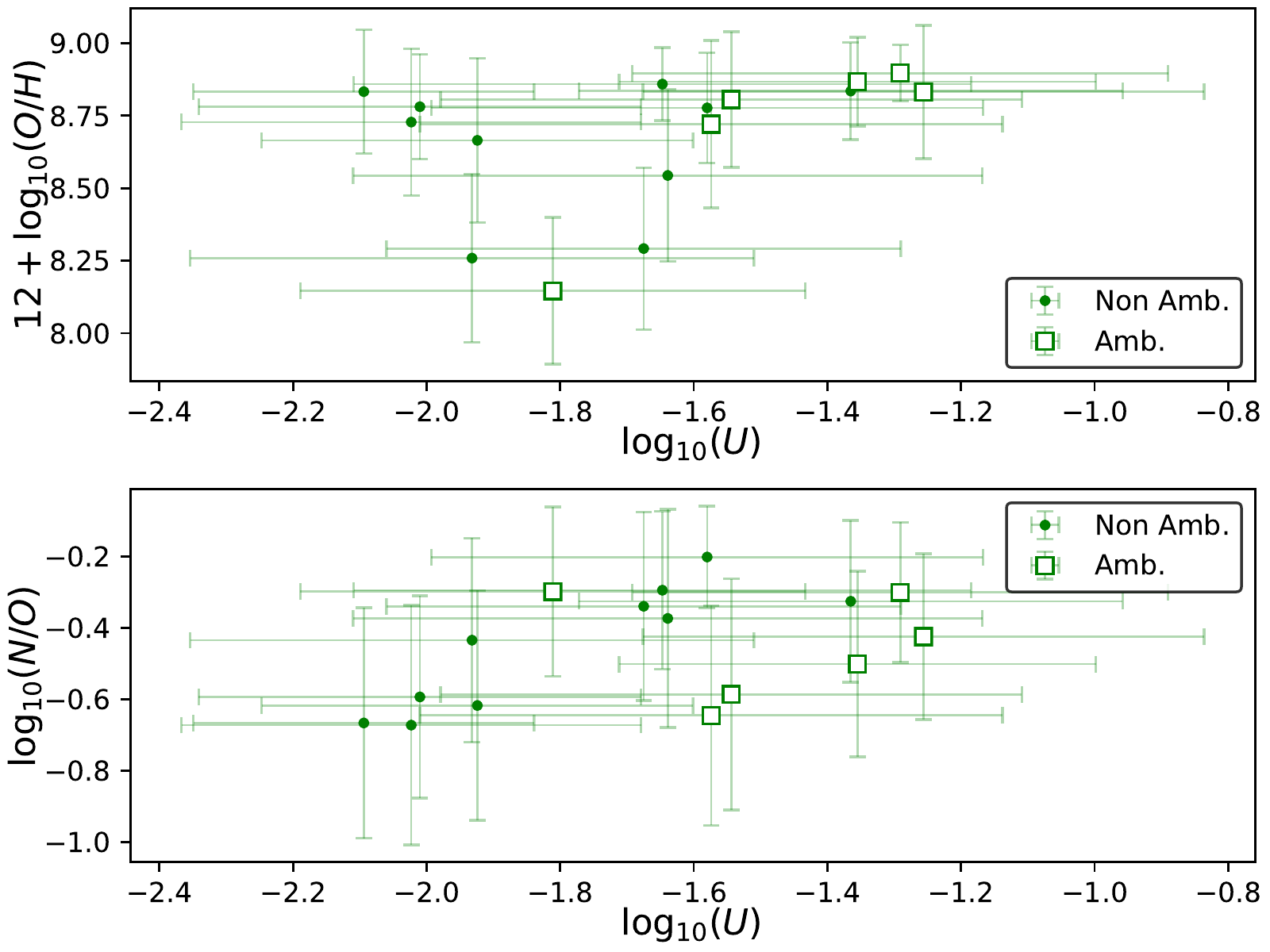}} \vspace{-0.2in} \end{minipage} 
	\end{tabular}
	\caption{Same as Fig. \ref{Chem_SFG} but for Seyferts 2.}
	\label{Chem_S2}
\end{figure*}

\begin{figure*}
	%\resizebox{\hsize}{!}
	\begin{tabular}{cccc}
		\begin{minipage}{0.05\hsize}\begin{flushright}\textbf{(a)} \end{flushright}\end{minipage}  &  \begin{minipage}{0.4\hsize}\centering{\includegraphics[width=1\textwidth]{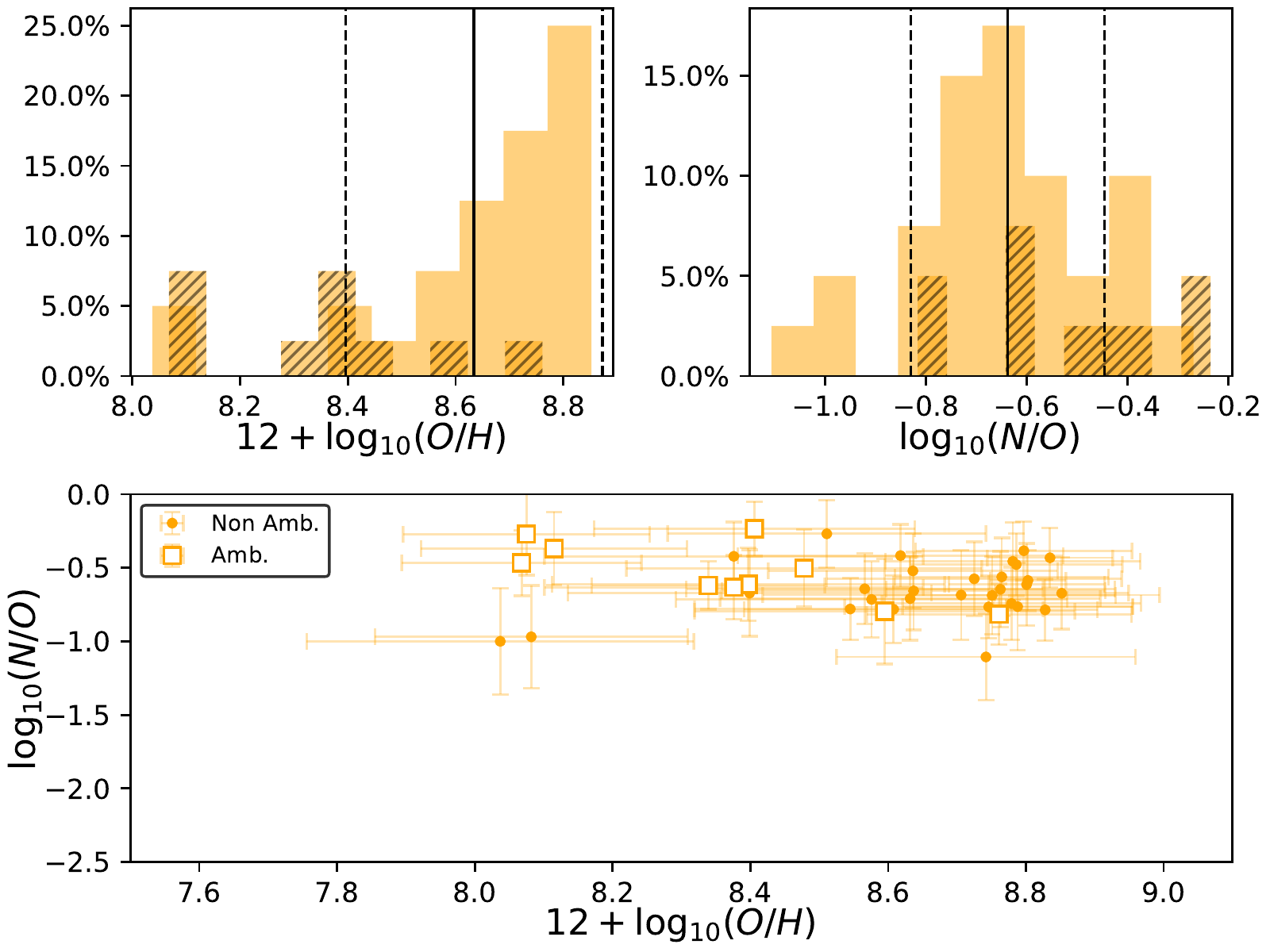}} \vspace{-0.2in} \end{minipage} & \begin{minipage}{0.05\hsize}\begin{flushright}\textbf{(b)} \end{flushright}\end{minipage}  &  \begin{minipage}{0.4\hsize}\centering{\includegraphics[width=1\textwidth]{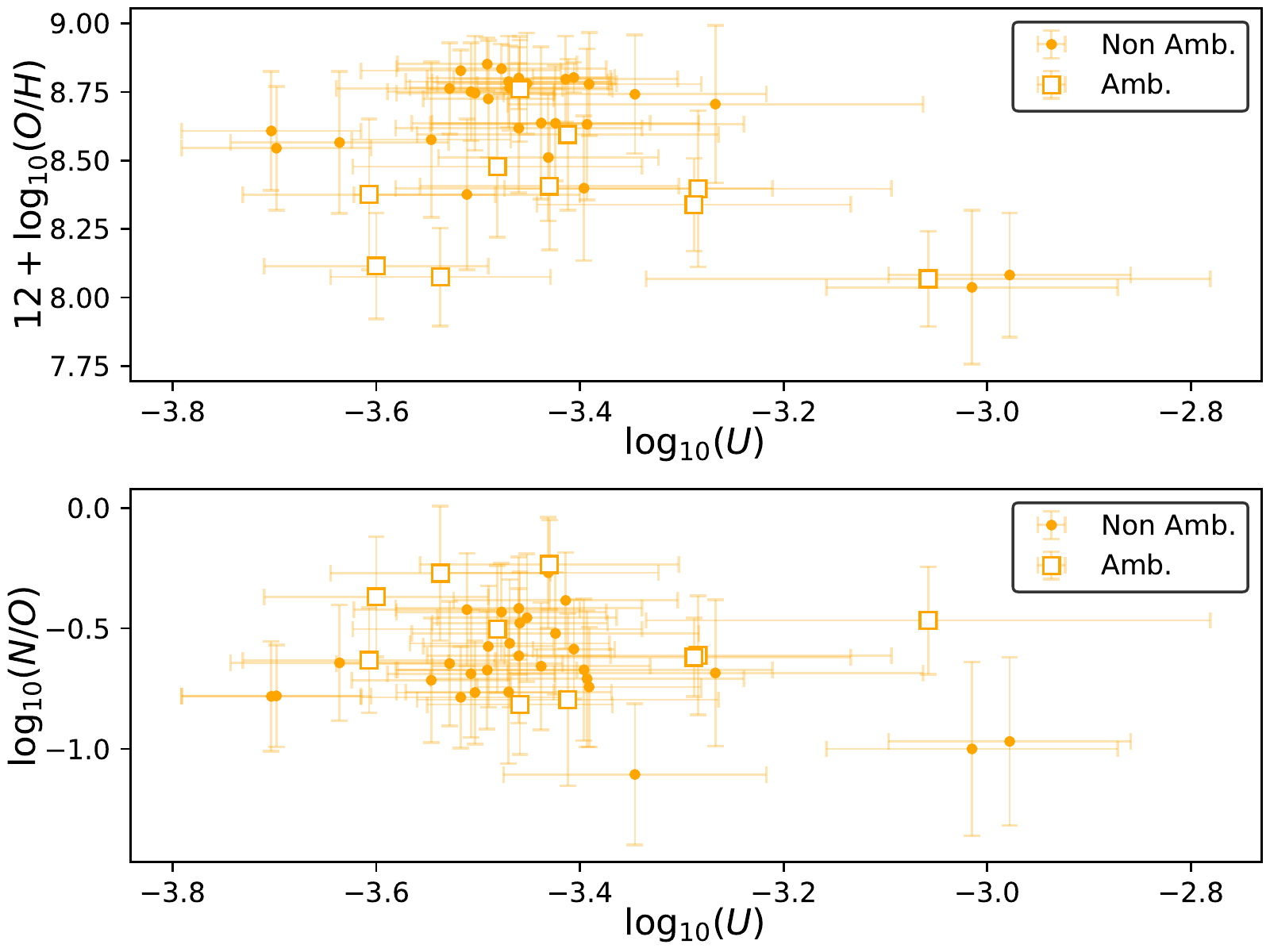}} \vspace{-0.2in} \end{minipage} 
	\end{tabular}
	\caption{Same as Fig. \ref{Chem_SFG} but for LINERs.}
	\label{Chem_L2}
\end{figure*}

%\begin{figure}
%	\centering
%	\includegraphics[width=\hsize]{Images/NOOH_relation_sfg.pdf}
%	\caption{Diagram $\log \left( N/O \right)$ vs $12+\log \left( O/H \right) $ for SFG. Dot symbols are used for non-ambiguous galaxies and square-empty symbols for ambiguous ones. We also present the relations obtained by \citet{Andrews_2013}, \citet{Perez-Montero_2014} and \citet{Belfiore_2015}.}
%	\label{NOOH_sfg}
%\end{figure}

\begin{table*}
	\caption{Statistics of the chemical abundances and ionization parameters for each spectral type of galaxies.}
	\label{Stats_chem}     
	\centering          
	\begin{tabular}{ll|lll|lll|lll}
		\multicolumn{2}{l}{} &
		\multicolumn{3}{|l}{\boldmath$12+\log_{10} \left( O/H \right) $} & \multicolumn{3}{|l}{\boldmath$\log_{10} \left( N/O \right) $} & \multicolumn{3}{|l}{\boldmath$\log_{10} \left( U \right) $}  \\ \hline \textbf{Type} & \textbf{N\boldmath$^{\circ}$} & \textbf{Median} & \textbf{Std. Dev.} & \textbf{Range} & \textbf{Median} & \textbf{Std. Dev.} & \textbf{Range} & \textbf{Median} & \textbf{Std. Dev.} & \textbf{Range} \\ \hline \textbf{SFG} & 87 & 8.71 & 0.16 & [7.85, 8.80] & -0.73 & 0.22 & [-1.41, -0.38] & -3.21 & 0.17 & [-3.43, -2.45]  \\  \textbf{Seyferts 2} & 16 & 8.78 & 0.23 & [8.15, 8.90] & -0.43 & 0.15 & [-0.67, -0.20] & -1.64 & 0.26 & [-2.09, -1.25] \\  \textbf{LINERs} & 40 & 8.63 & 0.26 & [8.04, 8.85] & -0.63 & 0.19 & [-1.11, -0.24] & -3.46 & 0.15 & [-3.70, -2.98] \\ \hline  
	\end{tabular}      
\end{table*}

\subsection{Biases in the selection of the sample}
\label{subsec22}

Since we do not consider the whole sample of galaxies from the Palomar Survey, we may introduce a bias in some of the host galaxy properties that we analyze, with respect to the parent sample. In this work, we consider morphological types, the absolute B-magnitude ($M_{B}^{0}$) and the stellar mass ($M_{*}$) for all spectral type of galaxies (SFG, Seyferts 2 and LINERs). Apart from the stellar mass (see Sec. \ref{subsec43} for more details), we retrieve the rest of host galaxy properties from \citet{Ho_III_1997, Ho_VII_2009}. Although they quantify the morphology with the de Vaucouleurs' number \citeyearpar{Vaucouleurs_1963}, we consider a total of eight morphological classes to guaranteed a detailed binning for our sample. These classes are defined in Tab. \ref{Class_tab} and they are based on Tab. 3 from \citet{deVau_1977}. The morphological distribution of our sample of galaxies is shown in Tab. \ref{morpho_distr}.

In order to quantify these possible biases, we perform a Kolmogorov-Smirnov test (hereafter KS-test, \citealt{Kolmogorov_1933}) for continuous variables (magnitude and stellar mass) and a Anderson-Darling test (hereafter AD-test, \citealt{Anderson_test}) for discrete variables (morphology). We assume as null hypothesis that both distributions are drawn from the same distribution, and we consider the hypothesis rejected for a p-value $< 0.05$. The AD-test shows that no bias was introduced in the morphological type during the construction of the sample (p-value $\approx 0.10$). The distribution of the absolute B-magnitude also shows that no bias was introduced (the KS-test returns p-value $\approx 0.26$). However, the KS-test for the stellar mass shows (p-value $\approx 0.015$) that a slight bias is introduced: there is a decrease in the number of massive galaxies ($M_{*} > 10^{9.6} \ M_{\odot }$) while there is an increase of galaxies with stellar masses in the range $10^{9} \ M_{\odot } < M_{*} < 10^{9.6} \ M_{\odot }$.  

Another important aspect related to the selection of our sample is the size of the region observed. A larger region observed implies that the contribution from the host galaxy is larger, and could affect the metallicities derived from their spectra. The KS-test shows that no bias was introduced during the selection of our final sample, obtaining p-value $\approx 0.24$.

%\textbf{Another important aspect related to the selection of our sample is the size of the region observed. A larger region observed implies that the contribution from the host galaxy is larger, and could affect the metallicities derived from their spectra. We show in Tab. \ref{Apertures} the statistics of the nuclear radius for each spectral type. Although there is a high dispersion of values, with standard deviations above $> 100$ pc and large ranges of values, the median values of each spectral type are close. Therefore, no biased is introduced among the three types of galaxies.}

Considering the distributions shown in Fig. \ref{Bias_prop} and the statistical results obtained for these distributions, we conclude that our selected galaxies constitute a representative sample of the parent sample.

\section{Chemical abundances and ionization parameters}
\label{sec3}
\label{s3}

\subsection{Results for each spectral type}
\label{subsec31}
Tab. \ref{Stats_chem} shows the median results of the chemical abundances in the nuclear region of each spectral type. The highest values of the chemical abundances ratios $12+\log \left( O/H \right) $ and $\log \left( N/O \right) $ are obtained for Seyferts 2, although the median values are the same within the errors in the case of $12+\log \left( O/H \right) $. The chemical abundance ratio $\log \left( N/O \right) $ is 2$\sigma $ higher for Seyferts 2 than for SFG, and $1\sigma $ higher than for LINERs.

The ionization parameter $\log \left( U \right) $ shows the highest values for Seyferts 2, that present a range of values that does not overlap with the range of SFG or LINERs. Chemical abundances and ionization parameters for the nuclear region of each galaxy in our sample are listed in Tab. \ref{TabA3}.

\subsection{Relation between O/H and N/O}
\label{subsec32}

Fig. \ref{Chem_SFG} (a), bottom,  shows that there is a correlation between the chemical abundance ratios $12+\log \left( O/H \right) $ and $\log \left( N/O \right) $. At low  Oxygen abundances ($12+\log \left( O/H \right) < 8.6$), the chemical abundance ratio $\log \left( N/O \right) $ shows a high dispersion of values. For higher Oxygen abundances, both ratios seem to scale but the Pearson's correlation coefficient is low ($r = 0.56$). In Fig. \ref{Chem_SFG} (a) bottom plot we present this relation and we compare it to previous results in the literature. Particularly, we consider three relations: that obtained by \citet{Andrews_2013} using galaxies from SDSS; that obtained by \citet{Perez-Montero_2014} for the grids of models used in \textsc{HCm} by using a sample of galaxies from \citet{Marino_2013} whose chemical abundance determinations were made applying the direct method; and that obtained by \citet{Belfiore_2015} considering a sample of galaxies from MaNGA.

Bottom plots in Fig. \ref{Chem_S2} (a) and Fig. \ref{Chem_L2} (a) show that there is no correlation in neither Seyferts 2 nor LINERs: the Pearson's correlation coefficients show that both chemical abundances are independent in Seyferts 2 ($r = -0.22$) and in LINERs ($r = -0.11$).

\subsection{Relations between the chemical abundances and ionization parameters}
\label{subsec33}
Fig. \ref{Chem_SFG} (b) top shows that the chemical abundance $12+\log \left( O/H \right) $ is anti-correlated with the ionization parameter $\log \left( U \right) $ (the Pearson's correlation coefficient is $r = -0.92$). The chemical abundance ratio $\log \left( N/O \right) $ shows a concentration of values in the range of $\left[ -1.0, -0.5 \right] $ for the lowest ionization parameters $\log \left( U \right) \in \left[ -3.5, -3.1 \right] $ (see Fig. \ref{Chem_SFG} (b) bottom).

For AGN (see Fig. \ref{Chem_S2} and Fig. \ref{Chem_L2} (b) top), the chemical abundance $12+\log \left( O/H \right) $ does not show any correlation with the ionization parameter for neither Seyferts 2 ($r = 0.36$) or LINERs ($r = -0.41$). A similar behavior is observed in Fig. \ref{Chem_S2} and Fig. \ref{Chem_L2} (b) bottom for the chemical abundance ratio $\log \left( N/O \right) $, with $r = 0.49$ for Seyferts 2 and $r = -0.25$ for LINERs.
\section{Relation between chemical abundances and properties of the host galaxies}
\label{sec4}

\begin{figure}
	\centering
	\begin{tabular}{cc}
		\begin{minipage}{0.05\hsize}\begin{flushright}\textbf{(a)} \end{flushright}\end{minipage}  &  \begin{minipage}{0.93\hsize}\centering{\includegraphics[width=1\textwidth]{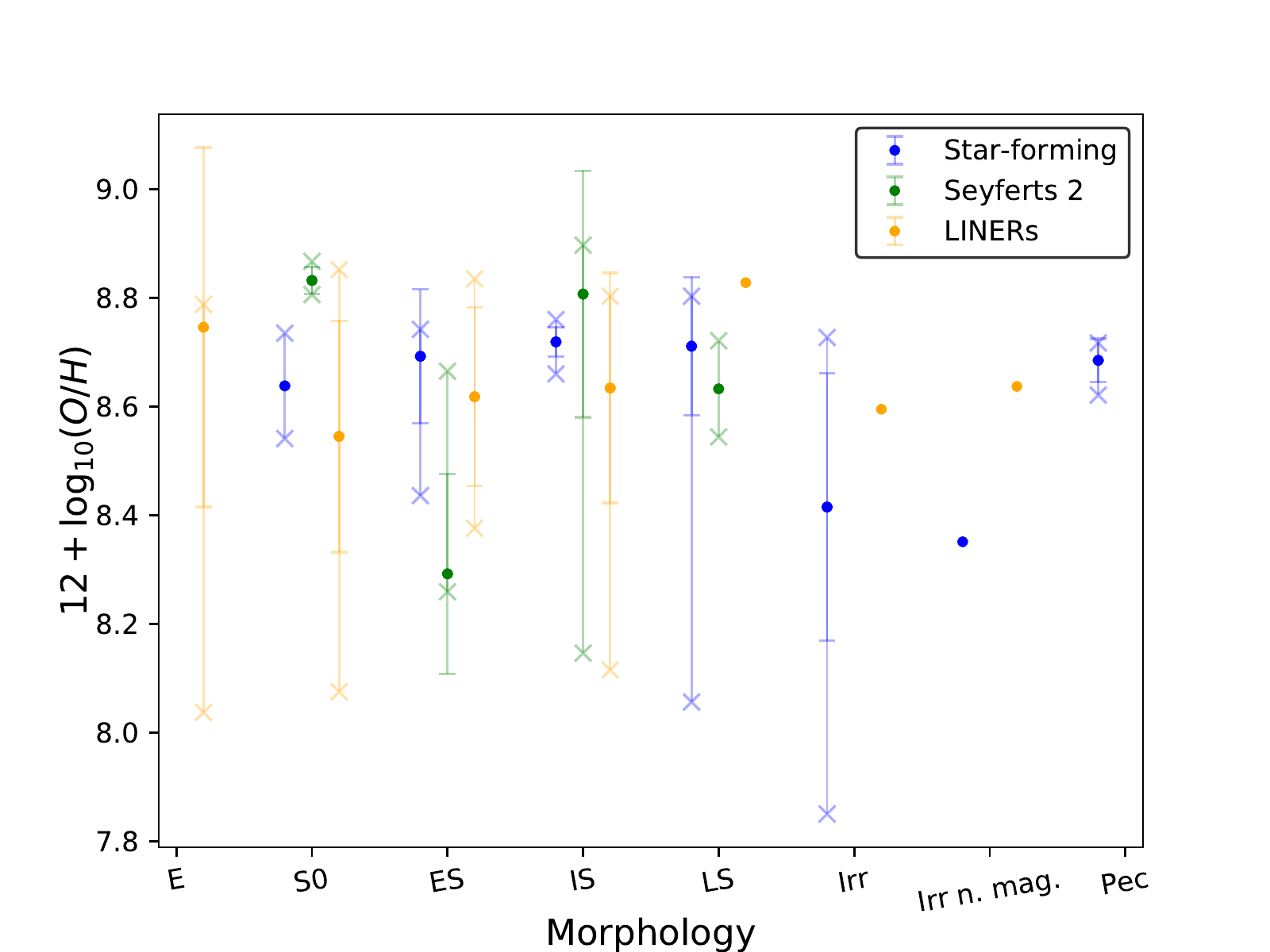}} \vspace{-0.15in} \end{minipage} \\
		\begin{minipage}{0.05\hsize}\begin{flushright}\textbf{(b)} \end{flushright}\end{minipage}  &  \begin{minipage}{0.93\hsize}\centering{\includegraphics[width=1\textwidth]{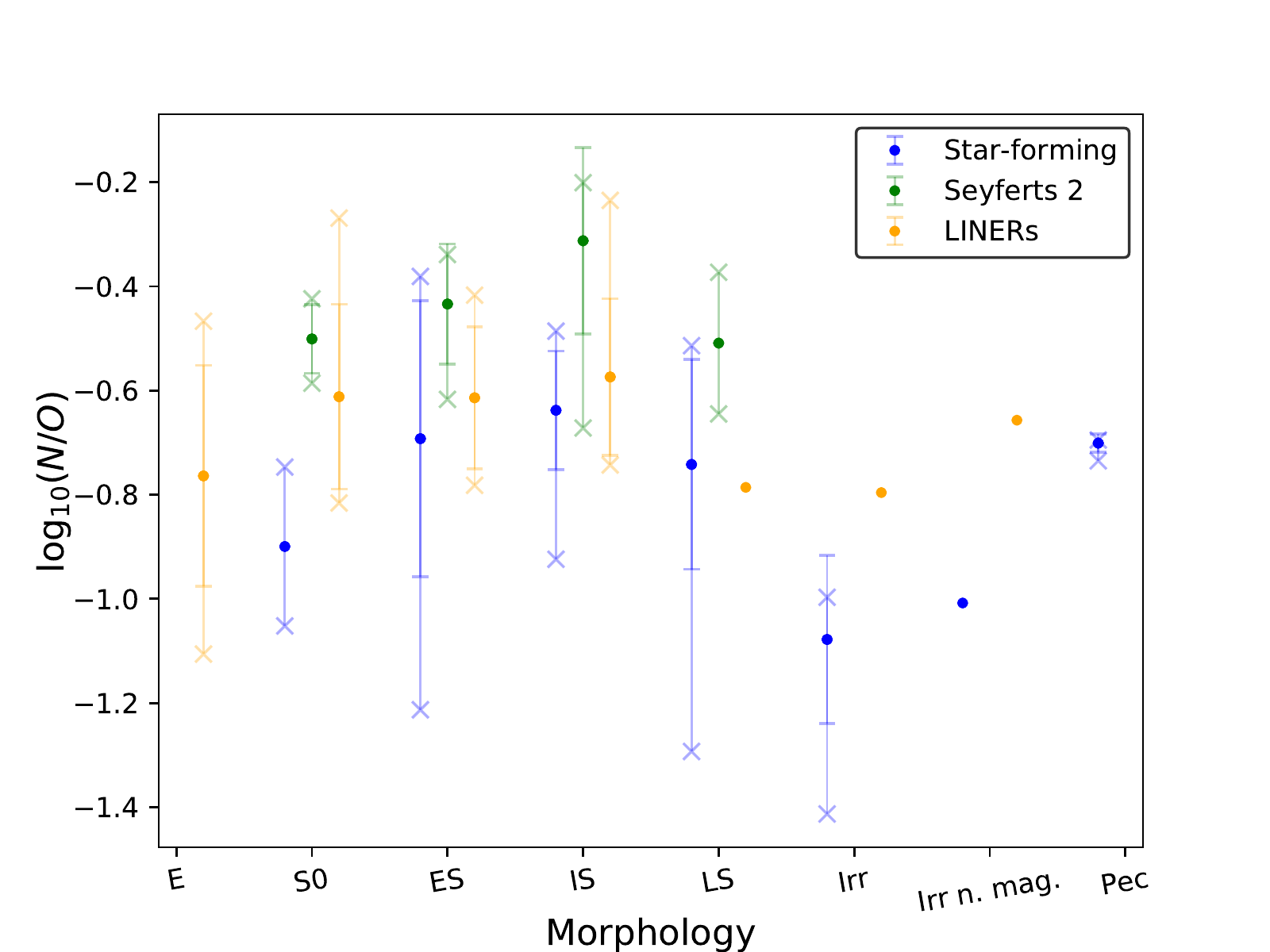}} \vspace{-0.2in} \end{minipage} \\
	\end{tabular}
	\caption{Chemical abundances $12+\log_{10} \left( O/H \right) $ (a) and $\log_{10} \left( N/O \right) $ (b) for each morphological type. The points correspond to the median value, the error bars marked with ``-'' denote the standard deviation (calculated only for classes with 3 or more objects) and the error bars marked with ``x'' the minimum and maximum value (calculated only for classes with 2 or more objects) for each morphology.}
	\label{Morpho_met}
\end{figure}

We study the dependency of the chemical abundances in the nuclear region with different host galaxy properties such as morphology, luminosity, stellar mass and mass of the Supermassive Black Hole. We use SFG as a reference sample by obtaining the same relations between chemical abundances and host galaxy properties already reported in the literature.
\subsection{Morphology}
\label{subsec41}
In Fig. \ref{Morpho_met} (a) and (b) we show the median values of the chemical abundances for each spectral type and for each morphology. The chemical abundance $12+\log \left( O/H \right) $ shows little variation for the morphological type in Seyferts 2 or LINERs. There is a remarkable decrease of the Oxygen abundance for irregular SFG, but there are only 8 galaxies with this morphology (7 as Irr. galaxies and 1 as Irr. n. mag.). Fig. \ref{Morpho_met} (b) shows that the same decrease is observed for the chemical abundance ratio $\log \left( N/O \right) $.

In general, Seyferts 2 present higher median values of the chemical abundance ratio $\log \left( N/O \right) $ than SFG or LINERs, although the results for the three spectral types are compatible within the errors.

\begin{figure*}
	%\resizebox{\hsize}{!}
	\begin{tabular}{cccc}
		\begin{minipage}{0.02\hsize}\begin{flushright}\textbf{(a)} \end{flushright}\end{minipage}  &  \begin{minipage}{0.45\hsize}\centering{\includegraphics[width=1\textwidth]{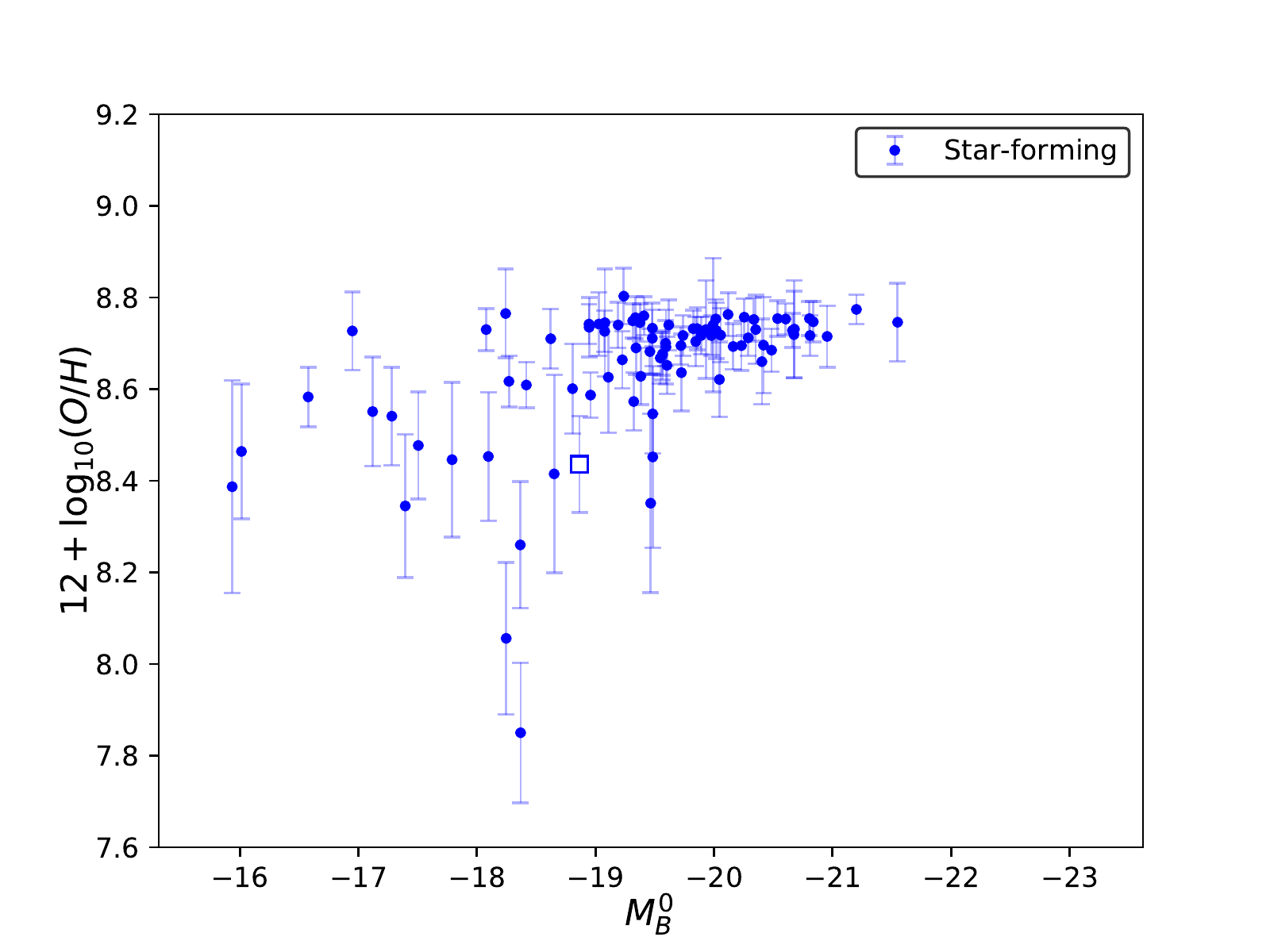}} \vspace{-0.15in} \end{minipage} & \begin{minipage}{0.02\hsize}\begin{flushright}\textbf{(b)} \end{flushright}\end{minipage}  &  \begin{minipage}{0.45\hsize}\centering{\includegraphics[width=1\textwidth]{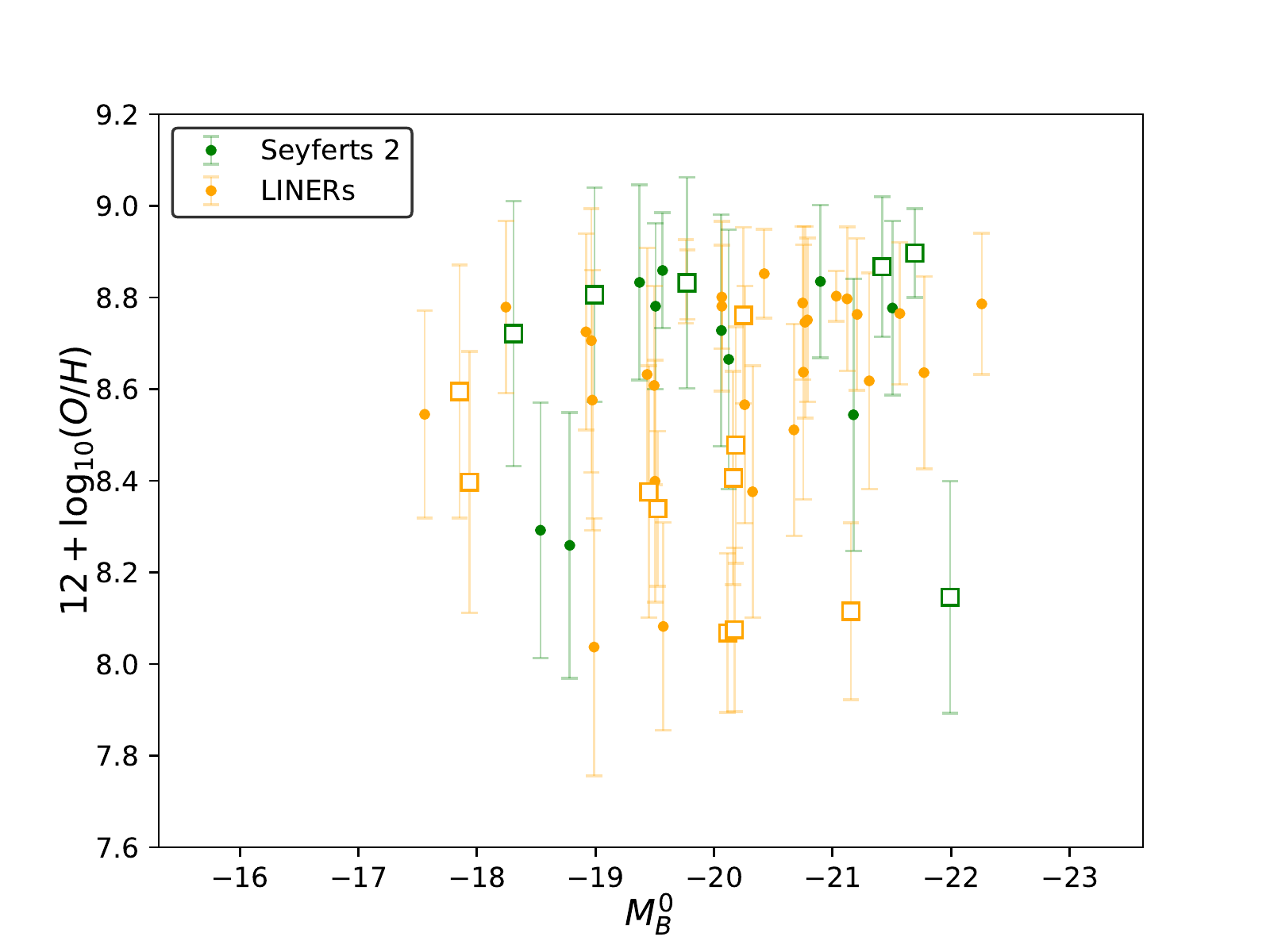}} \vspace{-0.15in} \end{minipage} \\ %& \\
		\begin{minipage}{0.02\hsize}\begin{flushright}\textbf{(c)} \end{flushright}\end{minipage}  &  \begin{minipage}{0.45\hsize}\centering{\includegraphics[width=1\textwidth]{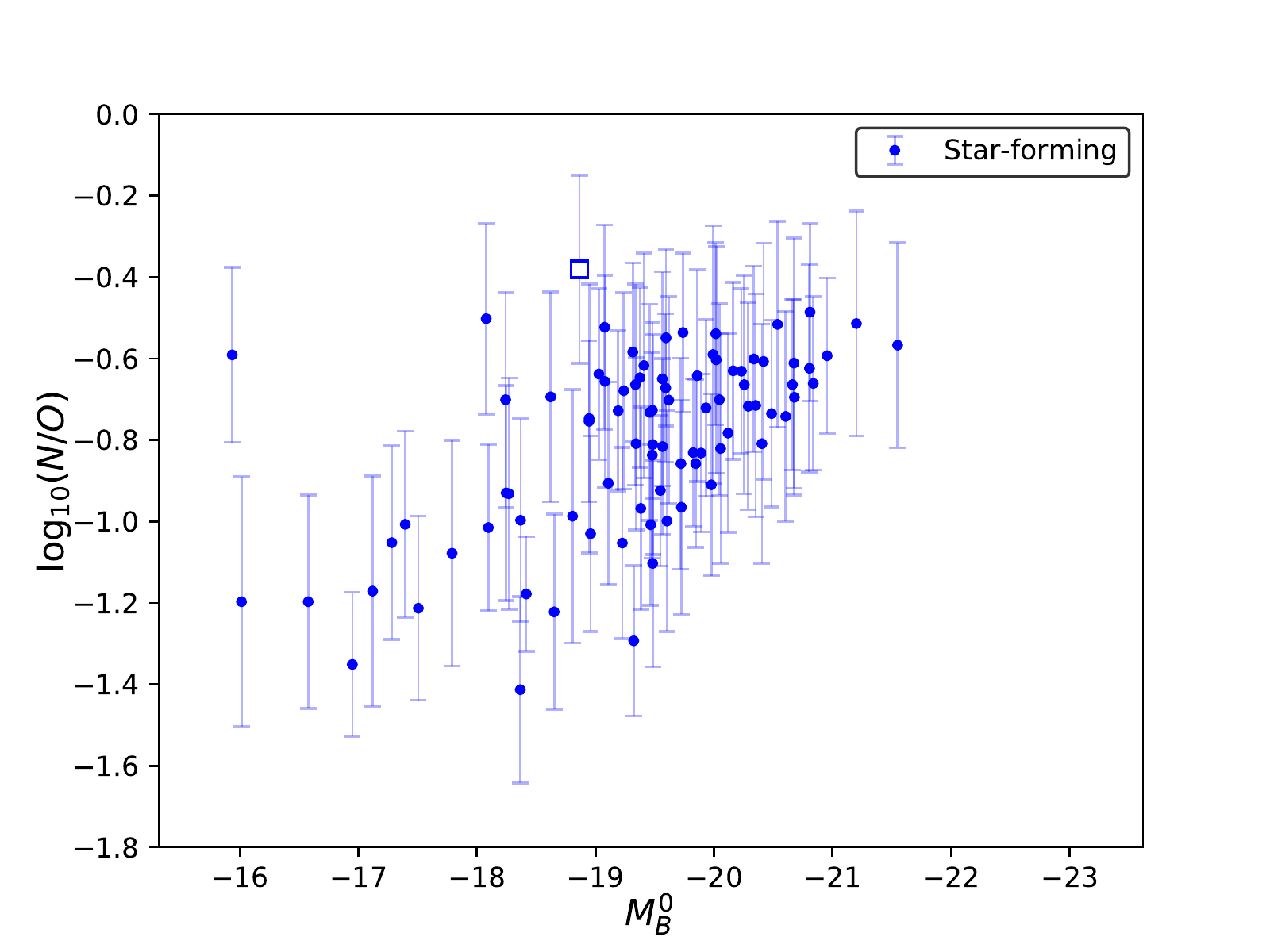}} \vspace{-0.2in} \end{minipage} & \begin{minipage}{0.02\hsize}\begin{flushright}\textbf{(d)} \end{flushright}\end{minipage}  &  \begin{minipage}{0.45\hsize}\centering{\includegraphics[width=1\textwidth]{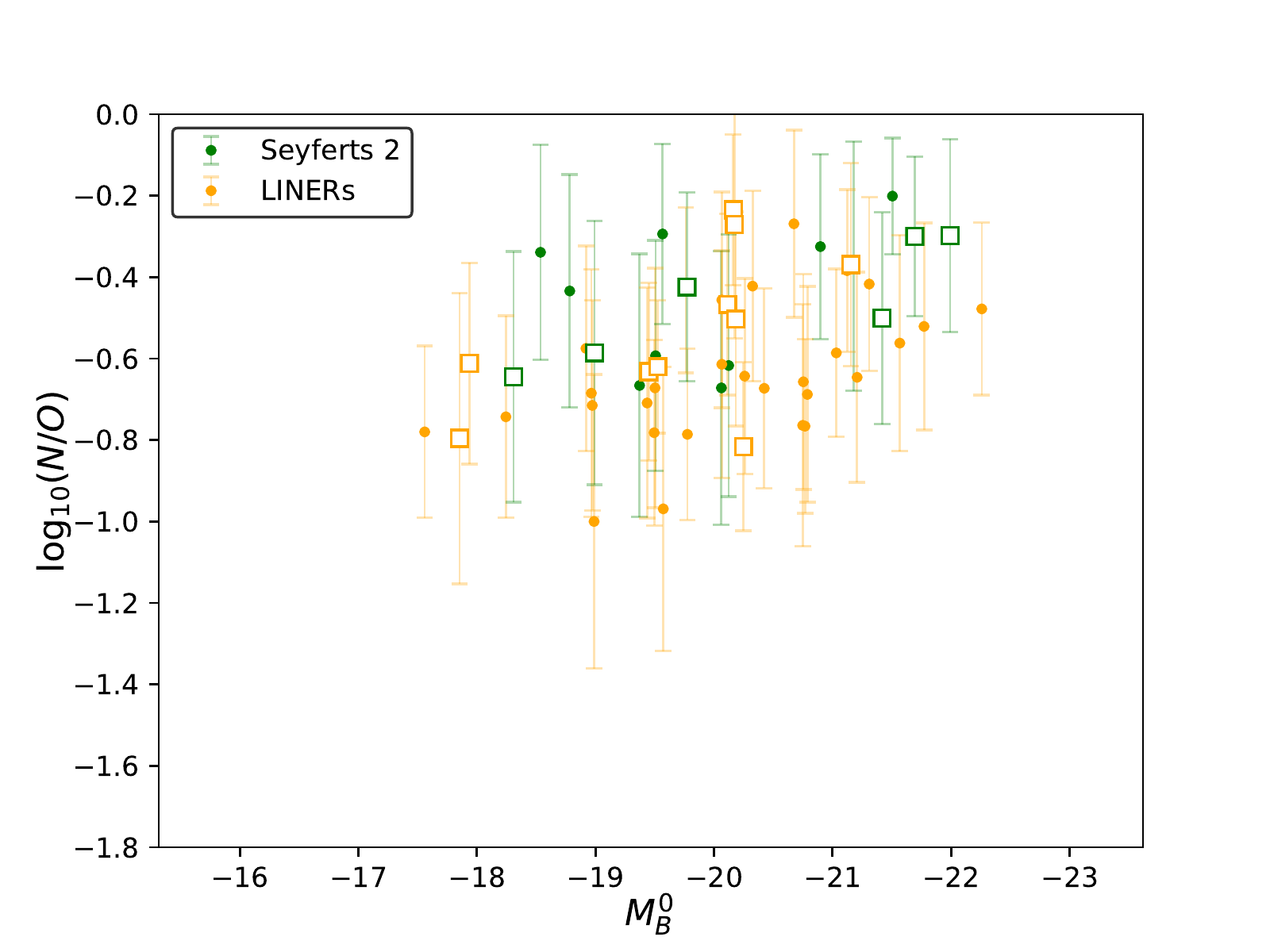}} \vspace{-0.2in} \end{minipage} \\ 
	\end{tabular}
	\caption{Chemical abundances $12+\log_{10} \left( O/H \right) $ (a) and (b), and $\log_{10} \left( N/O \right) $ (c) and (d) vs the corrected absolute B-magnitude ($M_{B}^{0}$). The left column shows the results for star-forming galaxies and the right column those for AGN (Seyferts 2 and LINERs).}
	\label{Magb_met}
\end{figure*}

\begin{figure*}
	%\resizebox{\hsize}{!}
	\begin{tabular}{cccc}
		\begin{minipage}{0.02\hsize}\begin{flushright}\textbf{(a)} \end{flushright}\end{minipage}  &  \begin{minipage}{0.45\hsize}\centering{\includegraphics[width=1\textwidth]{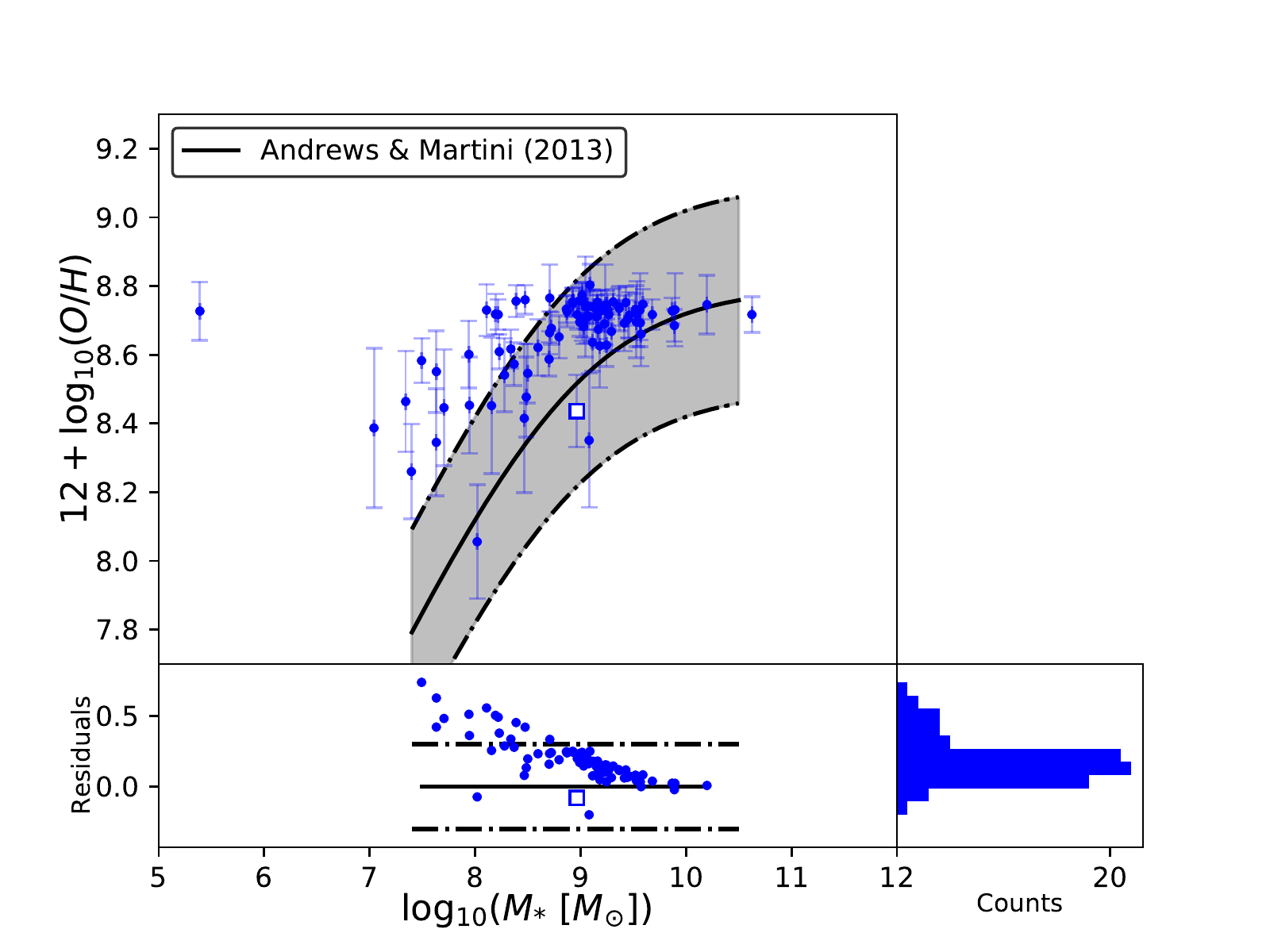}} \vspace{-0.15in} \end{minipage} & \begin{minipage}{0.02\hsize}\begin{flushright}\textbf{(b)} \end{flushright}\end{minipage}  &  \begin{minipage}{0.45\hsize}\centering{\includegraphics[width=1\textwidth]{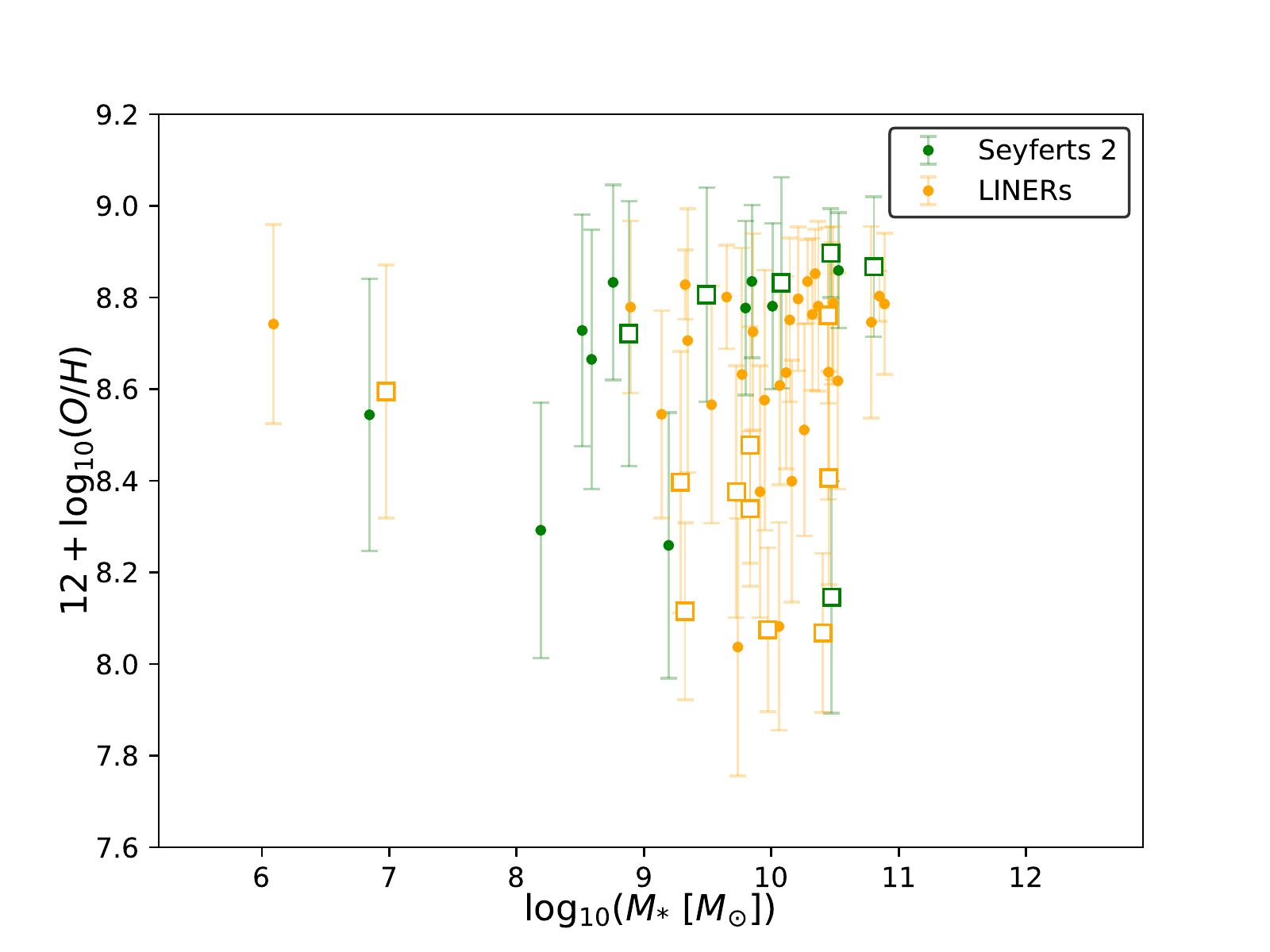}} \vspace{-0.15in} \end{minipage} \\ %\vspace{0.005in} & \vspace{0.005in} \\
		\begin{minipage}{0.02\hsize}\begin{flushright}\textbf{(c)} \end{flushright}\end{minipage}  &  \begin{minipage}{0.45\hsize}\centering{\includegraphics[width=1\textwidth]{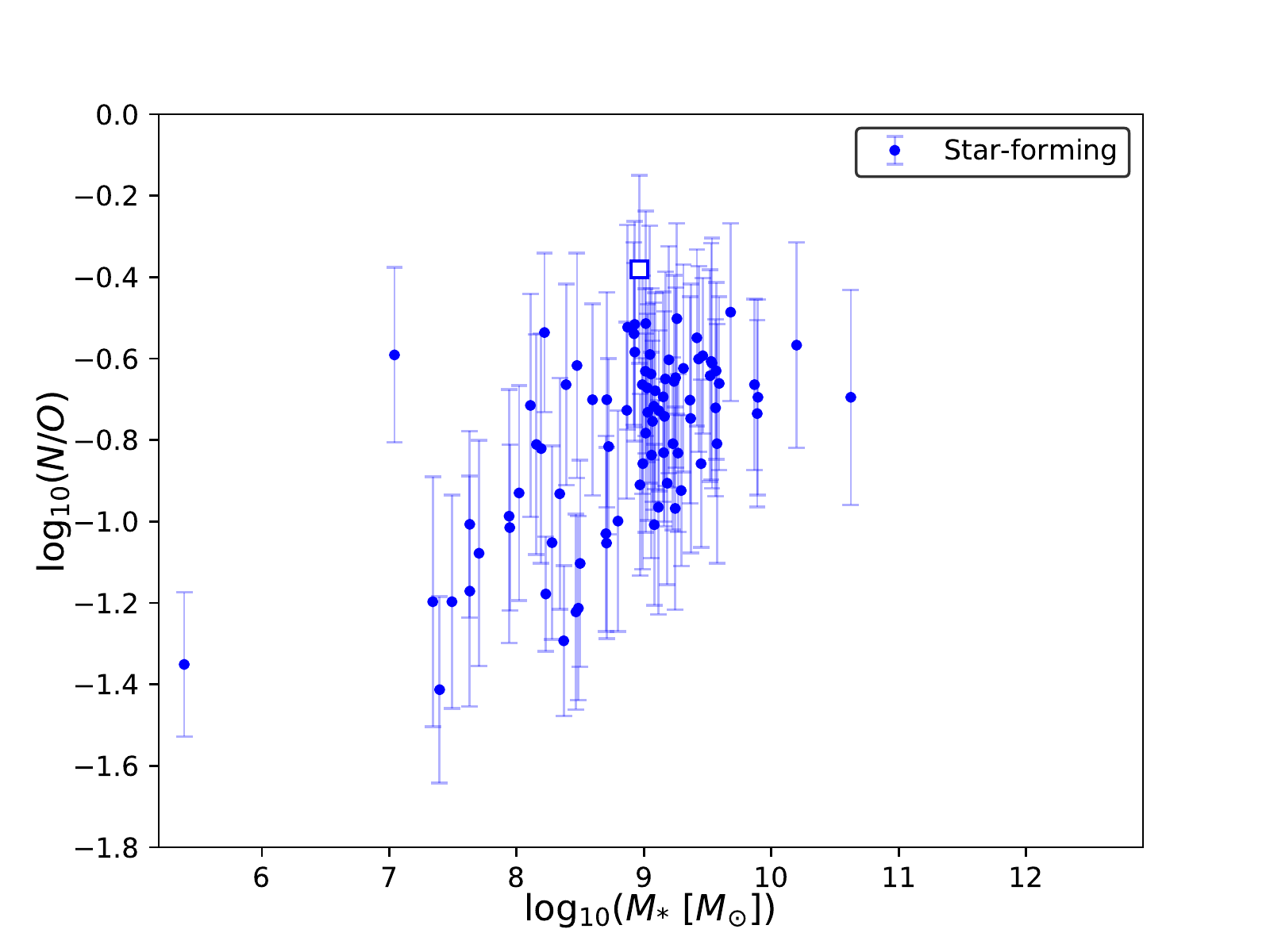}} \vspace{-0.2in} \end{minipage} & \begin{minipage}{0.02\hsize}\begin{flushright}\textbf{(d)} \end{flushright}\end{minipage}  &  \begin{minipage}{0.45\hsize}\centering{\includegraphics[width=1\textwidth]{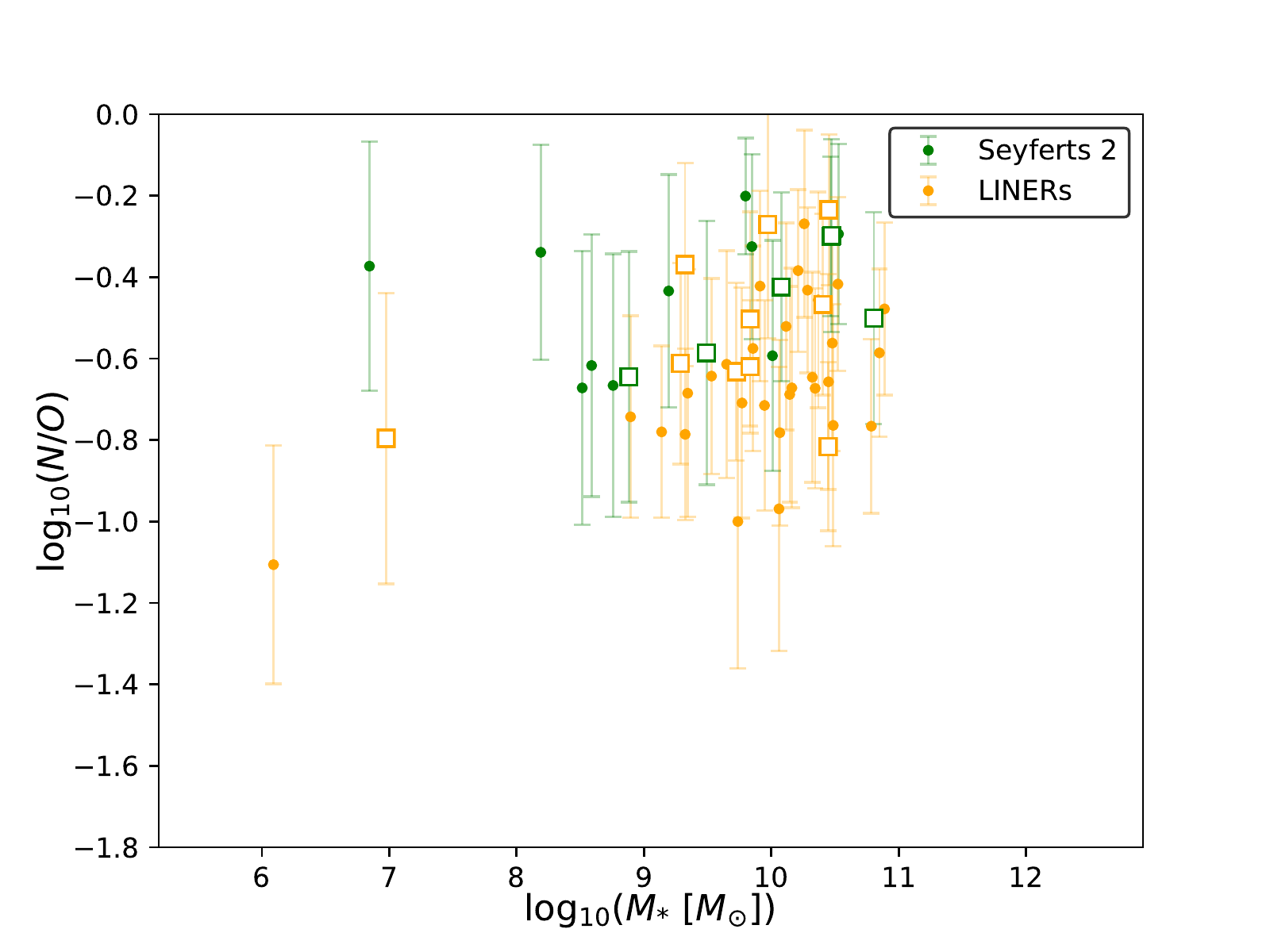}} \vspace{-0.2in} \end{minipage} \\ 
	\end{tabular}
	\caption{Chemical abundances $12+\log_{10} \left( O/H \right) $ (a) and (b), and $\log_{10} \left( N/O \right) $ (c) and (d) vs the stellar mass ($M_{*}$). The left column shows the results for star-forming galaxies and the right column those for AGN (Seyferts 2 and LINERs). In (a) we present the fit obtained by \citet{Andrews_2013} with a 5$\% $ of uncertainty and the corresponding residuals of our data.}
	\label{Ste_met}
\end{figure*}

\subsection{Absolute B-magnitude}
\label{subsec42}
To calculate the absolute B-magnitudes corrected from reddening ($M_{B}^{0}$), we retrieve the apparent B-magnitudes ($m_{B}$) and distances provided by \citet{Ho_III_1997} and we use the galactic HI maps presented by \citet{Burstein_1982, Burstein_1984} and the extinction curve by \citet{Howarth_1983}.

Analyzing SFG, Fig. \ref{Magb_met} (a) reveals that they cluster at $12 + \log \left( O/H \right) \approx 8.7$ for high luminosities ($M_{B}^{0} < -19.5$). Fainter SFG show higher dispersion of values. Fig. \ref{Magb_met} (b) shows that, for a given value of $M_{B}^{0}$, there is no significant differences between Seyferts 2 and LINERs, since both present a considerable dispersion of values. Moreover, the Pearson's correlation coefficients ($r = -0.076$ and $r = -0.11$, respectively) reveal that there is no correlation between this chemical abundance and the absolute B-magnitude. 

The chemical abundance ratio $\log \left( N/O \right) $ shows less dispersion of values for Seyferts 2 (being around $\sim -0.3$) than for LINERs (see Fig. \ref{Magb_met} (d)). In the case of SFG, $\log \left( N/O \right) $ seems to increase for high luminosities ($M_{B}^{0} < -19.5$), but it is not significant ($r = 0.33$).

\subsection{Stellar mass}
\label{subsec43}
To estimate the stellar mass ($M_{*}$) in our sample of galaxies, we use the empirical calibration presented by \citet{Cluver_2014}. This calibration was calculated by using a sample of galaxies whose stellar mass was estimated following the methodology from \citet{Taylor_2011}, consisted on the use of synthetic stellar population model (particularly those proposed by \citet{Bruzual_2003} to match aperture photometry, and assuming the Initial Mass Function from \citet{Chabrier_2003}. This calibration is based on using infrared magnitudes from the \textit{Wide-field Infrared Survey Explorer} (hereafter WISE):
\begin{equation}
\label{3} 
\log_{10} \left( M_{*} \left[ M_{\odot } \right]  \right) = -2.54 \left( M_{W_{1}} - M_{W_{2}} \right) -0.4 \left( M_{W_{1}} - 3.24 \right) -0.17
\end{equation}
being $M_{W_{1}}$ and $M_{W_{2}}$ the absolute magnitude in the WISE-bands $W_{1}$ (at $\lambda = 3.4 \ \mu\mathrm{m}$) and $W_{2}$ (at $\lambda = 4.6 \ \mu\mathrm{m}$). We obtained the values of the absolute magnitudes measured in WISE after a crossmatching between our sample of galaxies and the data presented in the \textsc{Infrared Science Archive}\footnote{IRSA: \url{https://irsa.ipac.caltech.edu/frontpage/}.}. The galaxy IC 10, classified as SFG, was not found in the data base so it is omitted in our study. 

The behavior shown in Fig. \ref{Ste_met} (a) reveals that for low stellar masses, $10^{7}-10^{9} \ M_{\odot }$, the chemical abundance $12+ \log \left( O/H \right) $ scales with $M_{*}$ in SFG. For higher masses ($M_{*} > 10^{9} \ M_{\odot }$) the chemical abundance essentially remains constant. For low stellar mass SFG, the chemical abundance ratio $\log \left( N/O \right) $ also scales with $M_{*}$, as it can be seen in Fig. \ref{Ste_met} (c).

Analyzing Seyferts, the chemical abundance $12+\log \left( O/H \right) $ practically remains constant, but three galaxies (NGC 660, NGC 3185 and NGC 3646), do not follow this behavior (see Fig. \ref{Ste_met} (b)). The chemical abundance ratio $\log \left( N/O \right) $ does not show any particular behavior with $M_{*}$ (see Fig. \ref{Ste_met} (d)).

Finally, analyzing LINERs, $12+\log \left( O/H \right) $ and $M_{*}$ (see Fig. \ref{Ste_met} (b)) are not correlated (the Pearson's coefficient correlation is $r = 0.038$). This lack of correlation is also presented between $\log \left( N/O \right) $ and $M_{*}$.

\subsection{Mass of the Supermassive Black Hole in AGN}
\label{subsec45}
To estimate the mass of the Supermassive Black Hole, we use the empirical relation given by \citet{McConnell_2013}, which relates $M_{SMBH}$ and the stellar velocity dispersion $\sigma_{*}$ in the galactic center (within the effective radius of the galaxy) as:
\begin{equation}
\label{2} \log_{10} \left( \frac{M_{SMBH}}{M_{\odot }} \right) = 8.32 + 5.64 \log_{10} \left( \frac{\sigma_{*}}{200 \ km\cdot s^{-1}} \right)
\end{equation}
The above relation can only be applied to AGN hosted by galaxies that present a velocity dispersion supported structure (bulge or elliptical core). Therefore, we omit from this study the LINER NGC 5363, hosted by a galaxy with irregular morphology \citep{Ho_III_1997}. We retrieve the values of $\sigma_{*}$ for our AGN sample from \citet{Ho_VII_2009}. 

We also omit the LINER NGC 185 that has a low-massive black hole ($M_{\mathrm{SMBH}} \sim 10^{2.7} \ M_{\odot }$) due to its low stellar velocity dispersion $\sigma_{*} \approx 19.9 \ \mathrm{km}\cdot\mathrm{s}^{-1}$ \citep{Ho_VII_2009}. This low stellar velocity dispersion might be explained because this dwarf elliptical galaxy presents several bright knots that may be identified as its nucleus \citep{Lira_2007}.

\begin{figure*}
	%\resizebox{\hsize}{!}
	\begin{tabular}{cccc}
		\begin{minipage}{0.02\hsize}\begin{flushright}\textbf{(a)} \end{flushright}\end{minipage}  &  \begin{minipage}{0.45\hsize}\centering{\includegraphics[width=1\textwidth]{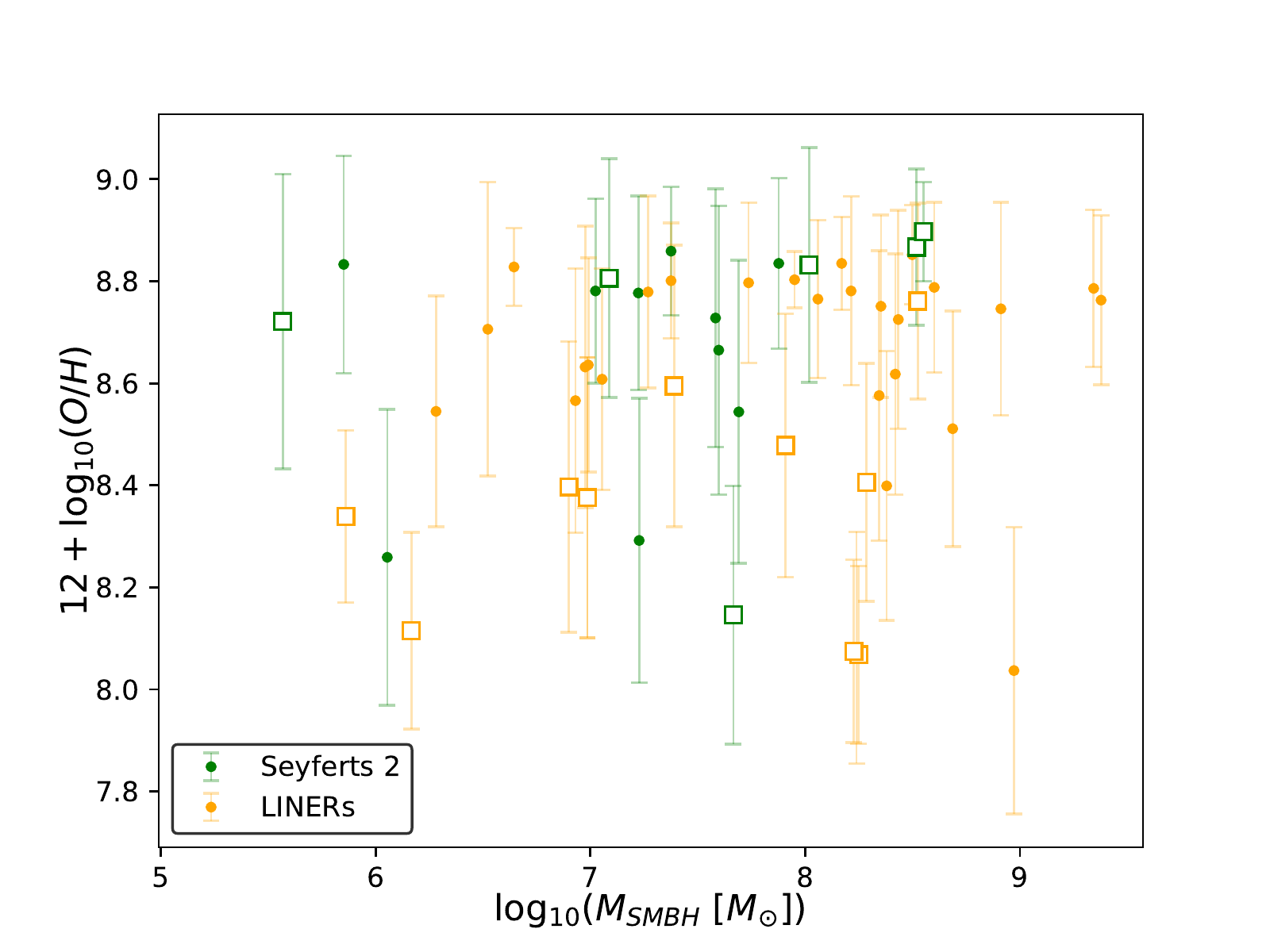}} \vspace{-0.2in} \end{minipage} & \begin{minipage}{0.02\hsize}\begin{flushright}\textbf{(b)} \end{flushright}\end{minipage}  &  \begin{minipage}{0.45\hsize}\centering{\includegraphics[width=1\textwidth]{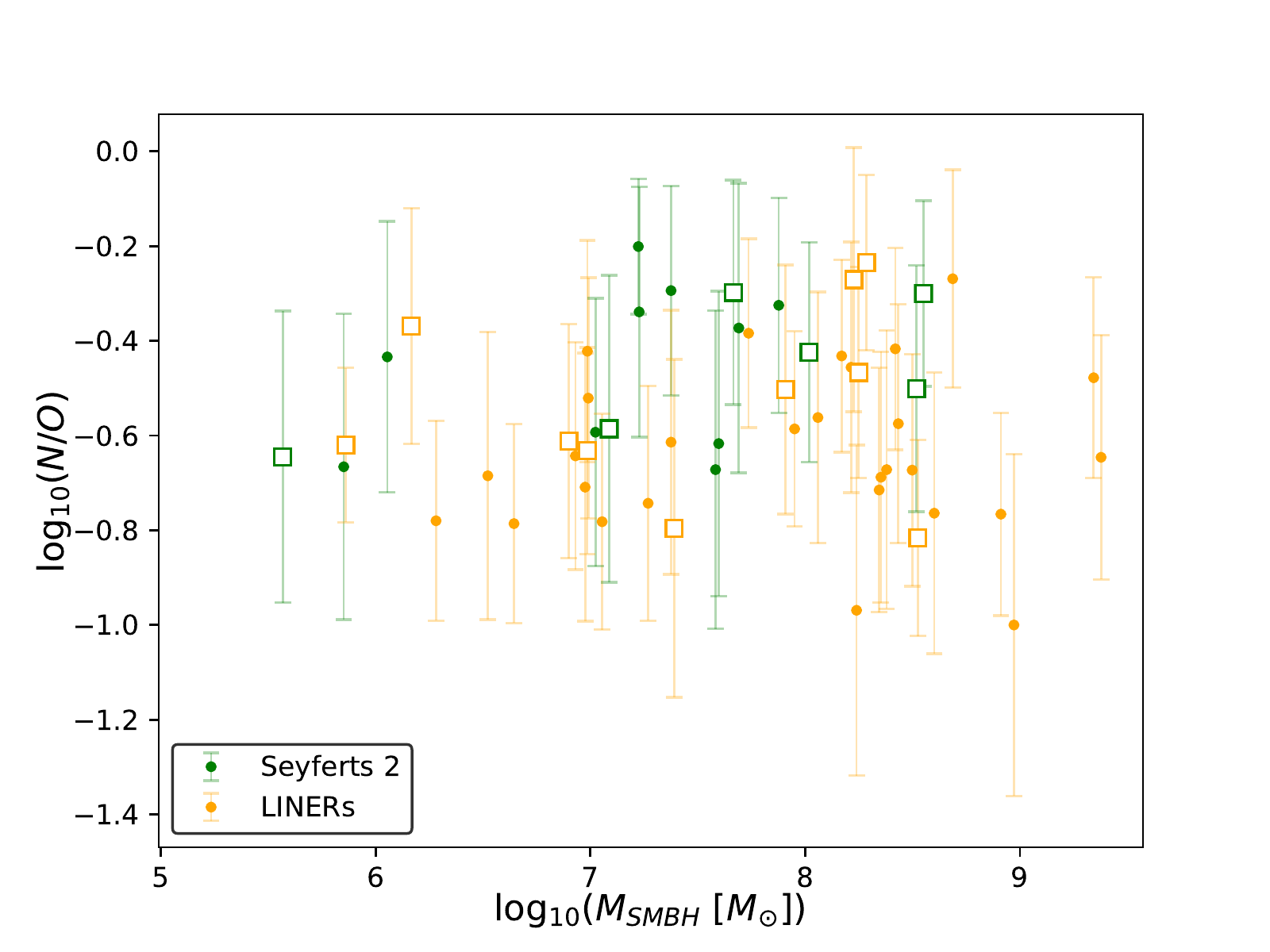}} \vspace{-0.2in} \end{minipage} \\ 
	\end{tabular}
	\caption{Chemical abundances $12+\log_{10} \left( O/H \right) $ (a) and $\log_{10} \left( N/O \right) $ (b) vs the SMBH mass $M_{SMBH}$ for AGN.}
	\label{bh_met}
\end{figure*}

The chemical abundance $12+\log \left( O/H \right) $ does not show any correlation with $M_{SMBH}$ for neither AGN type (see Fig. \ref{bh_met} (a)), being the Pearson's correlation coefficients $r = 0.21$ (Seyferts 2) and $r = 0.0304$ (LINERs). A similar behavior is presented in the chemical abundance ratio $\log \left( N/O \right) $ (see Fig. \ref{bh_met} (b)), although the correlation coefficients are slightly higher, $r = 0.43$ and $r = 0.30$ respectively.

\section{Discussion}
\label{sec5}

We recall that we study the nuclear region of the galaxies from the Palomar Survey, that covers a range between  6.8 pc and 1055.0 pc, below the radius of normalization $r \sim 3$ kpc proposed by \citet{Zaritsky_1994}.

\subsection{Star-forming galaxies}
\label{ssubsec51}

Although our main goal is to analyze chemical abundances in LLAGN, we also perform a study of gas-phase Z in SFG to probe the validity of our methodology.

The range of derived Oxygen abundance values for our subsample of SFG (with a median value close to the solar abundance) is in good agreement with previous studies based on the $T_{e}$-method \citep{Vila-Costas_1993, Andrews_2013}, calibrations based on strong emission lines \citep{Perez-Montero_2013, Dors_2015} or photoionization models \citep{Perez-Montero_2016, Vincenzo_2016}. 

Since the code \textsc{HCm} calculates O/H and N/O independently, we can explore the relation between both abundance ratios. A relation between these chemical abundance ratios has been reported in the literature \citep{Vila-Costas_1993, Andrews_2013, Perez-Montero_2014, Belfiore_2015, Curti_2017}. For low values $12+\log \left( O/H \right) < 8.5$ the chemical abundance ratio $\log \left( N/O \right) $ remains constant, which might be explained by a primary origin of the metals and the variations in $12+\log \left( O/H \right) $ with inflows of gas to the center acting as fuel for the star formation.  For $12+\log \left( O/H \right) > 8.5$, both O/H and N/O seem to  scale, which is the expected behavior if the metals are already present in the gas during the star formation process (secondary origin). While the turning point in this behavior for SFG has been reported at $12+\log \left( O/H \right) \sim 8.4$ \citep{Vincenzo_2016} or $8.5$ \citep{Vila-Costas_1993, Andrews_2013}, we locate it at $\sim 8.6$ (all of these values are compatible considering the uncertainties). In Fig. \ref{Chem_SFG} (a) bottom plot it is shown that for abundances $12+\log \left( O/H \right) > 8.6$, both N/O and O/H scales. However, instead of obtaining a practically constant value for low abundances, we obtain a high dispersion of values (the Pearson's coefficient correlation is $r=-0.14$). This might be explained due to the low number of SFG in our sample with chemical abundances below $12+\log \left( O/H \right) ) = 8.6$. However, it should be noticed that there are some galaxies reported in the literature that do not follow this relation between N/O and O/H, as it is the case of NGC 4670 \cite{Kumari_2018}. Nevertheless, the majority of our SFG present values $12+\log \left( O/H \right) > 8.6$ and Fig. \ref{Chem_SFG} (a) bottom plot shows that the relation between N/O and O/H is well reproduced by the fit proposed by \citet{Perez-Montero_2014} (shaded area delimited by solid lines). This fit (calculated with a sample of galaxies with chemical estimations from the direct method) is expressed in terms of the grids of \textsc{HCm} since it is used when it is impossible to calculate the nitrogen abundance from the emission lines (which is not our case).

We also analyze if the difference found for the  derived Z may come from the properties of their host galaxies. The first property that we analyze is the morphology, as suggested by \citet{Calura_2009}. We find the morphology introduces little changes in the chemical abundances: there is a drop in the chemical abundance $12+\log \left( O/H \right) $ for irregular galaxies, likely caused due to the mass-metallicity relation reported in Sec. \ref{subsec43} since the median stellar mass for this galaxies is low ($\log_{10} \left( M_{*} \left[ M_{\odot } \right] \right) = 7.4$ with $\sigma = 0.9$). It is important to notice that \citet{Kirby_2013} also obtained lower chemical abundances in their sample of dwarf galaxies despite considering the same range of stellar masses as other SFG galaxies. Nevertheless, it remains practically constant for S0, spiral (Sa, Sb and Sc) and peculiar galaxies. This last result is still compatible with that obtained by \citet{Calura_2009}, since they reported that morphology affects metallicities for galaxies at higher redshifts (from $z \sim 0.7 -3.2$), beyond the limits of our sample.

The behavior of $12+\log \left( O/H \right) $ with $M_{B}^{0}$ that we obtain for SFG agrees with previous studies \citep{Lequeux_1979, Contini_2002, Melbourne_2002, Lamareille_2004, Tremonti_2004, Izotov_2015}, showing that there is a luminosity-metallicity relation. However, we report a high dispersion of values for low chemical abundances (i.e. for low luminosities). This relation seems to be also present in the chemical abundance ratio $\log \left( N/O \right) $. Following the O/H and N/O relation in SFG, the brighter SFG are the ones with a secondary origin of the metals.

Finally, we analyze the influence of the stellar mass. It has been reported for many decades the existence of a mass-metallicity relation. Our study confirms this relation when we analyze the nuclear abundances of galaxies, showing that the curve seems to flatten for $M_{*} > 10^{9} \ M_{\odot }$ (as also reported by \citealt{Andrews_2013}), which contrasts with previous studies where the change occurs at higher masses, $M_{*} > 10^{10} \ M_{\odot} $ \citep{Contini_2002, Tremonti_2004}, range of mass for which we do not have a significant number of galaxies. The fit obtained by \citet{Andrews_2013}, whose validity is set in the range $7.4 < \log \left( M_{*} \left[ M_{\odot } \right] \right) < 10.5$, is still compatible with our results (see Fig. \ref{Ste_met} (a)) for $\log \left( M_{*} \left[ M_{\odot } \right] \right) > 8.5$. At lower stellar masses, the fit by \citet{Andrews_2013} underestimates our results of O/H, which might be explained due to the spectra coming from larger physical apertures, since they considered galaxies from SDSS, i.e., they analyzed a higher central region. Several models have been proposed in order to explain this metallicity-stellar mass relation in star-forming galaxies. Some of them propose a selective loss of metals from galaxies with shallow gravitational potentials through galactic outflows \citep{Tremonti_2004}. Others contemplate that, on average, high mass galaxies evolve faster than low mass galaxies, so the ISM in the former ones gets richer in metals \citep{Erb_2006_b}. There are also models that suggest that the recycled gas from previous periods of star-formation can be richer in metals in high mass galaxies \citep{Ma_2016, Vincenzo_2016_b}. There are some claims that other important aspects play roles in this relation such as the redshift \citep{Maiolino_2008, Calura_2009, Arabsalmani_2018, Huang_2019} and the morphology \citep{Calura_2009, Huang_2019}. Nevertheless, the origin of this relation is still an open question. A thorough and recent review of this topic can be found in \citet{Maiolino_2019}.

We report that several SFG show a saturate Oxygen abundance, around the solar value. Considering that we are analyzing chemical abundances in the nuclear region ($r \lesssim 1 $ kpc), this behavior is in good agreement with previous studies of metallicity gradients in SFG, showing a saturation in the nuclear regions \citep{Marquez_2002, Pilyugin_2004}.

In summary, by using the code \textsc{HCm} our study of the chemical abundances in SFG replicates the results obtained during the last decades for these galaxies, probing the soundness our methodology for determining chemical abundances.
\subsection{Active Galactic Nuclei}
\label{subsec52}

Although the number of Seyferts 2 in our study is small (16 galaxies), we also analyzed a significant number of LINERs (40 galaxies).

The AGN nature of LINERs is still an open question \citep{Marquez_2017}. To check their nuclear activity, we crossmatch our list of LINERs with those studied by \citet{Omaira_2009} using multiwavelength observations. From our sample of 40 LINERs, 14 are studied by \citet{Omaira_2009}, revealing that 13 of them (92.9$\% $) show characteristic AGN emission: only the galaxy NGC 4321 is classified as non-AGN. Thus, we statistically assume that our sample of LINERs is composed mainly by galaxies hosting an AGN.

The range of values of $12+\log \left( O/H \right) $ for Seyferts 2 is in good agreement with studies based on optical strong lines calibrations \citep{Storchi_1998, Dors_2015, Castro_2017, Dors_2019} and are slightly lower than those reported by \citet{Thomas_2019}, using photoionization models with \textsc{NebulaBayes}. Despite many of these studies report oversolar abundances for some Seyferts 2 ($12+\log \left( O/H \right) > 9.0$), only three galaxies in our sample reach that limit (within the errors), namely: NGC 3982, NGC 4138 and NGC 4477. As pointed out by \citet{Maiolino_2019}, these high oxygen abundances in Seyferts might be explained with two models that are not incompatible: 1) dust destruction in the NLR causing the release of metals into the ISM \citep{Nagao_2006, Matsouka_2009}; and, 2) AGN-driven outflows which are originated in the central metal-rich regions. In the case of LINERs, we obtain a range of Oxygen abundances pretty similar to that for SFG, although it is slightly lower.

Contrary to SFG, we do not find any relation between the chemical abundance ratios $12+\log \left( O/H \right)$ and $\log \left( N/O \right) $ neither for our samples of Seyferts 2 or LINERs (see Fig. \ref{Chem_S2} and \ref{Chem_L2} (a) bottom), obtaining Pearson's coefficients of $r=-0.22$ and $r=-0.11$, respectively. This result contrasts with the assumptions made in optical strong lines calibrations \citep{Storchi_1998, Castro_2017, Maiolino_2019}. \citet{Perez-Montero_2019} also found a relation between O/H and N/O, using \textsc{HCm} code, for Seyferts 2. Therefore, this discrepancy might be explained due to the limited sample of Seyferts 2. Further studies on Seyferts must be performed so as to check our results.

We do not observe neither any relation between the morphological type and the chemical abundance $12+\log \left( O/H \right) $ for Seyferts 2 or LINERs, although there is slight decrease in Seyferts 2 hosted by galaxies with an ES morphology. This decrease, only reported in Seyferts 2 hosted by ES galaxies, might be explained due to the small sample of Seyferts 2 hosted by ES galaxies (only 3 galaxies, NGC 660, NGC 2273 and NGC 3185) or rejuvenation due to feedback on the nucleus. For all morphological types, the chemical abundance ratio $\log \left( N/O \right) $ is higher in Seyferts 2 than in LINERs.

While we obtain for SFG a relation between $M_{B}^{0}$ and the chemical abundance ratios, this is not observed in the case of AGN. The lower values of the Pearson correlation coefficients indicate that this host galaxy property seems not to be correlated with the metallicity (see Fig. \ref{Magb_met} (b) and (d)) in our sample of AGN. This result contrasts with the study by \citet{Nagao_2006} for quasars at high redshift, reporting the existence of a luminosity-metallicity relation. The small sample of Seyferts 2 requires a re-examination of this result in larger samples of AGN. Such a relation is completely absent for LINERs, for which a high dispersion of values is obtained.

The mass-metallicity relation seems to be present in Seyferts 2 and, although the sample number is small, shows a flattened behavior within the errors (see Fig. \ref{Ste_met} (b) and (d)) as reported by \citet{Thomas_2019}.  However, there are three galaxies, namely NGC 660, NGC 3185 and NGC 3646, that present lower chemical abundances ($12+\log \left( O/H \right) < 8.4$) while other Seyferts 2 seem to cluster at $12+\log \left( O/H \right) = 8.8$ for $\log \left( M_{*} \left[ M_{\odot } \right] \right) > 8$. In the case of LINERs, no relation is found, showing a clearly scatter ($\sim 0.5$ dex) in their chemical abundances.

%The H$_{\alpha }$ luminosity is related to the optical continuum luminosity in AGN, with a slope between the logarithm of both quantities close to unity \citep{Shuder_1981, Greene_2005}, favoring the photoionization model for AGN. Our studies shows that there is no relation between $\log \left[ L \left( H_{\alpha } \right)_{corr} \right] $ and $12+ \log \left( O/H \right) $ or $\log \left( N/O \right) $ for both Seyferts 2 and LINERs.

The mass of the Supermassive Black Hole plays an important role in understanding the AGN phenomena. AGNs with prominent jets are usually associated to higher masses of the black hole \citep{Heckman_Best_2014}. In addition, $M_{SMBH}$ is related with the Eddington ratio \citep{Netzer_2015} and this ratio is proposed to be related to the Unified Model of AGN \citep{Antonucci_1993, Urry_1995} when LINERs are taken into account \citep{Marquez_2017}. Nevertheless, we find that neither $12+\log \left( O/H \right) $ or $\log \left( N/O \right) $ are related to this mass for both types of AGN (see Fig. \ref{bh_met}).

\subsection{Galaxies in the same stellar mass range}
\label{subsec53}

The existence of a mass-metallicity relation for SFG implies that the study of the chemical abundances for the three spectral types (SFG, Seyferts 2 and LINERs) should be done for the same range of stellar masses, to avoid any bias on the stellar mass. Several studies show that the host galaxies of the different spectral types have different stellar masses, being more massive LINERs than Seyferts 2 followed by SFG \citep{Kauffmann_2003, Kewley_2006, Vitale_2013, Thomas_2019}. We obtain a median value of the stellar mass $\log \left( M_{*} \left[ M_{\odot } \right] \right) = 9.03\pm 0.75$ for SFG, $9.64\pm1.03$ in Seyferts 2 and $10.07 \pm 0.91$ for LINERs, implying that, at face values, LINERs are, in fact, hosted by the most massive galaxies as previously reported.

In order to decontaminate from the eventual impact of the mass-metallicity relation and the difference of mass in the host galaxies of different spectral types, we perform a study of the chemical abundances selecting the same range of stellar masses. We established the range $ \log \left( M_{*} \left[ M_{\odot } \right] \right) \in \left[ 9.00, 10.62 \right] $, where the low limit represent the least massive galaxies hosting LINERs and the upper limit the most massive SFG.

\begin{figure*}
	%\resizebox{\hsize}{!}
	\begin{tabular}{cccc}
		\begin{minipage}{0.05\hsize}\begin{flushright}\textbf{(a)} \end{flushright}\end{minipage}  &  \begin{minipage}{0.4\hsize}\centering{\includegraphics[width=1\textwidth]{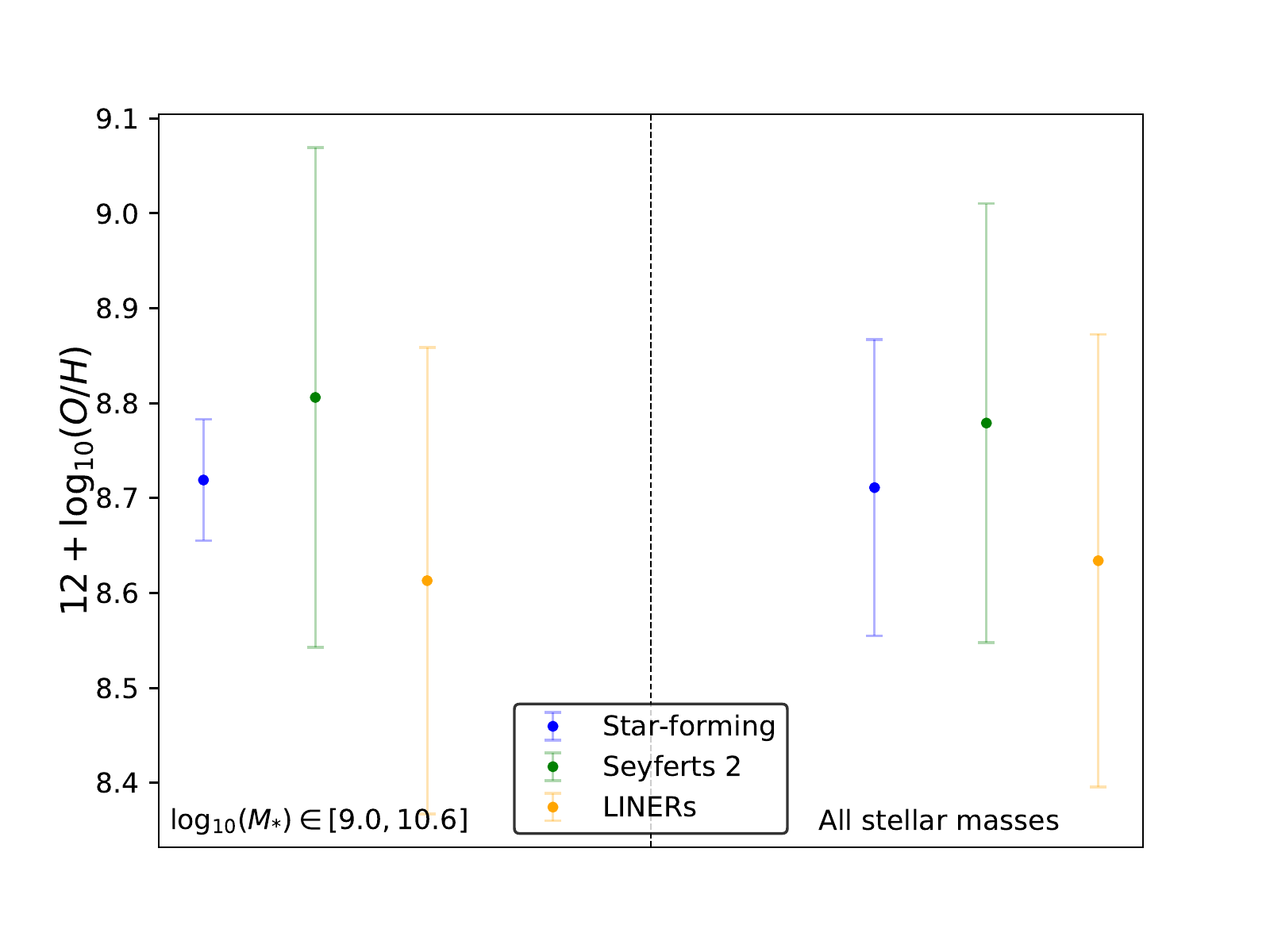}} \vspace{-0.2in} \end{minipage} & \begin{minipage}{0.05\hsize}\begin{flushright}\textbf{(b)} \end{flushright}\end{minipage}  &  \begin{minipage}{0.4\hsize}\centering{\includegraphics[width=1\textwidth]{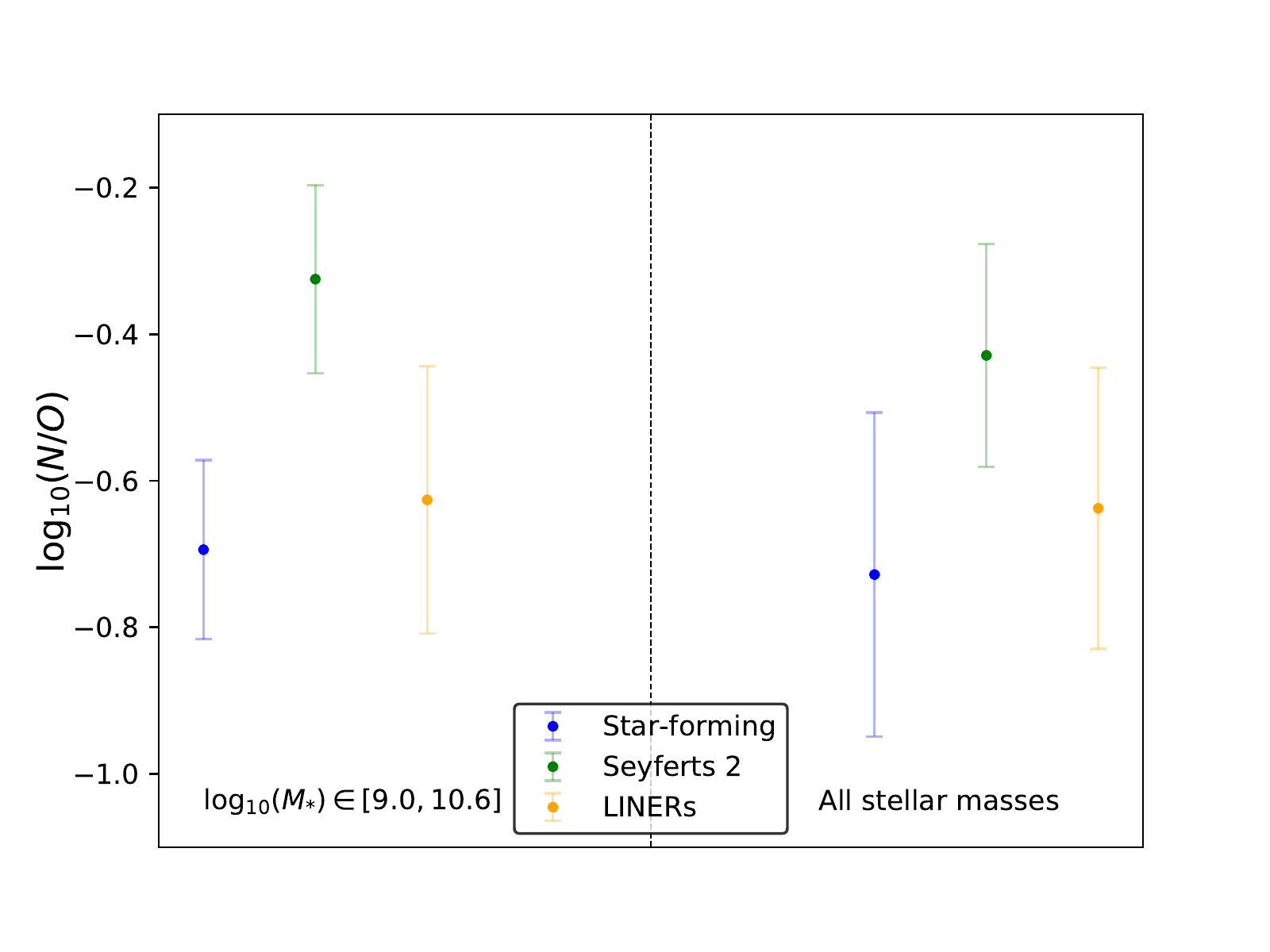}} \vspace{-0.2in} \end{minipage} 
	\end{tabular}
	\caption{Median values and standard deviations of the chemical abundance $12+ \log_{10} \left( O/H \right) $ (a) and the chemical abundance ratio $\log_{10} \left( N/O \right) $ (b) for different ranges of the stellar mass.}
	\label{Met_ste_mass}
\end{figure*}

\begin{figure}
	%\resizebox{\hsize}{!}
	\begin{tabular}{cccc}
		\begin{minipage}{0.05\hsize}\begin{flushright}\textbf{(a)} \end{flushright}\end{minipage}  &  \begin{minipage}{0.93\hsize}\centering{\includegraphics[width=1\textwidth]{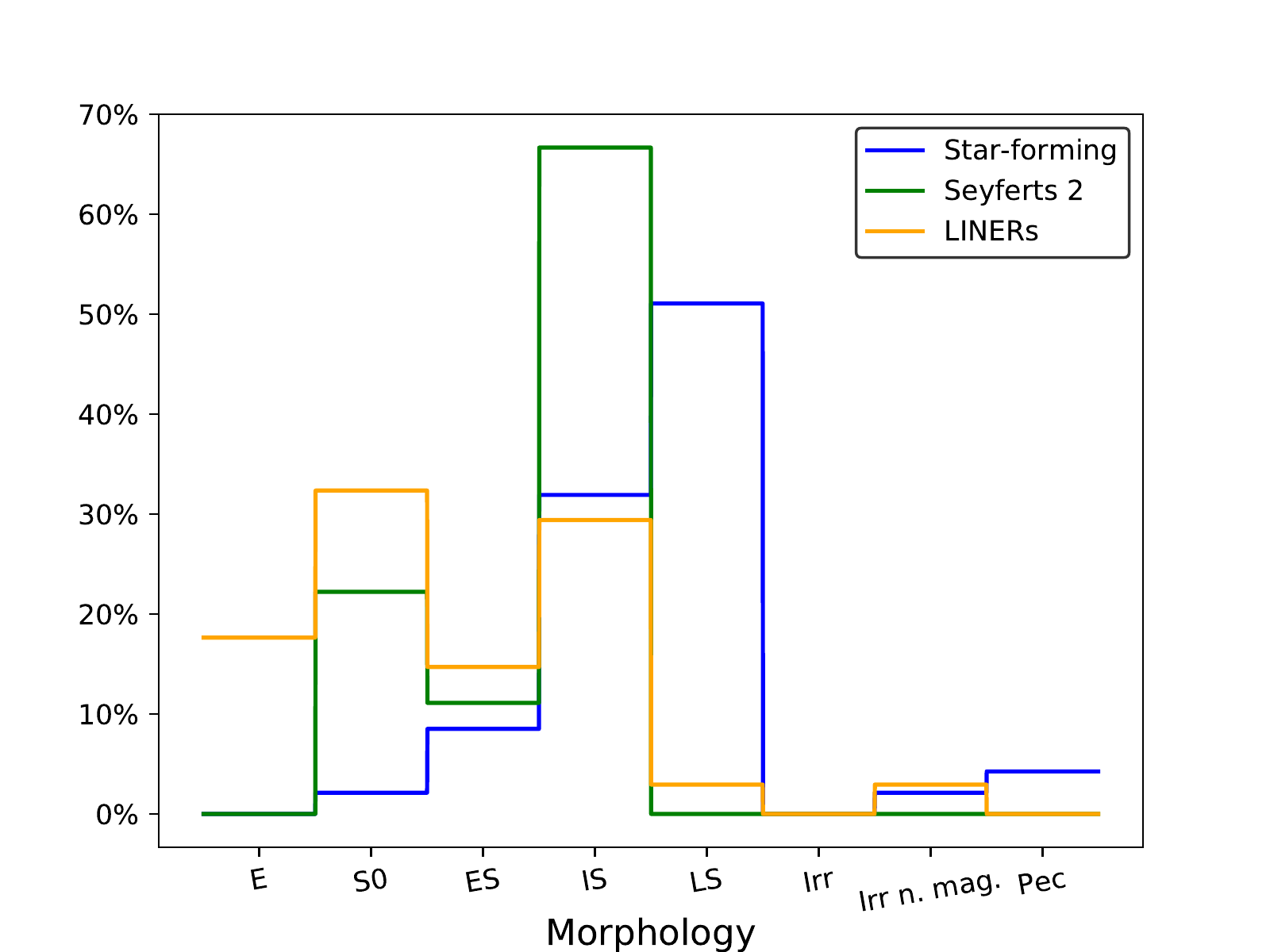}} \vspace{-0.15in} \end{minipage} \\ \begin{minipage}{0.05\hsize}\begin{flushright}\textbf{(b)} \end{flushright}\end{minipage}  &  \begin{minipage}{0.93\hsize}\centering{\includegraphics[width=1\textwidth]{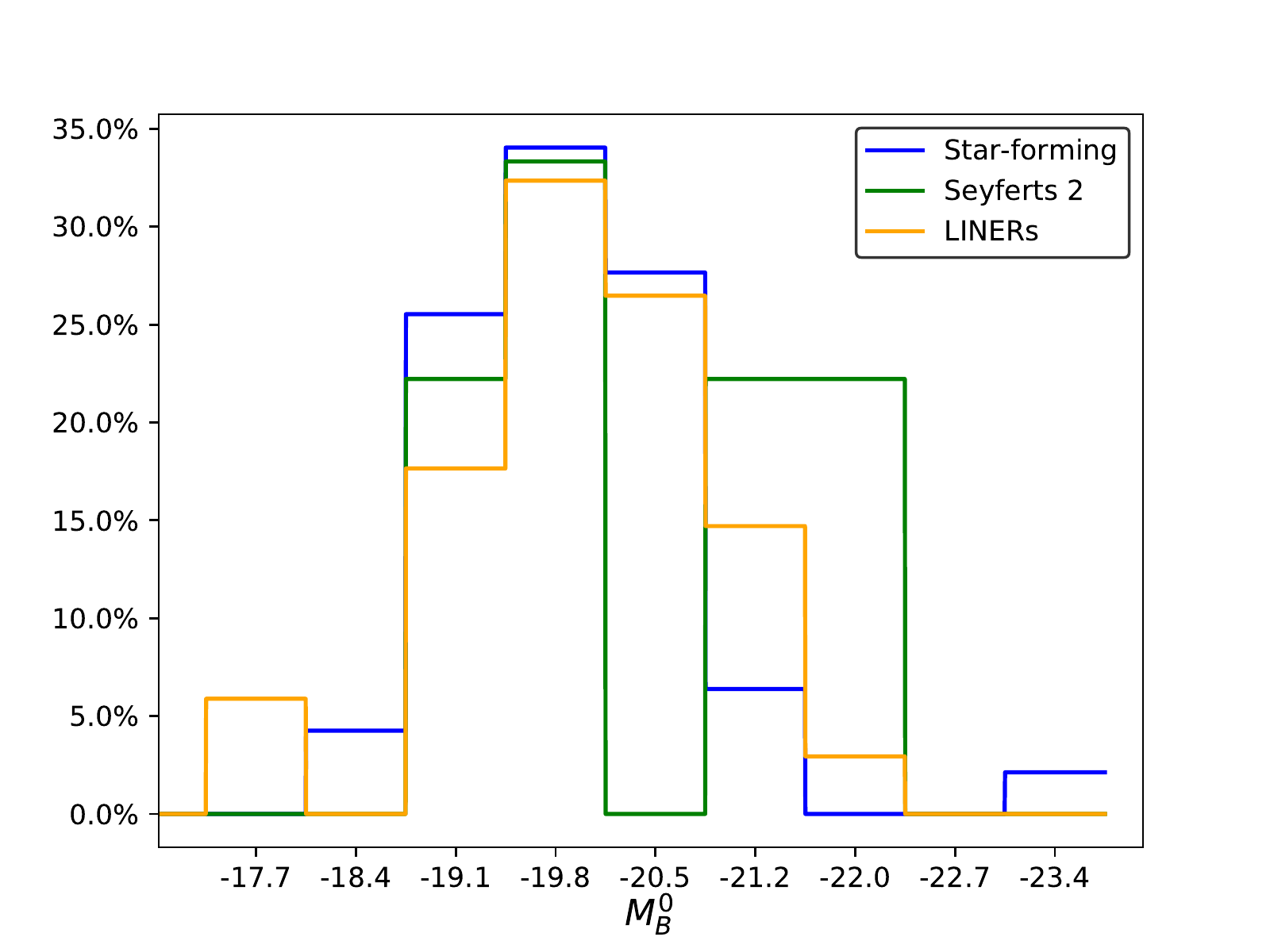}} \vspace{-0.15in} \end{minipage} \\ 
		\begin{minipage}{0.05\hsize}\begin{flushright}\textbf{(c)} \end{flushright}\end{minipage}  &  \begin{minipage}{0.93\hsize}\centering{\includegraphics[width=1\textwidth]{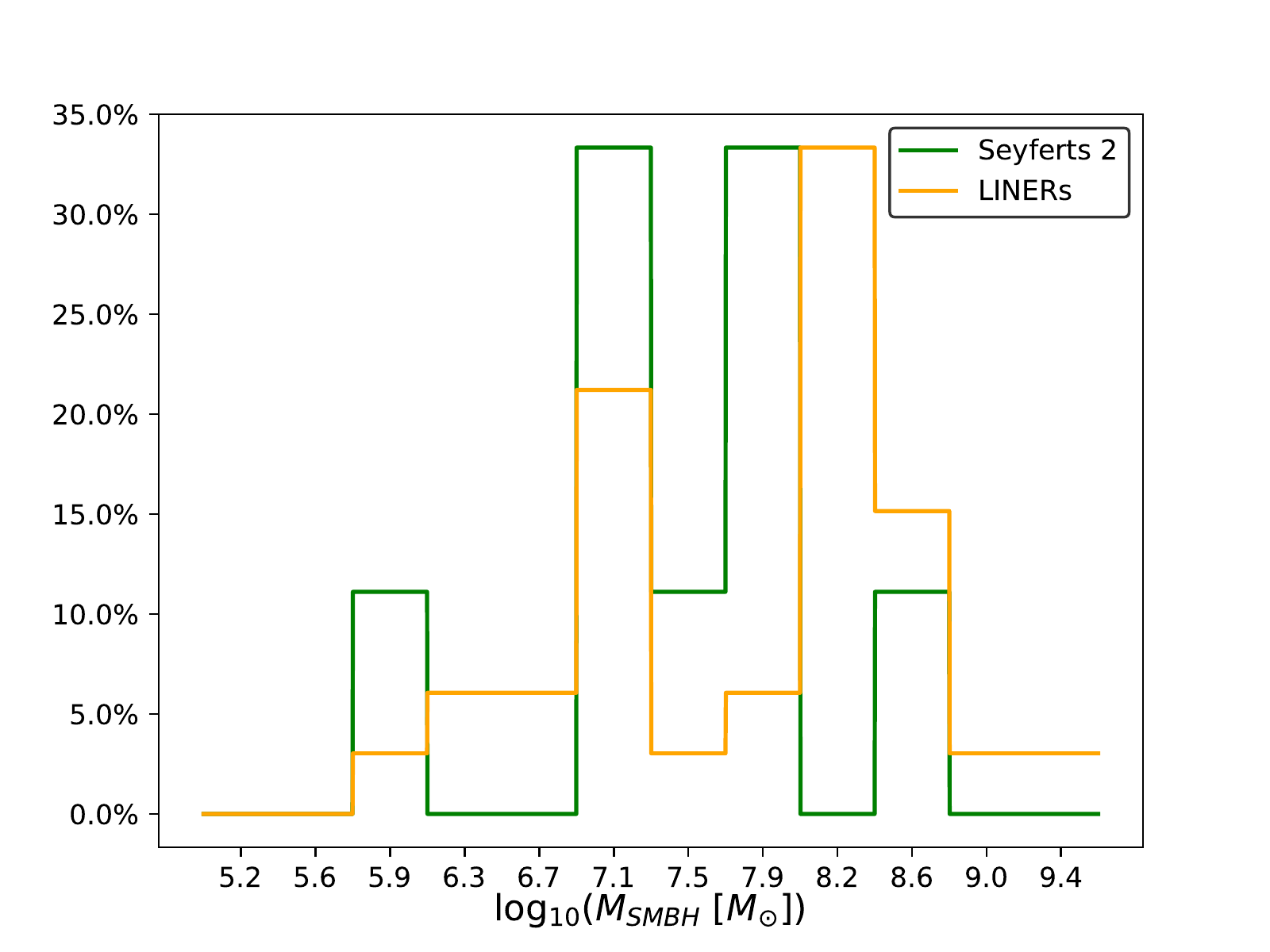}} \vspace{-0.2in} \end{minipage} \\ 
	\end{tabular}
	\caption{Relative distribution of different host galaxy properties for each spectral type considering only galaxies with $\log_{10} \left( M_{*} \left[ M_{\odot } \right] \right) \in \left[ 9.00, 10.62 \right] $. (a) Distribution of the morphology using the classes defined in Tab. \ref{Class_tab}. (b) Distribution of the corrected absolute B-magnitude $M_{B}^{0}$. (c) Distribution of the mass of the black hole $M_{SMBH}$.}
	\label{Ste_prop}
\end{figure}

In Fig. \ref{Met_ste_mass} (a) we observe that the median chemical abundance $12+\log (O/H)$ obtained for all SFG of the selected sample (left-hand plot, 8.71 with a $\sigma$ of 0.16) is virtually the same as that obtained in this constrained common stellar mass range (right-hand plot, $8.72$ with a $\sigma$ of 0.06). The same trend is observed for Seyferts 2, as the median value in the whole range (8.78 with a $\sigma $ of 0.23) is equivalent to the median value in the constrained range (8.81 with a $\sigma$ of 0.26), and for LINERs, for which the median value for all selected galaxies (8.63 with a $\sigma $ of 0.24) remains constant in relation to the median value in the constrained range (8.61 with a $\sigma$ of 0.25). An analogous result is observed in Fig. \ref{Met_ste_mass} (b) for $\log \left( N/O \right) $.

%of each spectral type practically remain constant if the range of stellar masses is constrained. If we restrict to this stellar mass range, only SFG with low chemical abundances are omitted (due to the mass-metallicity relation they have low stellar masses), so the median value of $12+\log \left( O / H \right) $ slightly increases, although with a considerable reduced standard deviation. From Fig. \ref{Met_ste_mass} (a), it is more clear that LINERs present slightly lower Oxygen abundances than SFG or Seyferts 2, but compatible within the errors. The only effect observed in $\log \left( N/O \right) $ is a decrease in the standard deviation for SFG, but the median value is essentially the same (see Fig. \ref{Met_ste_mass} (b)).

In order to explore the origin of these low chemical abundances in LINERs, which are also reported using the three different techniques explored in Appendix \ref{s: A}, we analyze the properties of the host galaxies in the stellar mass range $\log \left( M_{*} \left[ M_{\odot } \right] \right) = [9.00, 10.62]$. Fig. \ref{Ste_prop} (a) shows that the morphology distribution does not significantly change from that in Tab. \ref{morpho_distr}. Although there is a difference in the distribution of $M_{B}^{0}$ (having Seyferts 2 the highest luminosities, Fig. \ref{Ste_prop} (b)), our results from Sec. \ref{subsec42} show that the chemical abundances in AGN are not correlated with this property. In addition, the mass of the Supermassive Black Hole was probed not to have an effect on the chemical abundances, implying that the different distributions seen in Fig. \ref{Ste_prop} (c) cannot explain the subsolar values obtained for LINERs. 

We have also explored the possibility that wrong measurements in the auroral emission line [O\textsc{iii}]$\lambda$4363\r{A} could lead to wrong derivations in the chemical abundances. Therefore, we calculate the chemical abundances ratios and ionization parameters for the five galaxies with measurements of the auroral line, but omitting their value, i.e. considering that the auroral line was never measured. In the case of the SFG NGC 4532 and the Seyferts 2 NGC 2273 and NGC 3982, the chemical abundances and ionization parameters after omitting [O\textsc{iii}]$\lambda$4363\r{A} are compatible within the errors to those obtained with their auroral line. Only two LINERs in our sample present a measurement of the auroral line. A look at the spectrum of NGC 3226 (see Fig. 28 from \citealt{Ho_II_1995}) shows that the measurement of [O\textsc{iii}]$\lambda$4363\r{A} is more likely a statistical noise than a true emission line. In the case of NGC 4278, although the auroral line can be measured, this line might be affected by shocks located in obscured parts of the dusty torus in AGN \citep{Contini_Viegas_2001, Zakamska_2014}. In the case of LINERs NGC 3226 and NGC 4278, we obtain that the chemical abundance $12+\log \left( O/H \right) $ increases for both of them: for NGC 3226 this ratio increases to $8.7\pm 0.2$; and for NGC 4278 it reaches the value of $8.7\pm0.3$. Nevertheless, there are other LINERs (NGC 3193, NGC 3705, NGC 3953, NGC 4143, NGC 4216, NGC 4321 and NGC 4429) that still present low chemical abundances even the auroral line is not measured for them.

\subsection{Chemical abundances in a sample of luminous infrared LINERs}
\label{subsec5new}

We perform an additional test on the chemical abundances in the nuclear region of LINERs by selecting another sample of LINERs. We use the sample of luminous infrared LINERs (presenting a luminosity associated to the SFR $>10^{43}$ erg$\cdot$s$^{-1}$) observed by \citet{Povic_2016} to check whether the low metallicities in the LINERs from the Palomar sample may be related to their lower luminosities. Although the data from the host galaxies of these LINERs allow us to perform an analogous analysis to that for the Palomar Survey, it must be noticed that LINERs from \citet{Povic_2016} are more distant ($z \sim 0.04-0.11$), implying that the observations with slits ranging 1.2-1.5 arcsec trace a larger central region of the galaxies (the median radius of the region observed is $\sim$1839.9 pc, larger than the median region $r \sim 82.4$ pc observed in the Palomar Survey).

In order to perform a consistent comparison, we analyze if host galaxies from our sample have different properties than those in the sample in \citet{Povic_2016}. The KS-test shows that absolute B-magnitude (p-value $\sim 0.73$) and the stellar mass (p-value $\sim 0.64$) follow the same distribution in both samples. They only differ, as expected, in the mass of their SMBH, which is related to their AGN luminosity. However, as shown in Tab. \ref{Liners_corr}, this quantity does not affect the values of the chemical abundances. Therefore, we compare both samples of LINERs.

Following the same methodology explained in Sec. \ref{sec2}, we filter the sample of 42 LINERs from \citet{Povic_2016}. First of all, we check the accuracy of the emission lines (the relative error must be below 20$\% $). Only one galaxy, namely F24, is omitted in this process since has no measurement of other Balmer line than H$_{\alpha }$, implying that it is not possible to correct from extinction its emission lines. Secondly, we perform an spectral classification of the sample of LINERs using the diagnostic diagrams \citep{Veilleux_1987, Kauffmann_2003, Kewley_2006}. Due to the spectral coverage, only for 7 out of 42 LINERs the measurement of the emission line [O\textsc{i}]$\lambda $6300\r{A} is present. In addition, the LINERs F23, B04 and B14 do not present a measurement of the sulfur doublet. Therefore, we use to classify these galaxies the available diagrams for the set of emission lines of each galaxy. We identify a total of four galaxies that do not fall in the region of LINERs: the galaxy B20, identified as Seyfert 2; and the galaxies F20, B13 and B14 identified as Composite Galaxies. By applying these filters, the sample of LINERs from \citet{Povic_2016} is reduced to 37. The reddening correction of the emission line ratios was performed following Subsec. \ref{subsec24}. 

\textsc{HCm} can be used in different subsets of emission lines required. Due to the redshift of these new LINERs, some of them have measurements of the emission line [O\textsc{ii}]$\lambda $3727\r{A}, used by \textsc{HCm} and which is completely absent in the Palomar Survey. We perform a preliminary study of the influence of this emission line in the derivation of chemical abundances by using \textsc{HCm} before and after masking it. We obtain that there is no significant difference in using or not this emission line: the  median offset for $12+\log_{10} \left( O/H \right) $ is $0.00$ and RSME is $0.12$; and, for $\log_{10} \left( N/O \right) $ and $\log_{10} \left( U \right) $ the median offsets are $0.00$ while RSMEs are $0.10$ for both parameters. Applying the filter of presenting reliable results of chemical abundances and ionization parameters (see Subsec. \ref{subsec26}), the sample of 37 LINERs from \citet{Povic_2016} is reduced to 25. The results are listed in Tab. \ref{TabA4}.

\begin{figure*}
	%\resizebox{\hsize}{!}
	\begin{tabular}{cccc}
		\begin{minipage}{0.05\hsize}\begin{flushright}\textbf{(a)} \end{flushright}\end{minipage}  &  \begin{minipage}{0.4\hsize}\centering{\includegraphics[width=1\textwidth]{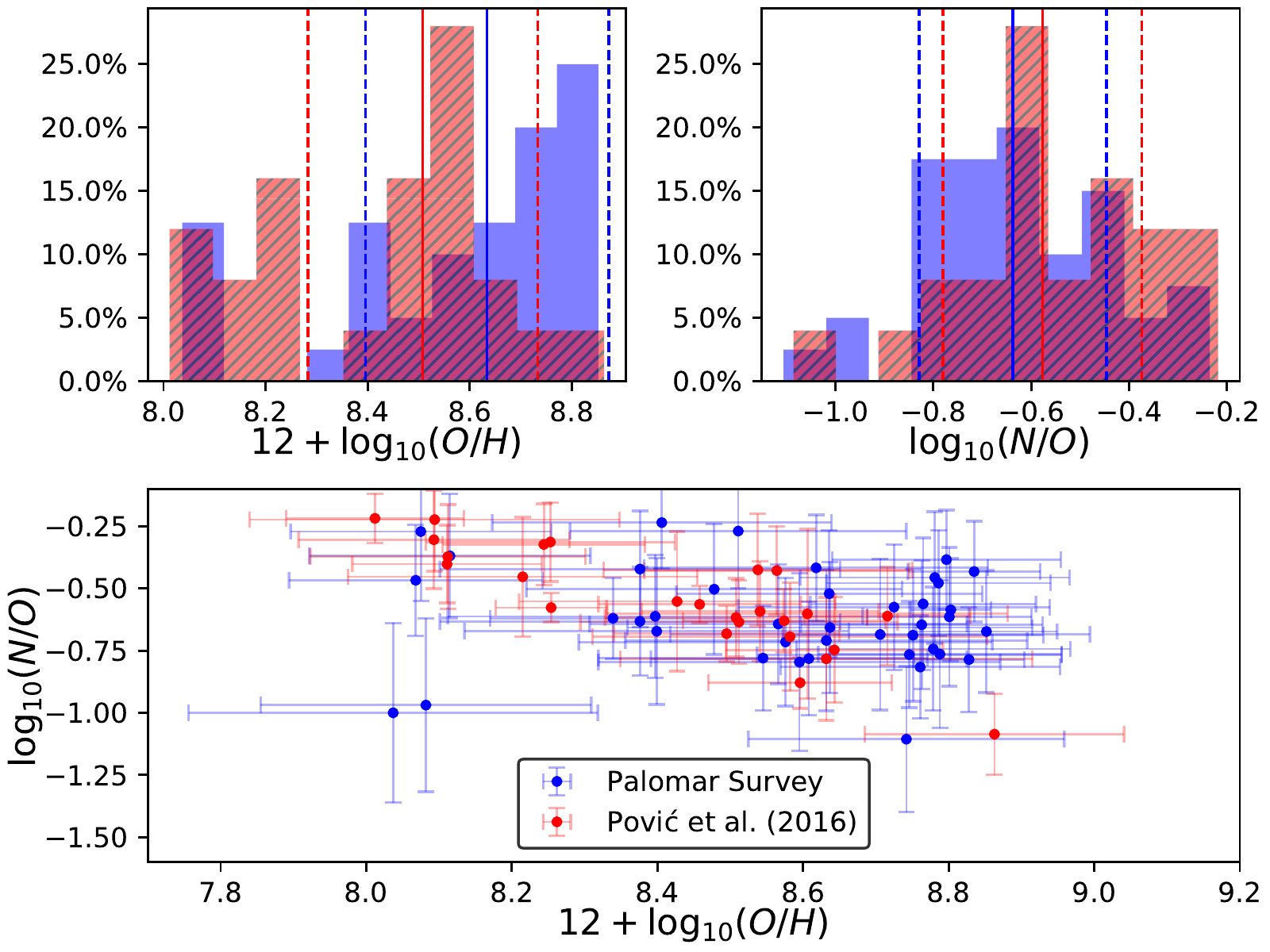}} \vspace{-0.2in} \end{minipage} & \begin{minipage}{0.05\hsize}\begin{flushright}\textbf{(b)} \end{flushright}\end{minipage}  &  \begin{minipage}{0.4\hsize}\centering{\includegraphics[width=1\textwidth]{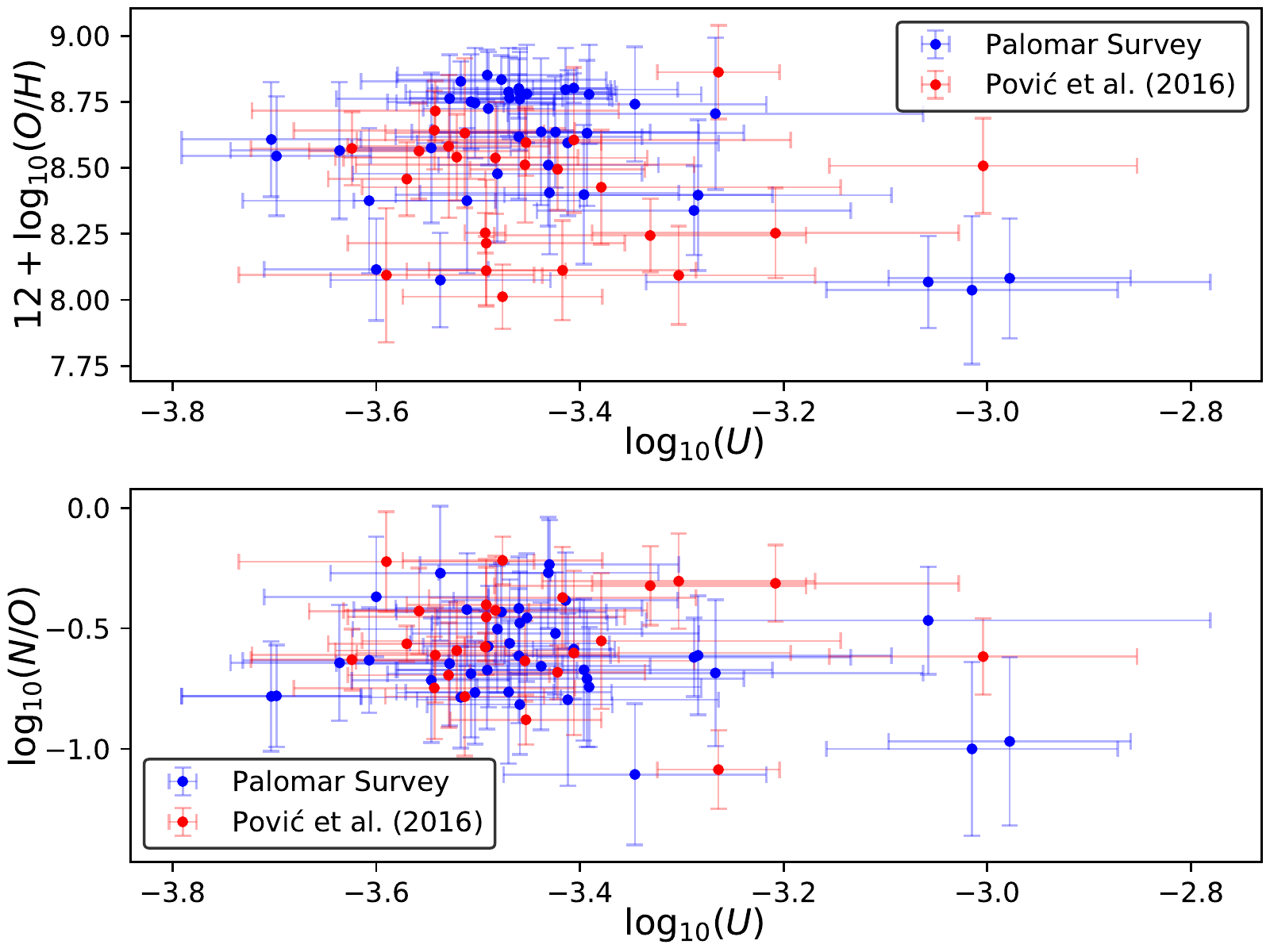}} \vspace{-0.2in} \end{minipage} 
	\end{tabular}
	\caption{Chemical abundances and ionization parameters obtained for LINERs from Palomar Survey (blue) and from \citet{Povic_2016} (red). (a) The top left plot shows the histogram of the chemical abundance $12 + \log_{10} \left( O/H \right) $. The top right plot shows the histogram of $\log_{10} \left( N/O \right) $. For both histograms, the solid lines represent the median values and the dashed lines the standard deviations. The bottom plot shows the relation between the chemical abundance ratio $\log_{10} \left( N/O \right) $ and $12 + \log_{10} \left( O/H \right) $. (b) Variation of the chemical abundances $12+\log_{10} \left( O/H \right) $ and $\log_{10} \left( N/O \right) $ with the ionization parameter $\log_{10} \left( U \right) $ in star-forming galaxies.}
	\label{Chem_LIN}
\end{figure*}

\begin{table*}
	\caption{Statistics of the chemical abundances and ionization parameters for LINERs in the Palomar Survey and from \citet{Povic_2016}.}
	\label{Stats_chem_lin}     
	\centering          
	\begin{tabular}{ll|lll|lll|lll}
		\multicolumn{2}{l}{} &
		\multicolumn{3}{|l}{\boldmath$12+\log_{10} \left( O/H \right) $} & \multicolumn{3}{|l}{\boldmath$\log_{10} \left( N/O \right) $} & \multicolumn{3}{|l}{\boldmath$\log_{10} \left( U \right) $}  \\ \hline \textbf{Type} & \textbf{N\boldmath$^{\circ}$} & \textbf{Median} & \textbf{Std. Dev.} & \textbf{Range} & \textbf{Median} & \textbf{Std. Dev.} & \textbf{Range} & \textbf{Median} & \textbf{Std. Dev.} & \textbf{Range} \\ \hline   \textbf{Palomar} & 40 & 8.63 & 0.26 & [8.04, 8.85] & -0.63 & 0.19 & [-1.11, -0.24] & -3.46 & 0.15 & [-3.70, -2.98] \\ \textbf{Povi\'{c}+2016} & 25 & 8.51 & 0.22 & [8.01, 8.86] & -0.58 & 0.20 & [-1.09, -0.22] & -3.48 & 0.13 & [-3.62, -3.00] \\ \hline  
	\end{tabular}      
\end{table*} 

Fig. \ref{Chem_LIN} and Tab. \ref{Stats_chem_lin} show that the chemical abundance ratios and ionization parameter present similar values in both samples of LINERs. In addition, little correlation between the chemical abundance ratios O/H and N/O and the ionization parameter U is found, being the coefficients $r = -0.05$ and $r = -0.03$ respectively, which reinforces the results obtained for LINERs in the Palomar Survey.

We also analyze the correlations between the chemical abundances and the host galaxy properties that we study in Sec. \ref{sec4}. Due to the redshift of the sample from \citet{Povic_2016}, it is not possible to use the morphological classes established in Tab. \ref{Class_tab}. We will classify the host galaxies in the same categories used by \citet{Povic_2016}: Ellipticals (which includes E and S0 galaxies), Spirals (which includes ES, IS and LS) and Irregulars (which includes Irr, Irr. Non. Mag. and Pec galaxies). Contrary to what is obtained in the Palomar Survey, LINERs hosted by elliptical galaxies presented slightly higher O/H abundances than those hosted by spirals. Nevertheless, as it was the case in the Palomar Survey, the results for these three morphological classes are compatible within the standard deviations. The behavior of N/O is similar to that reported for Palomar LINERs.

\begin{table}
	\caption{Pearson's correlation coefficients between the chemical abundance ratios and different host galaxy properties for both samples of LINERs.}
	\label{Liners_corr}     
	\centering          
	\begin{tabular}{l|cc|cc}
		
		\multicolumn{1}{c|}{} & \multicolumn{2}{c|}{\boldmath$12+\log_{10} \left( O/H \right) $} & \multicolumn{2}{c}{\boldmath$\log_{10} \left( N/O \right) $} \\ \hline \textbf{Prop.} & \textbf{Palomar} & \textbf{Povi\'{c}+2016} & \textbf{Palomar} & \textbf{Povi\'{c}+2016} \\ \hline
		\boldmath$M_{B}^{0}$ & $-0.11$ & $-0.13$ & $-0.56$ & $0.09$ \\ \boldmath$M_{SMBH}$ & $0.15$ & $-0.13$ & $0.04$ & $0.18$ \\  \boldmath$M_{*}$ & $0.04$ & $0.016$ & $0.46$ & $-0.05$ \\ \hline
	\end{tabular}      
\end{table}

We also analyzed the influence of host galaxy properties such as $M_{B}^{0}$, $M_{SMBH}$ and $M_{*}$. The absolute B-magnitude is calculated from the magnitudes in the filters $u$ and $g$ from SDSS and the relation from \citet{Jester_2005}. $M_{*}$  is calculated following the same procedure explained in Subsec. \ref{subsec42}, instead of using the values provided by \citet{Povic_2016}. These properties calculated in this work are listed in columns (2) and (3) from Tab. \ref{TabA4}. The values of $M_{SMBH}$ are calculated by \citet{Povic_2016} using the same methodology as here (see Subsec. \ref{subsec45}). The lack of correlation found between chemical abundances and these host galaxy properties for LINERs in the Palomar Survey is also found for infrared luminous LINERs from \citet{Povic_2016}.

These results towards the study in Subsec. \ref{subsec53} lead us to consider the low Oxygen abundance in LINERs as a consequence of: 1) LINERs present intrinsically lower chemical abundances than Seyferts 2; or, 2) the photoionization is not the only mechanism that must be considered in LINERs. This last possibility has been explored in the literature. Theoretical models postulate the possibility that the emission from LLAGN (such as LINERs) could arise in shocks of gas due to the interaction with jets or galactic winds \citep{Dopita_1996}, and it seems to be the case in some of them \citep{Groves_2006, Cazzoli_2020, Hermosa_2020}. The possibility that hot post-AGB stars and white dwarfs could be the ionization source in some LLAGN has also been proposed \citep{Stasinska_2008, Cid-Fernandes_2011}. Nevertheless, this is still under debate in the current study of AGN, especially in the low-luminosity ones \citep{Marquez_2017}. The non-AGN nature of NGC 4321 \citep{Omaira_2009} could be advocated for an explanation of its low chemical abundance ratio: if considered as LINER, we obtain $12+\log_{10} \left( O/H \right) = 8.11\pm0.19$ but if we consider that it is a SFG, this chemical abundance ratio increases to $8.53\pm0.04$. Nevertheless, from our sample of multiwavelength confirmed AGN LINERs, $15.4\% $ present low chemical abundances ($12+\log_{10} \left( O/H \right) < 8.3$).

\subsection{Ionization parameter}
\label{subsec54}
Focusing our attention on the ionization parameters, the highest values are obtained for Seyferts 2, which is the expected behavior \citep{Ferland_1983, Ho_1993, Ho_VI_2003, Almeida_2006, Kewley_2006}. Although the distribution of the ionization parameter $U$ in SFG and LINERs is quite similar, their peaks are separated by more than $0.2$ dex. Comparing chemical abundances and ionization parameter, the latter is the best tool for distinguishing between the three spectral types (SFG, Seyferts 2 and LINERs), which is in fact the underlying principle of work of the diagnostic diagrams \citep{Baldwin_1981, Veilleux_1987, Kauffmann_2003, Kewley_2006}. In addition, we obtained as an upper limit for the ionization parameter in SFG the value $\log \left( U \right) = -2.45$, which is essentially the same one found in the local Universe for young, massive super star clusters located in galaxies with intense star formation \citep{Smith_2006}. This upper limit suggests the existence of a mechanism (or mechanisms) that maintains the ionization below this limit, whereas this is still an open question \citep{Yeh_2012, Kewley_2019}.

We have also extended the analysis of the possible relations between the chemical abundances and the host galaxy properties to the ionization parameter. While the ionization parameters practically remains constant for each morphological class, a study of the possible correlation in continuous properties (see Tab. \ref{U_prop}) shows that it is independent from all the host galaxy properties analyzed in this study.

\begin{table}
	\caption{Pearson's correlation coefficients between the ionization parameter and different host galaxy properties for the three spectral types.}
	\label{U_prop}     
	\centering          
	\begin{tabular}{lccc}
		
		\multicolumn{4}{c}{\boldmath$\log_{10} \left( U \right) $} \\ \hline \textbf{Prop.} & \textbf{SFG} & \textbf{Seyferts2} & \textbf{LINERs} \\ \hline
		\boldmath$M_{B}^{0}$ & $-0.34$ & $-0.32$ & $0.16$ \\   \boldmath$M_{SMBH}$ & - & $0.55$ & $0.069$ \\  \boldmath$M_{*}$ & 0.22 & $0.39$ & $-0.049$ \\ \hline
	\end{tabular}      
\end{table}

\section{Summary and conclusions}
\label{sec6}
We study the chemical abundances in the nuclear regions of an heterogeneous sample of SFG and AGN in a sample of emission-line galaxies from the Palomar Spectroscopic Survey using the code \textsc{HCm}, as this is suitable for the analysis of large samples of galaxies attending to different reasons. Firstly, this program can be used with a set of strong emission lines that are available for optical spectroscopic surveys. Secondly, it can be applied for both, star-forming galaxies and AGN. And thirdly, it does not assume any underlying relation between the chemical abundance $12+\log_{10} \left( O/H \right) $ and the chemical abundance ratio $\log_{10} \left( N/O \right) $. In addition, we probe that chemical abundances in SFG estimated from \textsc{HCm} are compatible with those obtained with methods based on calibrations from strong optical emission lines. In the case of AGN there are discrepancies among the different methods analyzed.

The subsample of SFG is used as a test of the soundness of our methodology and, then, we apply the same techniques to the subsample of AGN. Our results for the different spectral types show:
\begin{itemize}
	\item \textbf{Star-forming galaxies}: The chemical abundance ratios of Oxygen to Hydrogen present a median value similar to that measured for the Sun, although there is a small group of galaxies with sub-solar abundances. This abundance is related to the chemical abundance ratio of Oxygen to Nitrogen, demonstrating the existence of two groups of SFG. For the first group, the chemical abundance ratio $\log_{10} \left( N/O \right) $ is practically constant, behavior explained by the production of carbon and oxygen in massive stars. For the second group, the chemical abundance ratio $\log_{10} \left( N/O \right) $ scales with $12+\log_{10} \left( O/H \right) $, explained due to the presence of an enriched ISM. Finally, the ionization parameter $\log_{10} \left( U \right) $ shows an anti-correlation with the chemical abundance $12+\log_{10} \left( O/H \right) $.
	
	\item \textbf{Seyferts 2}: The chemical abundance $12+\log_{10} \left( O/H \right) $ shows the highest values in our sample, with a median value slighty higher than the solar abundance. The chemical abundance ratio $\log_{10} \left( N/O \right) $ does not show any particular trend with $12+\log_{10} \left( O/H \right) $, contrasting with the behavior observed for SFG. The ionization parameter $\log_{10} \left( U \right) $ seems to be independent from both chemical abundances.
	
	\item \textbf{LINERs}: The median chemical abundance $12+\log_{10} \left( O/H \right) $ is slightly lower than in Seyferts 2 and SFG. A similar result is also found for the chemical abundance ratio $\log_{10} \left( N/O \right) $, but SFG show lower values for these parameters. As observed in Seyferts 2, both chemical abundances are independent one from each other and also from the ionization parameter $\log_{10} \left( U \right) $. These results are confirmed using an additional sample of luminous infrared LINERs at larger redshift, but in the same stellar mass range.
\end{itemize}

We also analyze the eventual differences due to the host galaxy properties of each spectral type. Although it was needed to reduce the original sample of galaxies to present reliable determinations of the chemical abundances (from 418 to 143 galaxies), our study of the host galaxy properties shows that no bias was introduced in almost all the host galaxy properties analyzed, despite the distribution of stellar masses is slightly modified.

The first host galaxy property analyzed is the morphology. From our study, the chemical abundance $12+\log_{10} \left( O/H \right) $ is not affected by the morphological type, although for SFG there is an important decrease in irregular galaxies (likely due to small size of the sample). The chemical abundance ratio $\log_{10} \left( N/O \right) $ is only affected by the spectral type: being higher in Seyferts 2 than in LINERs or SFG, that present similar values.

Our study of the absolute B-magnitude reveals that there is a slight correlation between this parameter and the chemical abundances in SFG. In the case of AGN, both abundances O/H and N/O are not correlated with the absolute B-magnitude. Then we analyze the stellar mass, which is the property more correlated with the chemical abundances for SFG. This is the expected behavior from the mass-metallicity relation that was previously observed for both spectral types. This correlation seems to be present for Seyferts 2, but we cannot conclude this due small sample of this type of AGN. Nevertheless, the chemical abundances in LINERs do not present such a relation with the stellar mass.

The mass of the SMBH does not introduce any change in the chemical abundances, that practically remain constant for Seyferts 2 and show high dispersion for LINERs.

The ionization parameter $\log_{10} \left( U \right) $ has been probed to be the best tool for distinguishing among SFG, Seyferts 2 and LINERs. Our study shows that this parameter is not affected by any host galaxy property, only the spectral type induces significant changes in $\log_{10} \left( U \right) $: the ionization parameter is higher for Seyferts 2, followed by SFG and then by LINERs.

\section*{Acknowledgements}

We acknowledge support from the Spanish MINECO grants AYA2016-76682C3-1-P, AYA2016-79724-C4 and PID2019-106027GB-C41. We also acknowledge  financial  support  from  the  State Agency for Research of the Spanish MCIU through the "Center of Excellence Severo Ochoa" award to the Instituto de Astrof\'{\i }sica de Andaluc\'{\i }a (SEV-2017-0709). We acknowledge the fruitful discussions with our research team. B.P.D. also acknowledges the support from CSIC grant JAEINT-19-00863. E.P.M acknowledges the assistance from his guide dog Rocko without whose daily help this work would have been much more difficult.

%%%%%%%%%%%%%%%%%%%%%%%%%%%%%%%%%%%%%%%%%%%%%%%%%%

\section*{Data Availability}

The data used for this work are available in the article and in its online supplementary material (see Appendix \ref{s: B}).



\bibliographystyle{mnras}
\bibliography{PalomarBib} % if your bibtex file is called example.bib

%\begin{thebibliography}{99}
%\end{thebibliography}

%%%%%%%%%%%%%%%%%%%%%%%%%%%%%%%%%%%%%%%%%%%%%%%%%%

%%%%%%%%%%%%%%%%% APPENDICES %%%%%%%%%%%%%%%%%%%%%

\appendix
\section{Comparison between different methods to derive oxygen abundances in SFG}
\label{s: A0}

We present here a comparison between our results from \textsc{HCm} and three methods based on strong optical emission lines to estimate chemical abundances in SFG.

Firstly, we consider the calibration proposed by \citet{Pilyugin_2016} based on the emission line ratios $R_{3} \equiv I \left( \right. $[O\textsc{iii}]$\lambda$4959\r{A}/H$_{\beta }\left.\right)$ + $I \left( \right. $[O\textsc{iii}]$\lambda$5007\r{A}/H$_{\beta }\left.\right)$, $N_{2} \equiv I \left( \right. $[N\textsc{ii}]$\lambda$6548\r{A}/H$_{\beta }\left.\right)$ + $I \left( \right. $[N\textsc{ii}]$\lambda$6584\r{A}/H$_{\beta }\left.\right)$ and $S_{2} \equiv I \left( \right. $[S\textsc{ii}]$\lambda$6717\r{A}/H$_{\beta }\left.\right)$ + $I \left( \right. $[O\textsc{ii}]$\lambda$6731\r{A}/H$_{\beta }\left.\right)$. This calibration, divided into two branches, is given for the low branch ($\log \left( N_{2} \right) < -0.6$) as:
\begin{equation}
\label{A0-1} 
\begin{aligned} 12 + \log \left( O/H \right)_{PG+16}  = & 8.072 + 0.789 \log \left( R_{3} / S_{2} \right) + 0.726 \log \left( N_{2} \right) \\ & + \left[ 1.069 -0.170 \log \left( R_{3} / S_{2} \right) \right. \\ & \left. + 0.022 \log \left( N_{2} \right) \right] \log \left( S_{2} \right)
\end{aligned}
\end{equation}
and for the upper branch ($\log \left( N_{2} \right) \geq -0.6$) as:
\begin{equation}
\label{A0-2} 
\begin{aligned} 12 + \log \left( O/H \right)_{PG+16}  = & 8.424 + 0.030 \log \left( R_{3} / S_{2} \right) + 0.751 \log \left( N_{2} \right) \\ & + \left[ - 0.349 + 0.182 \log \left( R_{3} / S_{2} \right) \right. \\ & \left. + 0.508 \log \left( N_{2} \right) \right] \log \left( S_{2} \right)
\end{aligned}
\end{equation}
Due to the spectral coverage of the Palomar Survey, the emission line ratio [O\textsc{iii}]$\lambda$4959\r{A}/H$_{\beta }\left.\right)$ is calculated by the theoretical relation $I \left( \right. $ [O\textsc{iii}]$\lambda$5007\r{A}$\left.\right) $/$I \left( \right. $ [O\textsc{iii}]$\lambda$4959\r{A}$\left.\right) $ = 2.98 \citep{Storey_2000}. In addition, since the spectroscopic information for our sample of galaxies does not include the measurement of the emission line [N\textsc{ii}]$\lambda$6548\r{A}, we also assume the theoretical relation $I \left( \right. $ [N\textsc{ii}]$\lambda$6584\r{A}$\left.\right) $/$I \left( \right. $ [N\textsc{ii}]$\lambda$6548\r{A}$\left.\right) $ = 3.05 \citep{Storey_2000}.

Secondly, we consider the calibration proposed by \citet{Curti_2017} based on the emission line ratio $O3N2 \equiv  I\left(\right. $[O\textsc{iii}]$\lambda$5007\r{A}/H$_{\beta }\left.\right) $/$I \left( \right.$[N\textsc{ii}]$\lambda$6584\r{A}/H$_{\beta }\left.\right) $, given by:
\begin{equation}
\label{A0-3}
\log \left( O3N2 \right) = 0.281 - 4.765 \left( x - 8.69 \right) - 2.268 \left( x - 8.69 \right) ^{2}
\end{equation}
where $x \equiv 12 + \log \left( O/H \right)_{Cu+17} $. This calibration is valid for the range $7.6 < 12+\log \left( O/H \right) < 8.85$.

Thirdly, to estimate the chemical abundance ratio $\log \left( N/O \right) $ we use the calibration proposed by \citet{Perez-Montero_2009} based on the emission line ratios $N2S2 \equiv \log \left( I \left( \right.\right.$[N\textsc{ii}]$\lambda$6584\r{A}$\left.\right) / I \left( \right.$[S\textsc{ii}]$\lambda$6717\r{A}+[S\textsc{ii}]$\lambda$6731\r{A}$\left.\right)\left.\right)$ given by:
\begin{equation}
\label{A0-4} \log \left( N/O \right)_{PC+09} = 1.26 N2S2 - 0.86
\end{equation}
We only consider this estimator for N/O since this is the most widely used calibration in the literature for $N2S2$. Although there are other calibrations based on the parameter $N2O2$, this involves the measurement of the emission line [O\textsc{ii}]$\lambda$3727\r{A}, which is not available in our sample due to the spectral coverage of the Palomar Survey.

\begin{figure}
	%\resizebox{\hsize}{!}
	\begin{tabular}{cc}
		\begin{minipage}{0.05\hsize}\begin{flushright}\textbf{(a)} \end{flushright}\end{minipage}  &  \begin{minipage}{0.91\hsize}\centering{\includegraphics[width=1\textwidth]{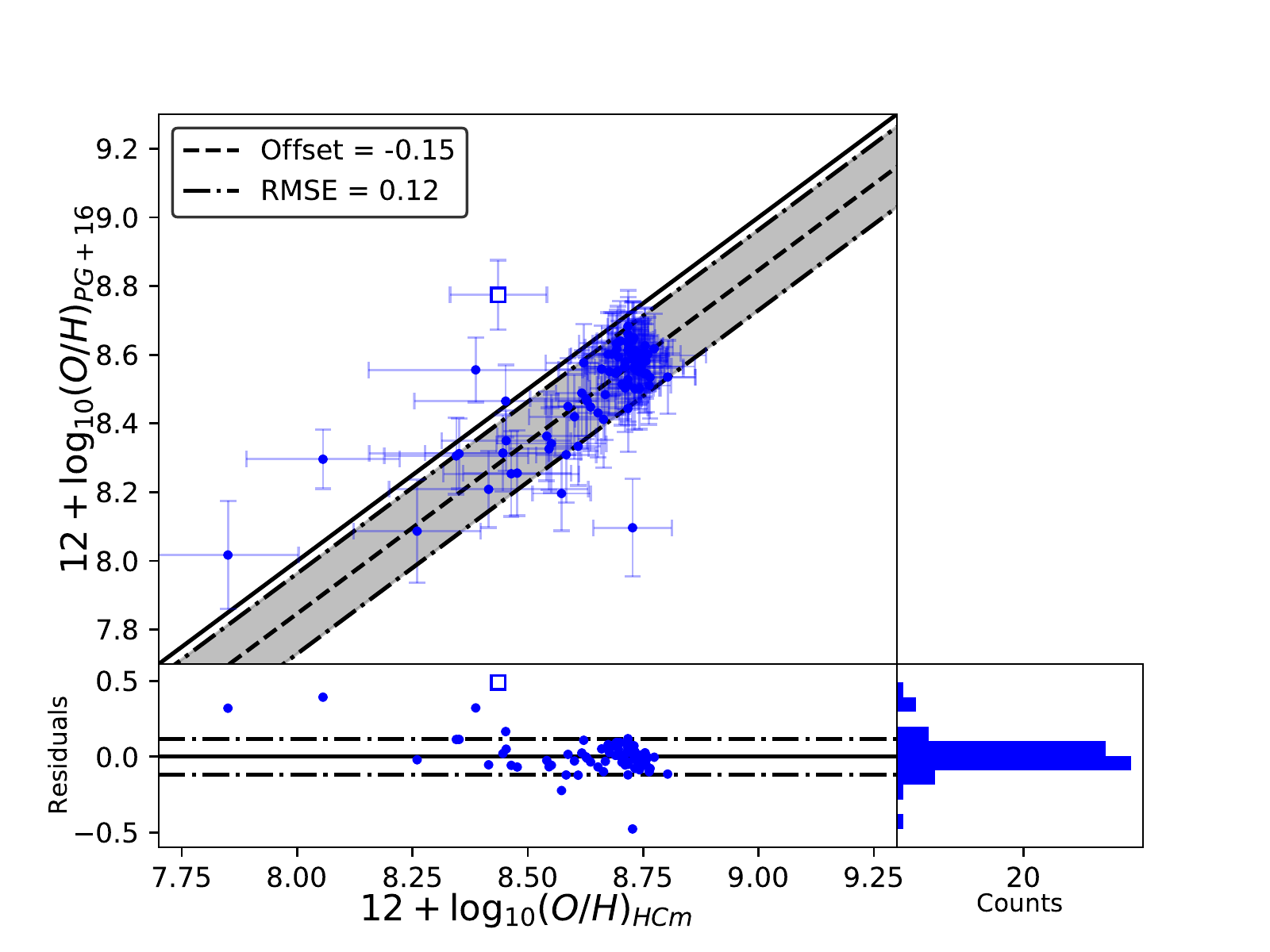}} \vspace{-0.2in} \end{minipage} \\ \begin{minipage}{0.05\hsize}\begin{flushright}\textbf{(b)} \end{flushright}\end{minipage}  &  \begin{minipage}{0.93\hsize}\centering{\includegraphics[width=1\textwidth]{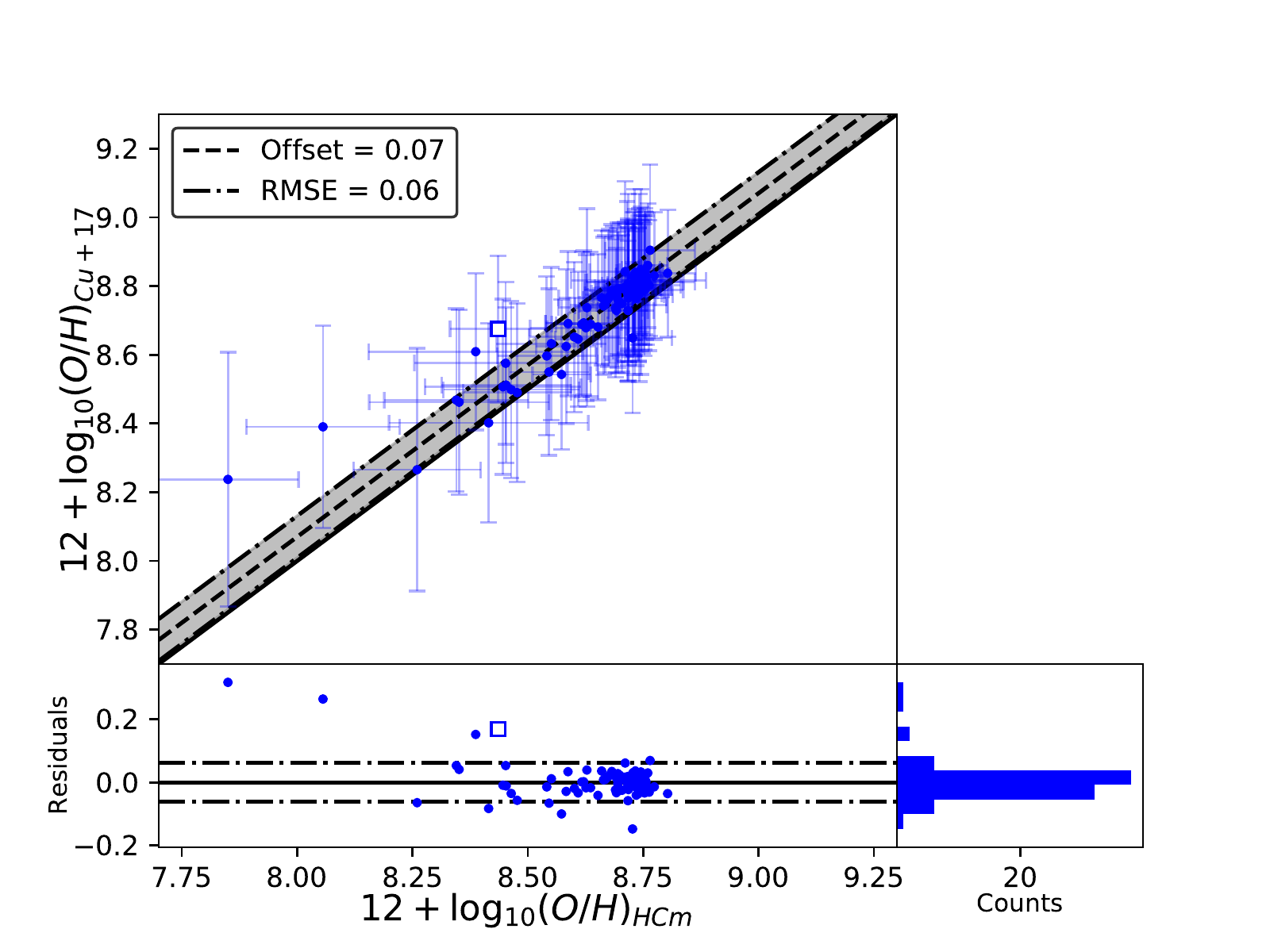}} \vspace{-0.2in} \end{minipage} \\ 
	\end{tabular}
	\caption{Comparison of oxygen abundances (O/H) derived with the calibrations from optical strong emission lines with \textsc{HCm} (x-axis). Square-empty symbols indicate galaxies initially classified as ambiguous (see Sec. \ref{subsec23}). (a) Results for the calibration proposed by \citet{Pilyugin_2016}, denoted as PG+16. (b) Results for the calibration proposed by \citet{Curti_2017}, denoted as Cu+17. The offsets are given using the median value (dashed line) and RMSE (dot-dashed lines). Bottom plots show the residuals from the offset and their distribution in a histogram (bottom-right plot).}
	\label{Offset_OH}
\end{figure}

\begin{figure}
	\centering
	\includegraphics[width=\hsize]{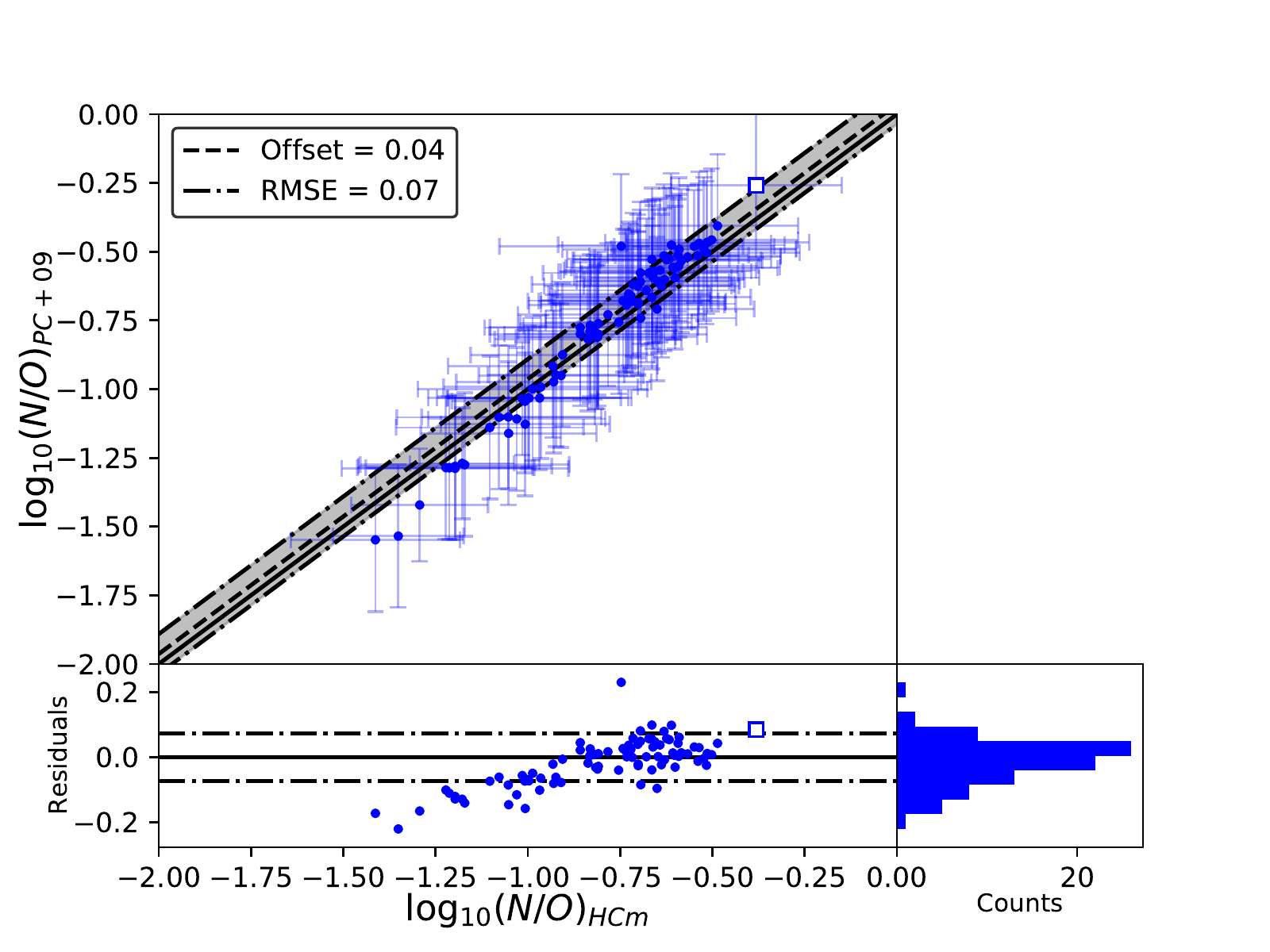}
	\caption{Comparison of chemical abundance ratios N/O derived with the calibration from \citet{Perez-Montero_2009} (y-axis), denoted as PC+09, with \textsc{HCm} (x-axis). Square-empty symbols indicate galaxies initially classified as ambiguous (see Sec. \ref{subsec23}). The offset is given using the median value (dashed line) and RMSE (dot-dashed lines). Bottom plots show the residuals from the offset and their distribution in a histogram (bottom-right plot).}
	\label{Offset_NO}
\end{figure}

We show in Fig. \ref{Offset_OH} (a) that the calibration proposed by \citet{Pilyugin_2016} underestimates (median offset of -0.15 dex and RMSE of 0.12 dex) in average the oxygen abundances in our sample of galaxies. This offset could arise in the sample of HII regions used by \citet{Pilyugin_2016}: the median value of oxygen abundances in their sample is $12+\log \left( O/H \right) \sim 8.2-8.3$, with few objects with $12+\log \left( O/H \right) > 8.6$, while our median value is $12+\log \left( O/H \right) \sim 8.7$.

Fig. \ref{Offset_OH} (b) shows that the calibration proposed by \citet{Curti_2017} and determinations from \textsc{HCm} are in good agreement, with an offset of 0.07 dex and RMSE of 0.06 dex. Although there are few galaxies at low chemical abundances ($12+\log\left( O/H \right) $) that present a higher offset, they are also obtained by \citet{Curti_2017}: they show that for chemical abundances below $12+\log \left( O/H \right) > 8.0$ the estimator $O3N2$ shows a dispersion of values of almost one order of magnitude.

Finally, we consider the calibration from \citet{Perez-Montero_2009} to estimate nitrogen abundances. As shown in Fig. \ref{Offset_NO}, there is a good agreement between this calibration and determinations from \textsc{HCm}, with a median offset of 0.04 dex and RSME of 0.07 dex.

	\section{Comparison between different methods to derive oxygen abundances in AGN}
\label{s: A}

\begin{figure*}
	%\resizebox{\hsize}{!}
	\begin{tabular}{cccc}
		\begin{minipage}{0.05\hsize}\begin{flushright}\textbf{(a)} \end{flushright}\end{minipage}  &  \begin{minipage}{0.4\hsize}\centering{\includegraphics[width=1\textwidth]{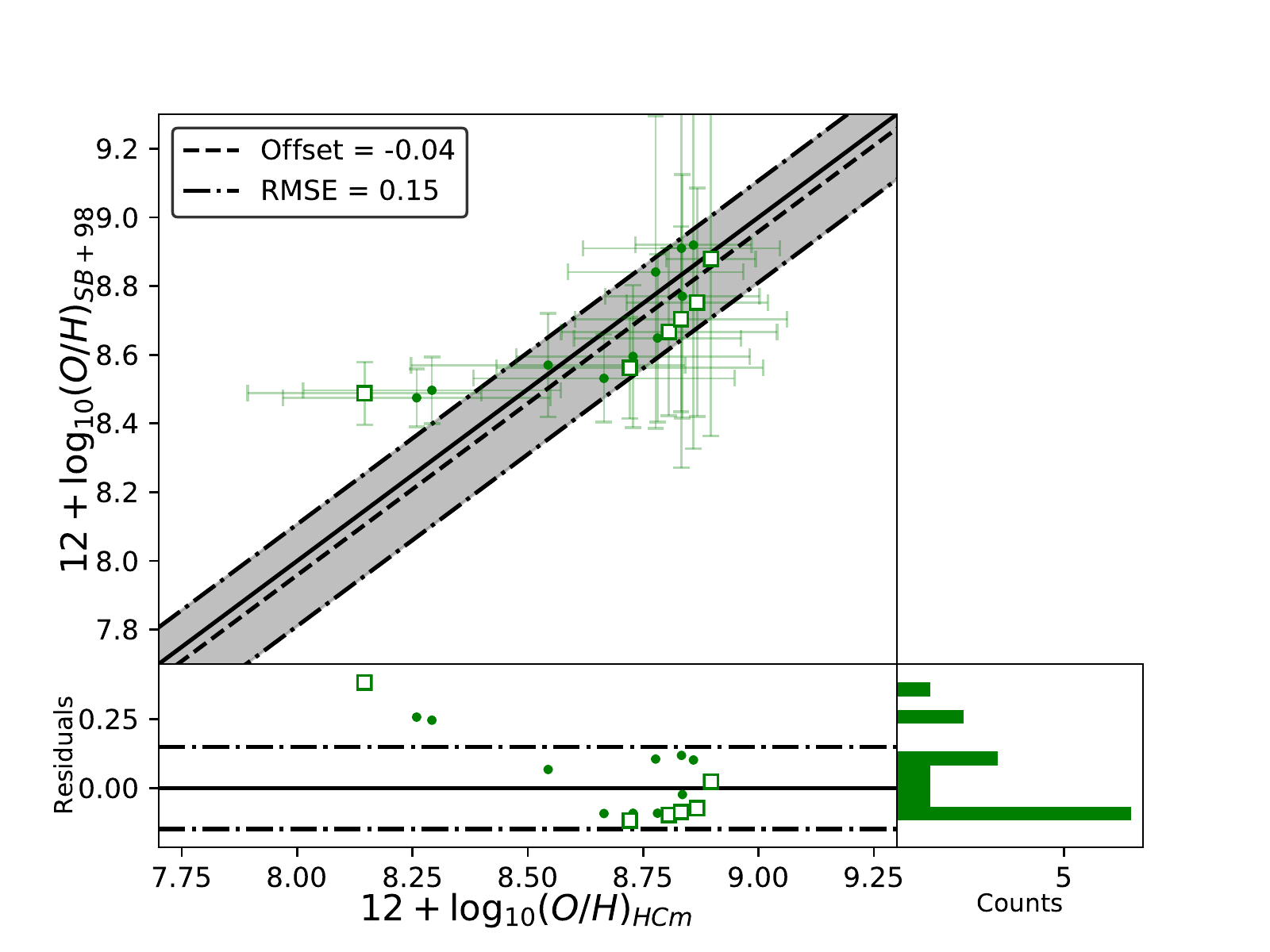}} \vspace{-0.2in} \end{minipage} & \begin{minipage}{0.05\hsize}\begin{flushright}\textbf{(b)} \end{flushright}\end{minipage}  &  \begin{minipage}{0.4\hsize}\centering{\includegraphics[width=1\textwidth]{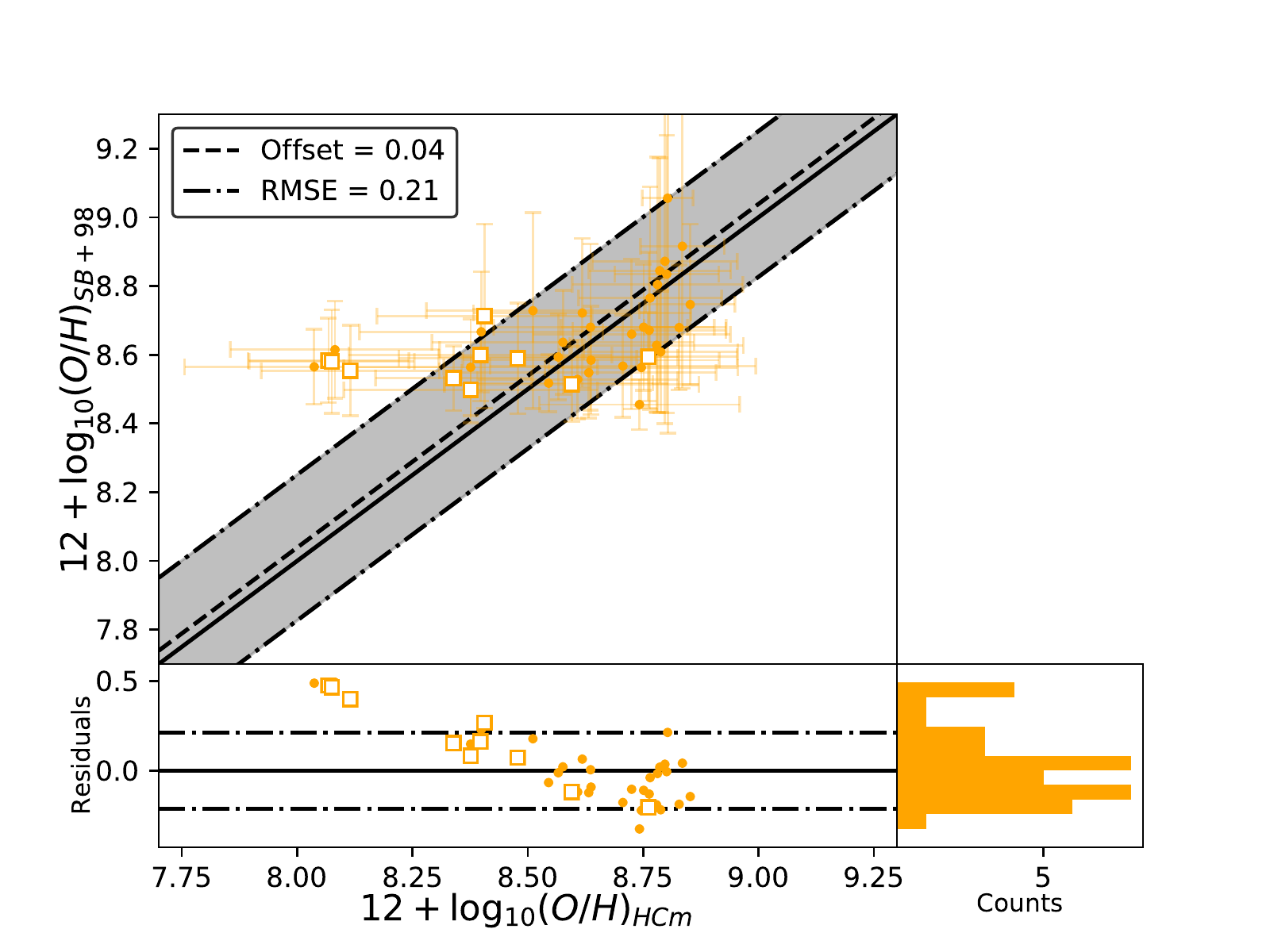}} \vspace{-0.2in} \end{minipage} 
	\end{tabular}
	\caption{Comparison of chemical abundances derived with the calibration from \citet{Storchi_1998} (y-axis), denoted as SB+98, with \textsc{HCm} (x-axis). Square-empty symbols indicate galaxies initially classified as ambiguous (see Sec. \ref{subsec23}). (a) Results for Seyferts 2 (green). (b) Results for LINERs (orange). The offsets are given using the median value (dashed line) and RMSE (dot-dashed lines). Bottom plots show the residuals from the offset and their distribution in a histogram (bottom-right plot).}
	\label{Offset_SB98}
\end{figure*}

\begin{figure*}
	%\resizebox{\hsize}{!}
	\begin{tabular}{cccc}
		\begin{minipage}{0.05\hsize}\begin{flushright}\textbf{(a)} \end{flushright}\end{minipage}  &  \begin{minipage}{0.4\hsize}\centering{\includegraphics[width=1\textwidth]{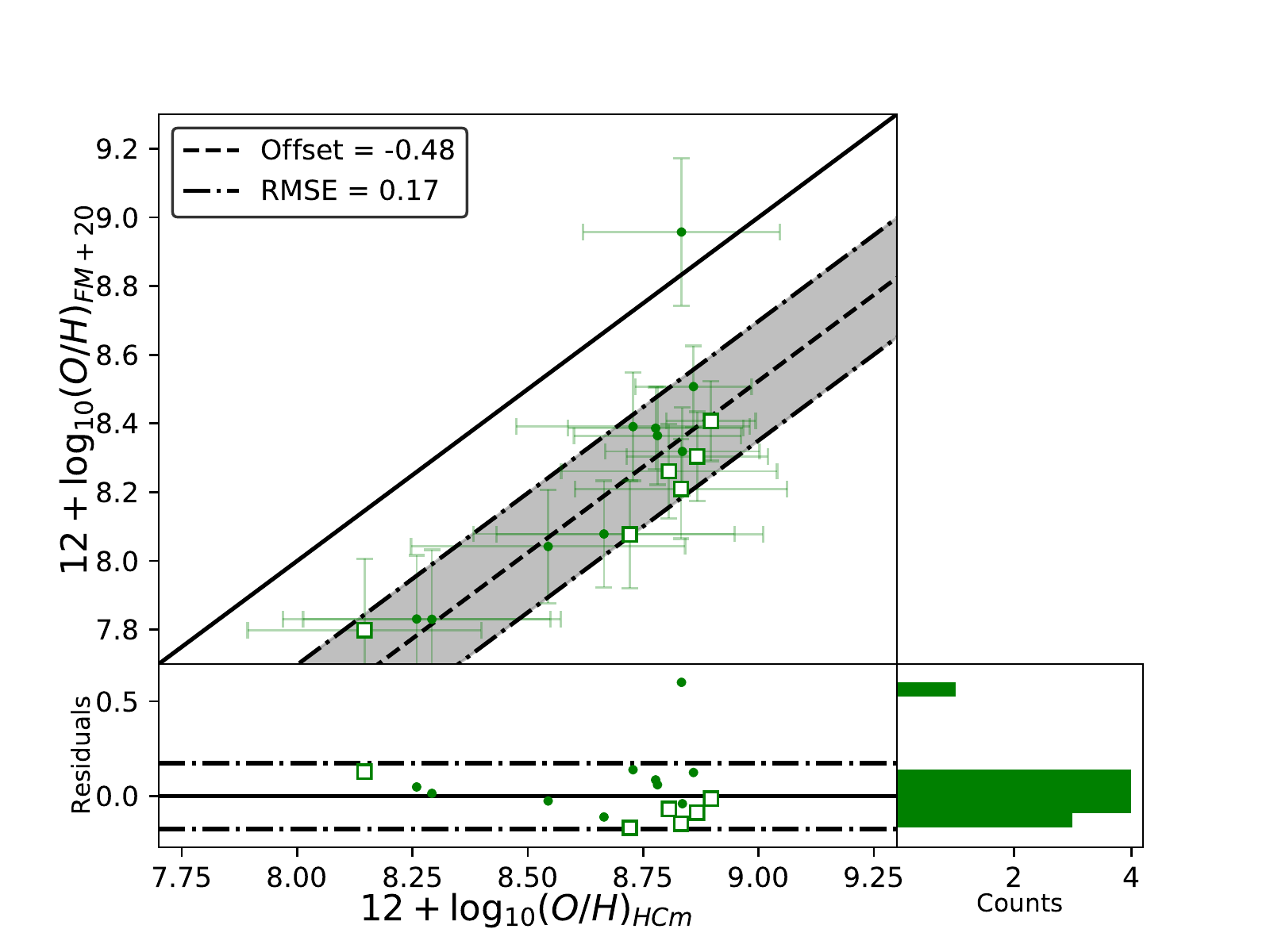}} \vspace{-0.2in} \end{minipage} & \begin{minipage}{0.05\hsize}\begin{flushright}\textbf{(b)} \end{flushright}\end{minipage}  &  \begin{minipage}{0.4\hsize}\centering{\includegraphics[width=1\textwidth]{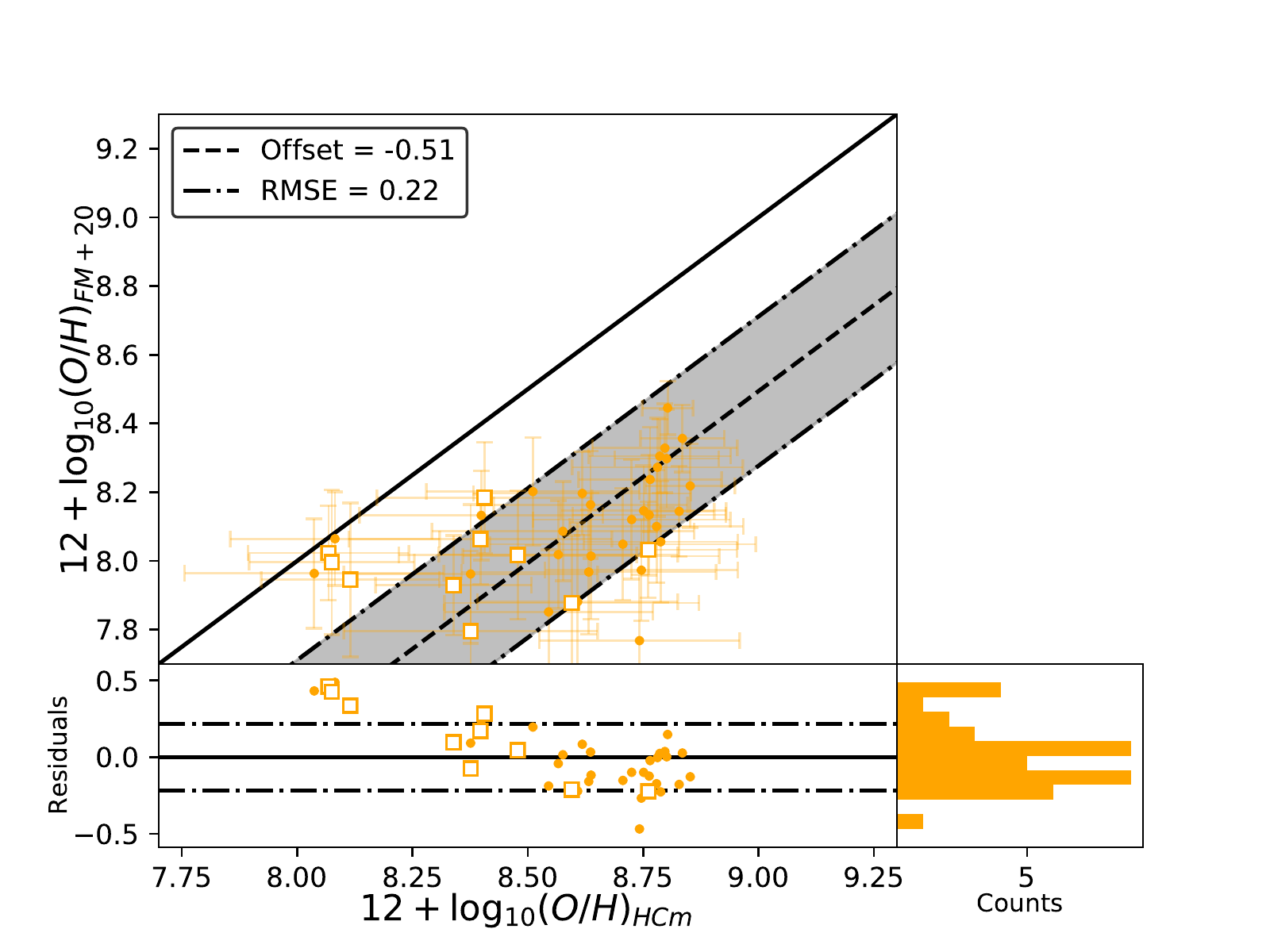}} \vspace{-0.2in} \end{minipage} 
	\end{tabular}
	\caption{Same as Fig. \ref{Offset_SB98} but the y-axis present the estimations using the calibration from \citet{Flury_2020} (y-axis), denoted as FM+20.}
	\label{Offset_FM20}
\end{figure*}

Here we compare our results for the chemical abundances obtained from \textsc{HCm} for AGN with those obtained using other techniques. \citet{Dors_2019} compare the results from $T_{e}$-method, calibrators such as those proposed by \citet{Storchi_1998} or \citet{Castro_2017} and photoionization models (\textsc{HCm}) in a sample of Seyferts 2 from the Sloan Digital Sky Survey. We extend this analysis for our sample of AGN, including both Seyferts 2 and LINERs.

We consider different semi-empirical calibrations for estimating the metallicity based on strong emission lines. Firstly, we use the calibration proposed by \citet{Storchi_1998}, using the emission lines H$_{\beta }$, [O\textsc{iii}]$\lambda$5007\r{A}, H$_{\alpha }$ and [N\textsc{ii}]$\lambda$6584\r{A}:
\begin{equation}
\label{A1}
\begin{aligned}
12+\log \left( O/H \right) _{SB+98} = & 8.34  + 0.212 x - 0.002 y + 0.007 xy  \\ & - 0.012 x^{2} + 6.52\cdot 10^{-4} y^{2} - 0.002 x^{2} y \\ &  + 2.27\cdot 10^{-4} xy^{2} + 8.87 \cdot 10^{-5} x^{2}y^{2}
\end{aligned}
\end{equation}
where $x =  I \left( \right. $ [N\textsc{ii}]$\lambda$6584\r{A}/H$_{\alpha } \left. \right) $ and $y = I \left( \right. $ [O\textsc{iii}]$\lambda$5007\r{A}/H$_{\beta } \left. \right) $. \textsc{HCm} considers grids of models for AGN with an electronic density $n_{e} \sim 500 \ \mathrm{cm}^{-3}$ \citep{Perez-Montero_2019}, so we apply an offset of $-0.022$ to account for the density correction \citep{Storchi_1998}. The validity of this calibration is set in the range $8.4 \leq 12 + \log \left( O/H \right) \leq 9.4$ \cite{Storchi_1998}.

We also use the modification to Eq. \ref{A1} given by \citet{Flury_2020}, who consider a more accurate photoionization model accounting for O$^{+3}$ and use a larger sample of galaxies. This modification is given by:
\begin{equation}
\label{A2}
\begin{aligned}
12 +  \log \left( O/H \right) _{FM+20} = & 7.863 + 1.170u + 0.027v - 0.406uv \\ & - 0.369u^{2} + 0.208v^{2} + 0.354u^{2}v \\ & - 0.333uv^{2} - 0.100u^{3} + 0.323v^{3}
\end{aligned}
\end{equation}
where $u = \log \left( I \left( \right. \right. $[N\textsc{ii}]$\lambda$6584\r{A}/H$_{\alpha } \left. \left. \right) \right) $ and $v = \log \left( I \left( \right. \right. $ [O\textsc{iii}]$\lambda$5007\r{A}/H$_{\beta } \left. \left. \right) \right) $. This calibration is valid for $7.5 \leq 12+ \log \left( O/H \right) \leq 9.0$ \citep{Flury_2020}.

\citet{Carvalho_2020} found a correlation between the emission line ratio [N\textsc{ii}]$\lambda$6584\r{A}/H$_{\alpha }$ and the metallicity Z in Seyferts 2, given by:
\begin{equation}
\notag Z/Z_{\odot } = a^{N2} + b
\end{equation}
where $N2 =  \log \left( I \left( \right. \right. $([N\textsc{ii}]$\lambda$6584\r{A}/H$_{\beta } \left. \left. \right) \right) $ , $a = 4.01 \pm 0.08$ and $b = -0.07 \pm 0.01$. In terms of the oxygen abundance, the calibration is given by\footnote{We assume here the solar abundances from \citet{Asplund_2009}}:
\begin{equation}
\label{A4} 12 + \log \left( O/H \right) _{Ca+20} = 8.69 - \log \left( 4.01^{N2} - 0.07 \right)
\end{equation}

Fig. \ref{Offset_SB98} shows a good agreement between the estimations from \textsc{HCm} and the calibration given by \citet{Storchi_1998} for both AGN, Seyferts 2 and LINERs. Although there are some discrepancies at low chemical abundances, these mainly appear for values from \textsc{HCm} of O/H below the limit of validity for Eq. \ref{A1}.

There is an offset ($\sim -0.5$ dex) between the estimations from the calibration by \citet{Flury_2020} and from \textsc{HCm}, which is present in both Seyferts 2 and LINERs (see Fig. \ref{Offset_FM20}) with RMSE of 0.17 and 0.22 dex respectively. \citet{Flury_2020} calculate their calibration from models where they apply the $T_{e}$-method in a model accounting for O$^{+3}$. Therefore, they use a sample of AGN that can be eventually biased to low metallicities since it is derived from galaxies with the detection of the auroral line [O\textsc{iii}]$\lambda$4363\r{A} This can be one possible reason for the systematic lower chemical abundances ($\sim$ -0.5 dex) obtained with Eq. \ref{A2}.

\begin{figure*}
	%\resizebox{\hsize}{!}
	\begin{tabular}{cccc}
		\begin{minipage}{0.05\hsize}\begin{flushright}\textbf{(a)} \end{flushright}\end{minipage}  &  \begin{minipage}{0.4\hsize}\centering{\includegraphics[width=1\textwidth]{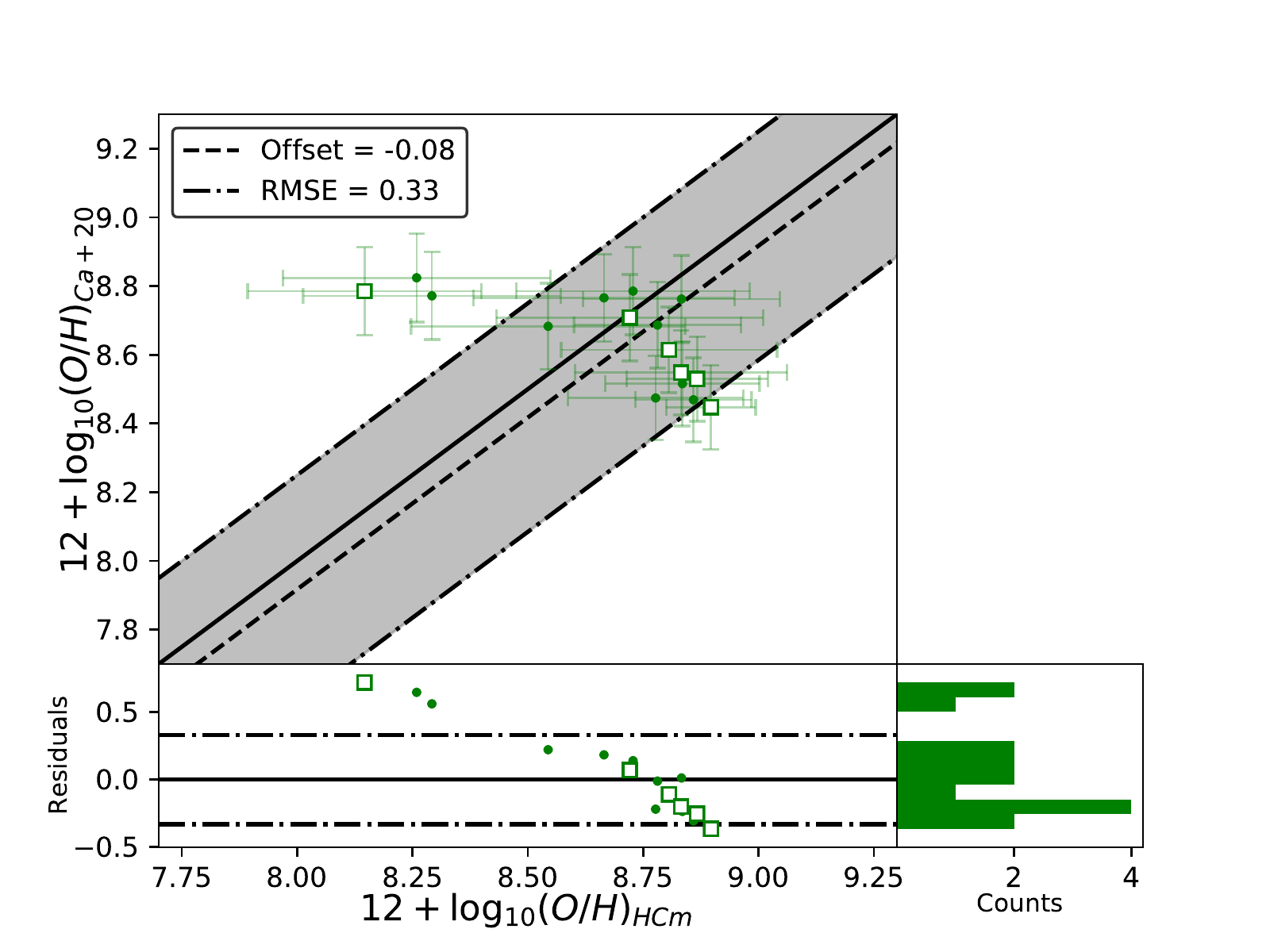}} \vspace{-0.2in} \end{minipage} & \begin{minipage}{0.05\hsize}\begin{flushright}\textbf{(b)} \end{flushright}\end{minipage}  &  \begin{minipage}{0.4\hsize}\centering{\includegraphics[width=1\textwidth]{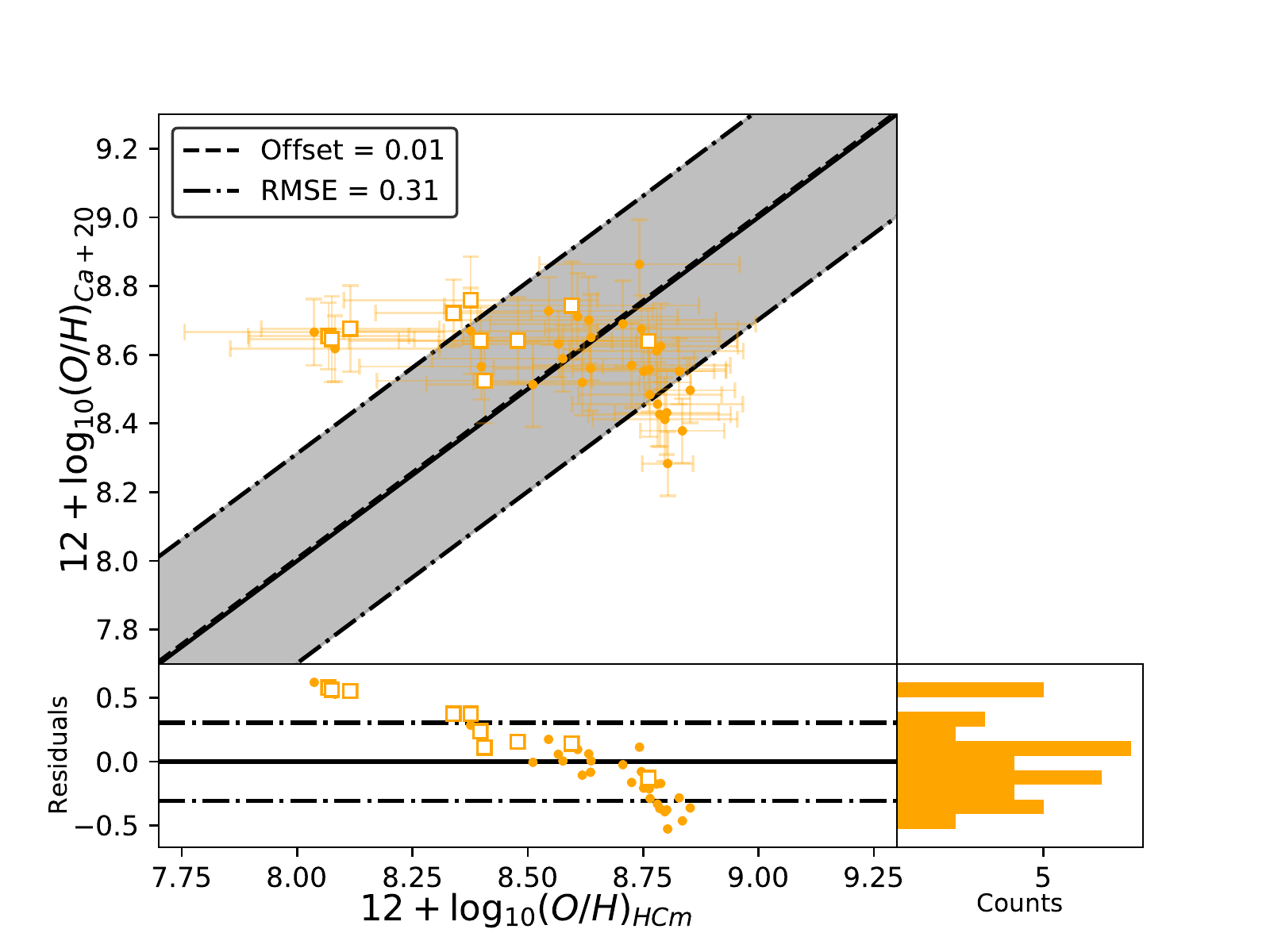}} \vspace{-0.2in} \end{minipage} 
	\end{tabular}
	\caption{Same as Fig. \ref{Offset_SB98} but the y-axis present the estimations using the calibration from \citet{Carvalho_2020} (y-axis), denoted as Ca+20.}
	\label{Offset_Ca20}
\end{figure*}

In Fig. \ref{Offset_Ca20} we observe that the estimations from Eq. \ref{A4} are in poor agreement with median offsets of -0.08 dex (Seyferts 2) and 0.01 dex (LINERs) and RMSE $\sim 0.32$ dex for both Seyferts 2 and LINERs (being the highest obtained in this study). As in the case of Fig. \ref{Offset_SB98}, the discrepancies are more important at low-chemical abundances. While in the case of Storchi-Bergmann's calibration \citeyearpar{Storchi_1998} this might be explained due to the validity range of the calibration, this is not the case for Carvalho's calibration \citeyearpar{Carvalho_2020}. It must be noticed that \citet{Carvalho_2020} assume a fixed relation between O/H and N/O, while \textsc{HCm} estimates them independently.

To end this comparison, we also use \textsc{NebulaBayes} \citep{Thomas_2018, Thomas_2019}, which estimates the chemical abundance ratio $12+\log \left( O/H \right) $ and the ionization parameter $\log \left( U \right) $ from a grid of models generated with \textsc{MAPPINGS} \citep{Sutherland_2017}. One important difference between \textsc{HCm} and \textsc{NebulaBayes} is that the models used by \textsc{HCm} assume $\alpha_{OX} = -0.8$ (see Sec. \ref{subsec25}), while \textsc{NebulaBayes} models are obtained for $\alpha_{OX} = -2.0$ \citep{Thomas_2018}. From \textsc{NebulaBayes}, we obtain high chemical abundances, clustering at $12+\log \left( O/H \right) \sim 9.1$, adding and offset of almost 0.8 dex in comparison with all the methods discussed above. In addition, the ionization parameter estimated from \textsc{NebulaBayes} results to be in the range $\left[ -3.6, -2.8 \right] $, more than 1 dex below the values obtained from \textsc{HCm} which are in good agreement with previous studies \citep{Ho_VI_2003, Kewley_2006}. 

%%%%%%%%%%%%%%%%%%%%%%%%%%%%%%%%%%%%%%%%%%%%%%%%%%

	\section{Data}
	\label{s: B}

\begin{landscape}	
	\begin{table}
		\caption{List of host galaxy properties used in the study of the Palomar Survey. Column (2): spectral type accoding to the classification in Sec. \ref{subsec23} (the * stands for galaxies initially classified as \textit{ambiguous}). Column (3): morphological type. Column (4): distance. Column (5): radius observed with the aperture of 4''. Column (6): stellar velocity dispersion in the galactic centers. Column (7): mass of the Supermassive Black Hole calculated by Eq. \ref{2}. We denote with ! the mass of the black hole of NGC 185 as explained in Sec. \ref{subsec45}. Column (8): galactic extinction. Column (9): apparent B-magnitude. Column (10): absolute B-magnitude corrected for galactic extinction. Columns (11) and (12): apparent magnitudes in the WISE-bands $W_{1}$ (at $\lambda = 3.4 \ \mu\mathrm{m}$) and $W_{2}$ (at $\lambda = 4.6 \ \mu\mathrm{m}$), respectively. Columns (13) and (14): absolute magnitudes in the WISE-bands. Column (15): stellar mass derived from Eq. \ref{3}. Columns (3), (4), (8) and (9) retrieved from \citet{Ho_III_1997}. Column (6) retrieved from \citet{Ho_VII_2009}. Columns (11) and (12) retrieved from the \textsc{Infrared Science Archive}. Full table is available online.}\label{TabA1}
		\centering
		\begin{tabular}{lcccccccccccccc}
			\textbf{Name} & \textbf{Type} & \textbf{T} & \textbf{d (Mpc)} & \textbf{r (pc)} & \boldmath$\sigma_{*} \ \left( \mathrm{km}\cdot\mathrm{s}^{-1} \right) $ & \boldmath$\log_{10} \left( M_{\mathrm{BH}} \left[ M_{\odot } \right] \right) $ & \boldmath$A_{g}$ & \boldmath$m_{B}$ & \boldmath$M_{B}^{0}$ & \boldmath$w_{1}$ & \boldmath$w_{2}$ & \boldmath$W_{1}$ & \boldmath$W_{2}$ & \boldmath$\log_{10} \left( M_{*} \left[ M_{\odot } \right] \right) $   \\ \textbf{(1)} & \textbf{(2)} & \textbf{(3)} & \textbf{(4)} & \textbf{(5)} & \textbf{(6)} & \textbf{(7)} & \textbf{(8)} & \textbf{(9)} & \textbf{(10)} & \textbf{(11)} & \textbf{(12)} & \textbf{(13)} & \textbf{(14)} & \textbf{(15)} \\ \hline
			IC 10 & SFG & 10.0 & 1.3 & 12.6 & 35.5 & - & 3.50 & 11.80 & -18.37 & - & - & - & - & - \\
			IC 1727 & LIN* & 9.0 & 8.2 & 79.5 & 136.8 & 7.39 & 0.27 & 12.07 & -17.85 & 14.68 & 14.64 & -14.89 & -14.93 & 6.98 \\
			IC 2574 & SFG & 9.0 & 3.4 & 33.0 & 33.9 & - & 0.07 & 10.80 & -16.95 & 15.67 & 15.46 & -11.99 & -12.20 & 5.39 \\
			NGC 185 & LIN & -5.0 & 0.7 & 6.8 & 19.9 & 2.67!  & 0.83 & 10.10 & -15.22 & 11.48 & 11.42 & -12.75 & -12.80 & 6.09 \\
			NGC 315 & LIN & -4.0 & 65.8 & 638.0 & 303.7 & 9.34 & 0.28 & 12.20 & -22.26 & 9.44 & 9.40 & -24.66 & -24.70 & 10.89 \\
		\end{tabular}
	\end{table}

	\begin{table}
		\caption{Emission line ratios for our sample of galaxies from the Palomar Survey corrected for reddening following the methodology in Sec. \ref{subsec24}. Column (2): extinction coefficient. Columns (3)-(7): emission line ratios corrected for reddening referred to the Balmer line H$_{\beta }$. Full table is available online.}\label{TabA2}
		\centering
		\begin{tabular}{lcccccc}
			\textbf{Name} & \boldmath$c \left( H_{\beta } \right) $ & \boldmath$\log_{10} \left[ L \left( H_{\alpha } \right) \right] $ & \textbf{[O\textsc{iii}]}\boldmath$ \lambda 4363$ \textbf{\r{A}} &  \textbf{[O\textsc{iii}]}\boldmath$ \lambda 5007$ \textbf{\r{A}} &  \textbf{[N\textsc{ii}]}\boldmath$ \lambda 6584$ \textbf{\r{A}} & \textbf{[S\textsc{ii}]}\boldmath$ \lambda \lambda 6717, 6731$ \textbf{\r{A}}  \\ 
			\textbf{(1)} & \textbf{(2)} & \textbf{(3)} & \textbf{(4)} & \textbf{(5)} & \textbf{(6)} & \textbf{(7)}  \\ \hline
			
			IC 10 & 2.0$\pm$0.3 & 38.95 & - & 3.7$\pm$1.1 & 0.11$\pm$0.04 & 0.15$\pm$0.05 \\
			IC 1727 & 0.7$\pm$0.3 & 38.17 & - & 2.2$\pm$0.6 & 2.9$\pm$1.0 & 3.3$\pm$1.0 \\
			IC 2574 & 0.9$\pm$0.3 & 37.56 & - & 0.21$\pm$0.06 & 0.20$\pm$0.07 & 0.7$\pm$0.2 \\
			NGC 185 & 0.5$\pm$0.3 & 35.06 & - & 3.1$\pm$0.9 & 1.9$\pm$0.7 & 4.5$\pm$1.4 \\
			NGC 315 & - & 39.55 & - & 2.1$\pm$0.6 & 8$\pm$2 & 4.6$\pm$1.1 \\
		\end{tabular}
	\end{table}

	\begin{table}
		\caption{Chemical abundances and ionization parameters for our sample of galaxies from the Palomar Survey using the code \textsc{HCm}. Column (2) shows the spectral type also listed in Tab. \ref{TabA1}. Full table is available online. }\label{TabA3}
		\centering
		\begin{tabular}{lcccc}
			\textbf{Name} & \textbf{Type} & \boldmath$12+\log_{10} \left( O/H \right) $ & \boldmath$\log_{10} \left( N/O \right) $ & \boldmath$\log_{10} \left( U \right) $ \\ \textbf{(1)} & \textbf{(2)} & \textbf{(3)} & \textbf{(4)} & \textbf{(5)} \\ \hline
			
			IC 10 & SFG & 7.858$\pm$0.15 & -1.0$\pm$0.2 & -2.45$\pm$0.13 \\
			IC 1727 & LIN* & 8.6$\pm$0.3 & -0.8$\pm$0.4 & -3.41$\pm$0.15 \\
			IC 2574 & SFG & 8.73$\pm$0.09 & -1.35$\pm$0.18 & -3.27$\pm$0.13 \\
			NGC 185 & LIN & 8.7$\pm$0.2 & -1.1$\pm$0.3 & -3.35$\pm$0.13 \\
			NGC 315 & LIN & 8.79$\pm$0.15 & -0.5$\pm$0.2 & -3.46$\pm$0.09 \\
		\end{tabular}
	\end{table}

	\begin{table}
		\caption{Chemical abundances and ionization parameters for sample of LINERs from \citet{Povic_2016} using the code \textsc{HCm}. Column (2) shows the absolute B-magnitude corrected from reddening calculated in this work. Column (3) shows stellar mass calculated using Eq. \ref{3}. Full table is available online.}\label{TabA4}
		\centering
		\begin{tabular}{lccccc}
			\textbf{Name} & \boldmath$M_{B}^{0}$ & \boldmath$\log_{10} \left( M_{*} \left[ M_{\odot } \right] \right) $ & \boldmath$12+\log_{10} \left( O/H \right) $ & \boldmath$\log_{10} \left( N/O \right) $ & \boldmath$\log_{10} \left( U \right) $ \\ \textbf{(1)} & \textbf{(2)} & \textbf{(3)} & \textbf{(4)} & \textbf{(5)} & \textbf{(6)} \\ \hline
			
			F01 & -20.30 & 8.78$\pm$0.10 & 8.24$\pm$0.14 & -0.32$\pm$0.16 & -3.33$\pm$0.15 \\
			F03 & -20.64 & 10.51$\pm$0.11 & 7.9$\pm$0.3 & -0.8$\pm$0.3 & -3.42$\pm$0.08 \\
			F04 & -19.69 & 10.46$\pm$0.13 & 8.6$\pm$0.3 & -0.7$\pm$0.2 & -3.51$\pm$0.10 \\
			F06 & -19.65 & 9.90$\pm$0.10 & 8.58$\pm$0.19 & -0.4$\pm$0.2 & -3.53$\pm$0.13 \\
			F13 & -19.50 & 9.76$\pm$0.12 & 8.1$\pm$0.2 & -0.6$\pm$0.3 & -3.4$\pm$0.2 \\
		\end{tabular}
	\end{table}
\end{landscape}

% Don't change these lines
\bsp	% typesetting comment
\label{lastpage}
\end{document}